\documentclass[apj,appendixfloats,numberedappendix]{emulateapj}

\usepackage{wasysym}
\usepackage{rotating}
\usepackage[dvips]{color}
\usepackage{verbatim}
\usepackage{soul}
\usepackage{hyperref}
\usepackage{subfigure}
\usepackage[T1]{fontenc} 
\usepackage{tipa} 

\shorttitle{Resolving the Geometry of the Innermost Relativistic Jets in AGN}

\shortauthors{Algaba et al.}

\begin{document}

\title{Resolving the Geometry of the Innermost Relativistic Jets in Active Galactic Nuclei}

\author{J. C. Algaba$^{1}$, M. Nakamura$^2$, K. Asada$^2$ and S. S. Lee$^1$}
\affil{$^1$Korea Astronomy \& Space Science Institute, 776, Daedeokdae-ro, Yuseong-gu, Daejeon, Republic of Korea 305-348 \url{algaba@kasi.re.kr}\\$^2$Academia Sinica, Institute of Astronomy and Astrophysics, 11F of Astronomy-Mathematics Building, AS/NTU. No.1, Sec. 4, Roosevelt Rd, Taipei 10617, Taiwan, R.O.C}

\begin{abstract}
In the current paradigm, it is  believed that the compact VLBI radio core of radio--loud AGNs represents the innermost upstream regions of relativistic outflows. These regions of AGN jets have generally been modelled by a conical outflow with roughly constant opening angle and flow speed. Nonetheless, some works suggest that a parabolic geometry would be more appropriate to fit the high energy spectral energy distribution properties and it has been recently found that, at least in some nearby radio--galaxies, the geometry of the innermost regions of the jet is parabolic.  We compile here multi--frequency core sizes of archival data to investigate the typically unresolved upstream regions of the jet geometry of a sample of 56 radio--loud AGNs.  Data combined from the sources considered here is not consistent with the classic picture of a conical jet starting in the vicinity of the super--massive black hole (SMBH), and may exclude a pure parabolic outflow solution, but rather suggest an intermediate solution with  quasi-parabolic streams, which are frequently seen in numerical simulations. Inspection of the large opening angles near the SMBH and the range of the Lorentz factors derived from our results support our analyses. Our result suggests that the conical jet paradigm in AGNs needs to be re-examined by mm/sub-mm VLBI observations.

\end{abstract}


\section{Introduction}

Extragalactic jets in active galactic nuclei (AGNs) are believed to be formed in the vicinity of accreting super--massive black holes (SMBHs) with $10^7 - 10^{10} M_\odot$. Magnetohydrodynamics (MHD) jet theory suggests that the differential rotation of the accretion flow and/or the frame dragging in the black hole ergosphere twist the poloidal magnetic field in the azimuthal direction so that the torsional Alfv\'en waves (TAWs) are excited continuously. Energy is then extracted from the accreting SMBHs in the form of the Poynting flux via the propagation of TAWs and the generated toroidal magnetic field is responsible for the jet bulk acceleration and collimation beyond the light cylinder \citep[e.g.][]{Begelman84,Meier01,Meier12}. Recent analytical and numerical examinations suggest that the external pressure plays an important role in shaping the jet geometry \citep{Zakamska08,Komissarov07,Komissarov09,Lyubarsky09}. The jet geometry can thereafter be approximated by a conical or paraboloidal expansion, with transverse radius $R\propto r^{\epsilon}$ ($r$: distance from the nucleus) if the jet is confined by the external medium with a power--law profile as a function of distance. The jet half opening angle $\theta_j=\arctan(R/r)$ is thus constant if $\epsilon=1$ (conical; the jet is ballistic) or $\theta_j$ decreases if $0 < \epsilon < 1$ (parabolic; the jet is being collimated). Note $\epsilon = 0$ corresponds to the cylindrical jet, in which we do not consider that the jet is under collimation in the present paper.

\cite{Ghisellini85} proposed an emission region with inner parabolic structure (to account for emission from infrared to X-ray bands) becoming conical (in order to explain the flat radio spectrum of several sources) as it reached outer regions, consistent with synchrotron self-Compton (SSC) physical models of the jet. They successfully applied their model to explain the spectral energy distribution (SED) of the objects PKS 2155-034 and PKS 0537-441. Although they demonstrated the importance of considering how and where the jet structural transition occurs, this puzzle has remained unsolved for decades, partly due to the large difficulties in resolving the transverse morphology of jets.

Assuming a conical structure, \cite{Jorstad05} found that a constant opening angle was a good approximation for the inner jet, although two BL Lac objects (3C~66A and 1803+784) displayed an increase of the opening angle on milli-arc second scales. Only recently, \cite{Asada12} found striking evidence of a transition from parabolical to conical shape in the M87 jet, maintaining the parabolic morphology upstream the jet for about $10^5$ Schwarzschild radii ($r_s$), up to the \mbox{HST-1} component. In \cite{Nakamura13}, they suggested that the MHD jet is initially confined by the balance between internal magnetic and external gas pressure within the region dominated by the super massive black hole potential, whereas beyond the sphere of gravitational influence (SGI) of the SMBH, the jet can freely adiabatically expand with a conical shape. Similar result is found very recently in NGC~6251 \citep{Tseng16}, suggesting that the jet structural transition at around $\sim 10^5\, r_{\rm s}$, where the SGI is located for the SMBH with $10^9-10^{10}\, M_{\odot}$, may be a fundamental process in AGN jets.

Recently, \cite{Potter13} developed a jet model which consists of a magnetically dominated parabolic jet under the bulk acceleration, but it transits into conical stream with a slow deceleration at $\sim 10^5\, r_{\rm s}$, inspired by recent M87 results \cite{Asada12,Nakamura13}. Their model successfully fits simultaneous multi-wavelength observations of the radio quasar PKS~0227--369, including low frequency radio bands. They suggested that one of the implications may be that the jet of PKS~0227--369 could be described by the same jet geometry as M87. However, if this is the typical trend for most AGNs and if M87 turns out to be the paradigm is still unclear. Unfortunately, other AGNs are not well resolved and it is currently very difficult to directly estimate the transverse morphology of the jet. 

An alternative method, which has already been successfully used in \cite{Nakamura13} and \cite{Hada13} consists on utilizing a combination of both the VLBI core shift \citep{Hada11} and size to determine the upstream characteristics of the jet and find out if the acceleration zones coincides with the collimation region \citep{Nakamura13,Asada14}. Although this method is typically limited by sparse data distributed over few limited orders of magnitude, it is nonetheless capable of probing on scales that would be unresolved and thus unavailable otherwise.

In this paper, we conduct an effort to explore the innermost structures of AGN jets with the VLBI cores in order to examine the classical conical jet hypothesis, which has been widely considered during decades. We will follow this latter approach. We  consider archival data containing information of core sizes at various frequencies, and combine this with core shift information to investigate the jet morphology for a sample of 56 sources. For the first time, we will be able to discuss the results of this methodology in a statistical sense, and analyze the physics of a whole sample. The contents is organized as follows: Section 2 describes the methodology for this study, Section 3 includes a discussion and Section 4 summarizes the conclusions.

\section{Methodology and Data Analysis}

Although one may think that the innermost regions of the distant blazar jets are not accessible for unresolved sources, we can still obtain valuable information from their VLBI cores. In the standard picture, the compact VLBI radio core is believed to correspond to the throat of a diverging jet \cite{BK79}. Alternatively, \cite{Daly88} proposed that the VLBI core can be identified as the first re--collimation shock. With this hypothesis, properties of the VLBI core can be attributed to the upstream regions of the jet.

At a given frequency, the VLBI core is considered to be located in the regions upstream the jet where the optical depth is $\tau=1$\footnote{At short mm-wavelengths the core may also be the first recolimation shock downstream of the $\tau=1$ surface instead of the surface itself \citep{MarscherGear85}. This does not affect our analysis, which uses longer wavelengths.}. In this case, the absolute position of the core would have a frequency dependence in the form $r_c=\nu^{-1/k_r}$ \citep{BK79,Konigl81}, where the value of $k_r$ depends on the electron energy spectrum and on both the magnetic field and particle density distributions in the jet. This means that the apparent position of the VLBI core shifts away from the central engine at lower and lower frequencies (so-called ``core shift'' effect). In the most general case, as we investigate higher frequencies, we will be probing core (=jet) properties, such as its (deconvolved) size, closer to the central engine. In particular, this means that measuring the core sizes as various frequencies, we can investigate the jet size, and hence its geometry via the core shift effect. 

We have looked into surveys in order to find information of VLBI core sizes at various frequencies. The available datasets are summarized in Table \ref{surveys}, where we indicate the frequency, instrument, epoch(s) of observation and references. We note that various surveys are not simultaneous and indeed several years can span among them, meaning that values obtained for a particular source at different frequencies will not be simultaneous.  Given that core shifts and/or core sizes may be subject to variability, we will treat our results with caution (see discussion below). 

\begin{table}
\begin{center}
\caption{VLBI surveys used in this work.\label{surveys}}
\begin{tabular}{cccc}
\tableline
\hline
Freq. (GHz) & Instrument & Epoch & References$^1$\\
(1)&(2)&(3)&(4)\\
\tableline
1.6 & 11-16$\times$VLBI$^2$ & 1990-1991 & P95,TH95\\
2.3 & VLBA &  1998-2003 & P12\\ 
5.0 & VSOP & 1997-2002 & S04,D08\\
8.6 & VLBA & 1998-2003 & P12\\ 
15 & VLBA & 1994-2003 & L05\\ 
22 & 6$\times$VLBI$^2$ & 1993 &M96\\
86 & GMVA & 2001-2002 & L08\\
\tableline
\end{tabular}
\end{center}
$^1$P95: \cite{Polatidis95}; TH95: \cite{Thakkar95}; P12: \cite{Pushkarev12b}; S04: \cite{Scott04}; D08: \cite{Dodson08}; L05: \cite{Lister05}; M96: \cite{Moellenbrock96}; L08: \cite{Lee08}\\
$^2$\footnotesize{A number of antennas participating in the VLBI sessions is indicated. See references for details.}\\
\end{table}

In general, most authors fit for the core size using an elliptical gaussian and find for the full width at half-maximum (FWHM) of its major ($\theta_{max}$) and minor ($\theta_{min}$) axis \citep{Pushkarev12,Lister05,Dodson08,Polatidis95}. \cite{Lee08} fits for a circular gaussian. \cite{Moellenbrock96} show the core flux $S_{core}$ and the brightness temperature $T_b$, so we solved for the size of the core using \citep[see e.g.][]{Kovalev05}
\begin{equation}
T_{b,core}=\frac{2\ln 2}{\pi k}\frac{S_{core}\lambda^2(1+z)}{\theta_{max} \theta_{min}}K,
\end{equation}
where $\lambda$ is the wavelength of observation, $z$ is the redshift, and $k$ is the Boltzmann constant. For simplicity in this case we set $\theta_{max}=\theta_{min}$. Whenever data for $T_b$, core size and core fluxes were available, we checked for consistency.

In order to obtain the transverse size of the upstream jet based on the core size we proceeded as follows. In the case of a circular gaussian, we simply took  its FWHM value as the width of the jet. In the case of an elliptical gaussian, we searched for the jet direction at 15~GHz from MOJAVE data \citep{Lister13} and obtained the FWHM of the gaussian transverse to such direction. If no available information of the jet direction for a particular source was found, we took $(2\theta_{max}+\theta_{min})/3$, which is a commonly used value to approximate the mean radius of an ellipsoid. As a conservative approach, we considered only resolved VLBI cores and did not include in our study upper limits. We converted the core size  from mas to parsecs using the scale factor in \cite{Pushkarev12}. For some sources (e.g. 0420--014, 0528+134, 1546+027, 1928+738 and 2251+158) which did not include this information, we obtained the scale factor from the NASA/IPAC Extragalactic Database (NED)\footnote{https://ned.ipac.caltech.edu/}. We note that the jet radius, as defined perpendicular to the jet axis, does not suffer from projection effects.

We also estimated the location of the core from the central engine $r_{\nu}$ at each frequency for every source using  the core--shift values in \cite{Pushkarev12}. We followed the approach of their Equation~5
\begin{equation}
r_{core}(\nu)=\frac{\Omega_{r \nu}}{\nu \sin \theta}\sim   \frac{\Omega_{r \nu}(1+\beta^2_{app})^{1/2}}{\nu},
\end{equation}
using the viewing angles $\theta$ from \cite{Hovatta09} whenever possible or, alternatively, the apparent velocities discussed in \cite{Pushkarev12} and \cite{Zamaninasab14}. 
We note that the approximation in Equation~5 of \cite{Pushkarev12} assumes that the viewing angle is in general close to the critical angle, which may not be valid for a significant number of sources. Indeed, when we had information for both $\theta$ and $\beta_{app}$ and were able to compare both methods, we found an average difference of about 40\% in the resulting $r_{\nu}$, which can be directly attributed to this approximation. Nonetheless, this difference is translated into a scaling factor for all points of the same source, and will thus not affect our fitting results.


In general, the literature does not provide errors associated with the core size fitting, except for \cite{Lee08} for 86~GHz. On the other hand, for 1.6, 2.3, 8.6 and 15~GHz, information about the map root mean square $\sigma_{rms}$ and peak fluxes $S_{peak}$ are given, so we can estimate the errors of the fitted size of the core $\sigma_d$ as \citep{Fomalont99,Lee08} 

\begin{equation}
\sigma_{d}=d\frac{\sigma_{rms}}{S_{peak}}\left( 1+\frac{S_{peak}}{\sigma_{rms}} \right) ^{1/2}
\end{equation}

For 5 and 22~GHz, the information provided relates to the maximum projected baseline in terms of the sky Fourier transformed plane (i.e, in units of $\lambda$).  From there, we estimated the corresponding resolution limit and used 1/5 of the beam size in the direction of the jet as an estimation of the error for the core size \citep[see e.g.][]{Lobanov05}. On the other hand, the maximum error associated with the core shift was estimated to be of the order of 0.05~mas in \cite{Pushkarev12}, and thus we will take this errors as representative for our core location error.

\section{Results}

From an initial sample of up to 441 objects, we found appropriate values for the core size and distance for at least four frequencies in a total of 56 objects. In Table \ref{data} we summarize the source parameters. Column 2 indicate the object class (Q=quasar, B=BL Lac, R=radio galaxy, U=unidentified); column 3 indicates the redshift; columns 4 and 5 the estimated black hole mass and the reference [S14=\cite{Zamaninasab14}, W02=\cite{WooUrry02}]; columns 6--12 indicate the inferred projected distance of the core from the central engine, columns 13--19 indicate the core transverse size found following the procedure as explained above for each frequency; column 20 shows the slope of the fit, corresponding to the jet geometry parameter $\epsilon$; and column 21 shows the goodness of the fit $\mathcal{R}^2$. For the estimation of the error in $\epsilon$, the error obtained from the weighted fit and the 10\% uncertainties arising from $k_r$ were added in quadrature.

\begin{sidewaystable*}
\begin{center}
\footnotesize
\hspace{9cm}
\vspace{-9cm}
\caption{Source Parameters.\label{data}}
\scalebox{0.68}{
\begin{tabular}{ccccccccccccccccccccc}
\tableline
\hline
Source	& Class & z & $\log(M_{BH})$& Ref. & $r_{1.6}$	& $r_{2.3}$	&	$r_{5.0}$	&	$r_{8.6}$	& $r_{15}$	&	$r_{22}$	&	$r_{86}$	& $R_{1.6}$	& $R_{2.3}$	&	$R_{5.0}$	&	$R_{8.6}$	& $R_{15}$	&	$R_{22}$	&	$R_{86}$	&	$\epsilon$	&	$\mathcal{R}^2$   \\[0.01in]
(1)&(2)&(3)&(4)&(5)&(6)&(7)&(8)&(9)&(10)&(11)&(12)&(13)&(14)&(15)&(16)&(17)&(18)&(19)&(20)&(21)\\
\tableline
0003-066	&	B	&	0.347	&	 --	&	 --	&$	17.23	\pm	0.25	$&$	11.99	\pm	0.25	$&$	5.51	\pm	0.25	$&$	3.21	\pm	0.25	$&$	1.84	\pm	0.25	$&$	1.25	\pm	0.25	$&$	0.32	\pm	0.25	$&$	 -			$&$	4.59	\pm	0.13	$&$	1.23	\pm	0.27	$&$	0.91	\pm	0.02	$&$	0.91	\pm	0.02	$&$	 -			$&$	0.10	\pm	0.01	$&$	0.01	\pm	0.08	$&	0.72	\\	[0.01in]
0106+013	&	Q	&	2.107	&	 --	&	 --	&$	97.10	\pm	0.42	$&$	67.55	\pm	0.42	$&$	31.07	\pm	0.42	$&$	18.06	\pm	0.42	$&$	10.36	\pm	0.42	$&$	7.06	\pm	0.42	$&$	1.81	\pm	0.42	$&$	 -			$&$	5.85	\pm	0.11	$&$	 -			$&$	0.64	\pm	0.01	$&$	0.55	\pm	0.01	$&$	0.49	\pm	0.22	$&$	0.15	\pm	0.03	$&$	0.29	\pm	0.03	$&	1.00	\\	[0.01in]
0119+115	$^{\dagger}$&	Q	&	0.570	&	 --	&	 --	&$	404.14	\pm	0.33	$&$	281.14	\pm	0.33	$&$	129.33	\pm	0.33	$&$	75.19	\pm	0.33	$&$	43.11	\pm	0.33	$&$	29.39	\pm	0.33	$&$	7.52	\pm	0.33	$&$	 -			$&$	4.91	\pm	0.10	$&$	4.17	\pm	0.27	$&$	1.41	\pm	0.03	$&$	0.35	\pm	0.01	$&$	 -			$&$	0.10	\pm	0.01	$&$	1.38	\pm	0.29	$&	0.86	\\	[0.01in]
0133+476	&	Q	&	0.859	&	8.73	&	W02	&$	196.16	\pm	0.38	$&$	136.46	\pm	0.38	$&$	62.77	\pm	0.38	$&$	36.49	\pm	0.38	$&$	20.92	\pm	0.38	$&$	14.27	\pm	0.38	$&$	3.65	\pm	0.38	$&$	4.08	\pm	0.09	$&$	3.79	\pm	0.08	$&$	0.72	\pm	0.32	$&$	0.72	\pm	0.01	$&$	0.17	\pm	0.01	$&$	 -			$&$	0.18	\pm	0.03	$&$	1.50	\pm	0.24	$&	0.94	\\	[0.01in]
0149+218	$^{\dagger}$&	Q	&	1.320	&	 --	&	 --	&$	344.14	\pm	0.42	$&$	239.40	\pm	0.42	$&$	110.12	\pm	0.42	$&$	64.03	\pm	0.42	$&$	36.71	\pm	0.42	$&$	25.03	\pm	0.42	$&$	6.40	\pm	0.42	$&$	 -			$&$	3.23	\pm	0.15	$&$	1.26	\pm	0.35	$&$	0.29	\pm	0.01	$&$	 -			$&$	 -			$&$	0.28	\pm	0.09	$&$	1.82	\pm	0.19	$&	0.95	\\	[0.01in]
0234+285	&	Q	&	1.207	&	9.22	&	S14	&$	377.49	\pm	0.42	$&$	262.60	\pm	0.42	$&$	120.80	\pm	0.42	$&$	70.23	\pm	0.42	$&$	40.27	\pm	0.42	$&$	27.45	\pm	0.42	$&$	7.02	\pm	0.42	$&$	 -			$&$	2.71	\pm	0.06	$&$	 -			$&$	1.13	\pm	0.02	$&$	 -			$&$	0.43	\pm	0.22	$&$	0.27	\pm	0.10	$&$	0.66	\pm	0.07	$&	0.96	\\	[0.01in]
0333+321	&	Q	&	1.259	&	9.25	&	S14	&$	675.83	\pm	0.42	$&$	470.14	\pm	0.42	$&$	216.27	\pm	0.42	$&$	125.74	\pm	0.42	$&$	72.09	\pm	0.42	$&$	49.15	\pm	0.42	$&$	12.57	\pm	0.42	$&$	 -			$&$	6.80	\pm	0.17	$&$	 -			$&$	1.60	\pm	0.04	$&$	1.16	\pm	0.01	$&$	0.25	\pm	0.22	$&$	0.13	\pm	0.04	$&$	0.93	\pm	0.12	$&	0.93	\\	[0.01in]
0336-019	&	Q	&	0.852	&	8.98	&	W02	&$	161.34	\pm	0.38	$&$	112.24	\pm	0.38	$&$	51.63	\pm	0.38	$&$	30.02	\pm	0.38	$&$	17.21	\pm	0.38	$&$	11.73	\pm	0.38	$&$	3.00	\pm	0.38	$&$	 -			$&$	1.90	\pm	0.03	$&$	 -			$&$	0.38	\pm	0.01	$&$	1.00	\pm	0.01	$&$	0.84	\pm	0.20	$&$	0.26	\pm	0.04	$&$	0.39	\pm	0.34	$&	0.59	\\	[0.01in]
0420-014	&	Q	&	0.915	&	9.03	&	W02	&$	677.50	\pm	0.38	$&$	471.30	\pm	0.38	$&$	216.80	\pm	0.38	$&$	126.05	\pm	0.38	$&$	72.27	\pm	0.38	$&$	49.27	\pm	0.38	$&$	12.60	\pm	0.38	$&$	 -			$&$	2.75	\pm	0.38	$&$	1.15	\pm	0.46	$&$	 -			$&$	 -			$&$	0.69	\pm	0.20	$&$	0.18	\pm	0.05	$&$	0.72	\pm	0.10	$&	0.96	\\	[0.01in]
0430+052	$^{\dagger}$&	G	&	0.033	&	7.36	&	S14	&$	6.86	\pm	0.03	$&$	4.77	\pm	0.03	$&$	2.20	\pm	0.03	$&$	1.28	\pm	0.03	$&$	0.73	\pm	0.03	$&$	0.50	\pm	0.03	$&$	0.13	\pm	0.03	$&$	 -			$&$	0.32	\pm	0.01	$&$	0.13	\pm	0.04	$&$	0.14	\pm	0.01	$&$	0.09	\pm	0.01	$&$	0.11	\pm	0.02	$&$	0.01	\pm	0.01	$&$	0.70	\pm	0.08	$&	0.88	\\	[0.01in]
0528+134	&	Q	&	2.060	&	9.03	&	 --	&$	508.34	\pm	0.41	$&$	353.63	\pm	0.41	$&$	162.67	\pm	0.41	$&$	94.58	\pm	0.41	$&$	54.22	\pm	0.41	$&$	36.97	\pm	0.41	$&$	9.46	\pm	0.41	$&$	 -			$&$	10.15	\pm	0.23	$&$	4.38	\pm	0.34	$&$	0.58	\pm	0.01	$&$	1.12	\pm	0.01	$&$	0.54	\pm	0.21	$&$	0.13	\pm	0.01	$&$	1.13	\pm	0.44	$&	0.81	\\	[0.01in]
0552+398	&	Q	&	2.363	&	 --	&	 --	&$	2764.52	\pm	0.41	$&$	1923.15	\pm	0.41	$&$	884.65	\pm	0.41	$&$	514.33	\pm	0.41	$&$	294.88	\pm	0.41	$&$	201.06	\pm	0.41	$&$	51.43	\pm	0.41	$&$	 -			$&$	4.37	\pm	0.08	$&$	1.63	\pm	0.37	$&$	1.50	\pm	0.02	$&$	1.01	\pm	0.01	$&$	1.70	\pm	0.31	$&$	0.14	\pm	0.01	$&$	0.80	\pm	0.09	$&	0.82	\\	[0.01in]
0605-085	&	Q	&	0.872	&	8.87	&	S14	&$	94.94	\pm	0.39	$&$	66.05	\pm	0.39	$&$	30.38	\pm	0.39	$&$	17.66	\pm	0.39	$&$	10.13	\pm	0.39	$&$	6.91	\pm	0.39	$&$	1.77	\pm	0.39	$&$	 -			$&$	4.40	\pm	0.14	$&$	0.93	\pm	0.38	$&$	1.08	\pm	0.03	$&$	 -			$&$	0.34	\pm	0.20	$&$	 -			$&$	1.07	\pm	0.11	$&	0.89	\\	[0.01in]
0607-157	$^{\dagger}$&	Q	&	0.324	&	8.68	&	S14	&$	53.78	\pm	0.23	$&$	37.41	\pm	0.23	$&$	17.21	\pm	0.23	$&$	10.01	\pm	0.23	$&$	5.74	\pm	0.23	$&$	3.91	\pm	0.23	$&$	1.00	\pm	0.23	$&$	 -			$&$	2.54	\pm	0.06	$&$	 -			$&$	0.52	\pm	0.01	$&$	0.26	\pm	0.01	$&$	 -			$&$	0.20	\pm	0.02	$&$	1.15	\pm	0.21	$&	0.82	\\	[0.01in]
0716+714	&	B	&	0.310	&	 --	&	 --	&$	70.06	\pm	0.23	$&$	48.74	\pm	0.23	$&$	22.42	\pm	0.23	$&$	13.03	\pm	0.23	$&$	7.47	\pm	0.23	$&$	5.10	\pm	0.23	$&$	1.30	\pm	0.23	$&$	3.69	\pm	0.13	$&$	0.73	\pm	0.03	$&$	0.80	\pm	0.16	$&$	0.21	\pm	0.01	$&$	 -			$&$	 -			$&$	0.08	\pm	0.03	$&$	1.44	\pm	0.37	$&	0.79	\\	[0.01in]
0748+126	$^{\dagger}$&	Q	&	0.889	&	9.06	&	W02	&$	155.11	\pm	0.39	$&$	107.90	\pm	0.39	$&$	49.64	\pm	0.39	$&$	28.86	\pm	0.39	$&$	16.55	\pm	0.39	$&$	11.28	\pm	0.39	$&$	2.89	\pm	0.39	$&$	 -			$&$	5.70	\pm	0.18	$&$	1.53	\pm	0.53	$&$	0.58	\pm	0.01	$&$	0.31	\pm	0.01	$&$	 -			$&$	0.09	\pm	0.02	$&$	1.51	\pm	0.20	$&	0.99	\\	[0.01in]
0804+499	&	Q	&	1.432	&	9.39	&	S14	&$	1996.40	\pm	0.42	$&$	1388.80	\pm	0.42	$&$	638.85	\pm	0.42	$&$	371.42	\pm	0.42	$&$	212.95	\pm	0.42	$&$	145.19	\pm	0.42	$&$	37.14	\pm	0.42	$&$	4.74	\pm	0.11	$&$	3.00	\pm	0.10	$&$	 -			$&$	0.80	\pm	0.02	$&$	0.28	\pm	0.01	$&$	 -			$&$	0.13	\pm	0.03	$&$	1.23	\pm	0.15	$&	0.98	\\	[0.01in]
0851+202	&	B	&	0.306	&	 --	&	 --	&$	45.38	\pm	0.23	$&$	31.57	\pm	0.23	$&$	14.52	\pm	0.23	$&$	8.44	\pm	0.23	$&$	4.84	\pm	0.23	$&$	3.30	\pm	0.23	$&$	0.84	\pm	0.23	$&$	 -			$&$	3.38	\pm	0.07	$&$	0.68	\pm	0.14	$&$	0.52	\pm	0.01	$&$	 -			$&$	0.38	\pm	0.12	$&$	0.06	\pm	0.01	$&$	0.40	\pm	0.10	$&	0.98	\\	[0.01in]
0906+015	$^{\dagger}$&	Q	&	1.018	&	8.55	&	S14	&$	381.09	\pm	0.40	$&$	265.10	\pm	0.40	$&$	121.95	\pm	0.40	$&$	70.90	\pm	0.40	$&$	40.65	\pm	0.40	$&$	27.72	\pm	0.40	$&$	7.09	\pm	0.40	$&$	 -			$&$	2.73	\pm	0.06	$&$	0.86	\pm	0.39	$&$	1.53	\pm	0.04	$&$	1.08	\pm	0.02	$&$	 -			$&$	0.16	\pm	0.01	$&$	0.50	\pm	0.06	$&	0.39	\\	[0.01in]
0923+392	&	Q	&	0.698	&	9.28	&	S14	&$	91.48	\pm	0.36	$&$	63.64	\pm	0.36	$&$	29.27	\pm	0.36	$&$	17.02	\pm	0.36	$&$	9.76	\pm	0.36	$&$	6.65	\pm	0.36	$&$	1.70	\pm	0.36	$&$	5.69	\pm	0.11	$&$	3.71	\pm	0.06	$&$	2.46	\pm	0.49	$&$	2.50	\pm	0.04	$&$	 -			$&$	1.28	\pm	0.19	$&$	 -			$&$	0.43	\pm	0.10	$&	0.94	\\	[0.01in]
0945+408	&	Q	&	1.252	&	8.99	&	W02	&$	110.99	\pm	0.42	$&$	77.21	\pm	0.42	$&$	35.52	\pm	0.42	$&$	20.65	\pm	0.42	$&$	11.84	\pm	0.42	$&$	8.07	\pm	0.42	$&$	2.06	\pm	0.42	$&$	4.84	\pm	0.16	$&$	5.06	\pm	0.13	$&$	 -			$&$	1.21	\pm	0.04	$&$	 -			$&$	 -			$&$	0.22	\pm	0.03	$&$	0.88	\pm	0.17	$&	0.99	\\	[0.01in]
1101+384	$^{\dagger}$&	B	&	0.031	&	 --	&	 --	&$	2.27	\pm	0.03	$&$	1.58	\pm	0.03	$&$	0.73	\pm	0.03	$&$	0.42	\pm	0.03	$&$	0.24	\pm	0.03	$&$	0.17	\pm	0.03	$&$	0.04	\pm	0.03	$&$	0.23	\pm	0.01	$&$	0.12	\pm	0.01	$&$	0.09	\pm	0.03	$&$	0.04	\pm	0.01	$&$	 -			$&$	 -			$&$	0.01	\pm	0.01	$&$	1.00	\pm	0.16	$&	0.92	\\	[0.01in]
1127-145	$^{\dagger}$&	Q	&	1.733	&	9.30	&	S14	&$	117.63	\pm	0.42	$&$	81.83	\pm	0.42	$&$	37.64	\pm	0.42	$&$	21.88	\pm	0.42	$&$	12.55	\pm	0.42	$&$	8.55	\pm	0.42	$&$	2.19	\pm	0.42	$&$	 -			$&$	1.86	\pm	0.11	$&$	1.69	\pm	0.46	$&$	1.54	\pm	0.11	$&$	1.34	\pm	0.03	$&$	 -			$&$	 -			$&$	0.18	\pm	0.02	$&	0.97	\\	[0.01in]
1150+812	$^{\dagger}$&	Q	&	1.250	&	 --	&	 --	&$	55.22	\pm	0.42	$&$	38.42	\pm	0.42	$&$	17.67	\pm	0.42	$&$	10.27	\pm	0.42	$&$	5.89	\pm	0.42	$&$	4.02	\pm	0.42	$&$	1.03	\pm	0.42	$&$	 -			$&$	3.06	\pm	0.06	$&$	2.09	\pm	0.42	$&$	0.53	\pm	0.01	$&$	0.26	\pm	0.01	$&$	 -			$&$	 -			$&$	1.33	\pm	0.14	$&	0.93	\\	[0.01in]
1156+295	&	Q	&	0.729	&	8.54	&	S14	&$	360.14	\pm	0.36	$&$	250.53	\pm	0.36	$&$	115.25	\pm	0.36	$&$	67.00	\pm	0.36	$&$	38.42	\pm	0.36	$&$	26.19	\pm	0.36	$&$	6.70	\pm	0.36	$&$	 -			$&$	5.04	\pm	0.13	$&$	0.63	\pm	0.39	$&$	0.44	\pm	0.01	$&$	 -			$&$	0.46	\pm	0.19	$&$	0.15	\pm	0.03	$&$	0.44	\pm	0.08	$&	0.86	\\	[0.01in]
1219+285	$^{\dagger}$&	B	&	0.102	&	8.69	&	S14	&$	3.62	\pm	0.09	$&$	2.52	\pm	0.09	$&$	1.16	\pm	0.09	$&$	0.67	\pm	0.09	$&$	0.39	\pm	0.09	$&$	0.26	\pm	0.09	$&$	0.07	\pm	0.09	$&$	 -			$&$	1.04	\pm	0.05	$&$	0.19	\pm	0.06	$&$	0.24	\pm	0.01	$&$	 -			$&$	0.17	\pm	0.05	$&$	0.03	\pm	0.01	$&$	0.96	\pm	0.12	$&	0.86	\\	[0.01in]
1228+126	&	G	&	0.014	&	9.82	&	S14	&$	0.30	\pm	0.01	$&$	0.21	\pm	0.01	$&$	0.10	\pm	0.01	$&$	0.06	\pm	0.01	$&$	0.03	\pm	0.01	$&$	0.02	\pm	0.01	$&$	0.01	\pm	0.01	$&$	 -			$&$	0.12	\pm	0.01	$&$	0.03	\pm	0.03	$&$	0.03	\pm	0.01	$&$	0.01	\pm	0.01	$&$	 -			$&$	0.01	\pm	0.01	$&$	1.43	\pm	0.20	$&	0.89	\\	[0.01in]
1308+326	&	Q	&	0.997	&	8.72	&	S14	&$	152.38	\pm	0.40	$&$	106.01	\pm	0.40	$&$	48.76	\pm	0.40	$&$	28.35	\pm	0.40	$&$	16.25	\pm	0.40	$&$	11.08	\pm	0.40	$&$	2.84	\pm	0.40	$&$	 -			$&$	3.65	\pm	0.07	$&$	0.80	\pm	0.36	$&$	0.56	\pm	0.01	$&$	0.38	\pm	0.01	$&$	0.27	\pm	0.21	$&$	0.14	\pm	0.04	$&$	0.66	\pm	0.07	$&	0.99	\\	[0.01in]
1334-127	$^{\dagger}$&	Q	&	0.539	&	7.98	&	S14	&$	200.24	\pm	0.32	$&$	139.30	\pm	0.32	$&$	64.08	\pm	0.32	$&$	37.25	\pm	0.32	$&$	21.36	\pm	0.32	$&$	14.56	\pm	0.32	$&$	3.73	\pm	0.32	$&$	 -			$&$	5.34	\pm	0.09	$&$	0.81	\pm	0.26	$&$	1.09	\pm	0.02	$&$	0.17	\pm	0.01	$&$	0.28	\pm	0.18	$&$	 -			$&$	1.83	\pm	0.33	$&	0.84	\\	[0.01in]
1508-055	$^{\dagger}$&	Q	&	1.191	&	9.32	&	S14	&$	312.79	\pm	0.41	$&$	217.59	\pm	0.41	$&$	100.09	\pm	0.41	$&$	58.19	\pm	0.41	$&$	33.36	\pm	0.41	$&$	22.75	\pm	0.41	$&$	5.82	\pm	0.41	$&$	 -			$&$	3.02	\pm	0.22	$&$	0.83	\pm	0.37	$&$	0.99	\pm	0.04	$&$	 -			$&$	 -			$&$	0.26	\pm	0.07	$&$	0.77	\pm	0.12	$&	0.88	\\	[0.01in]
1606+106	&	Q	&	0.672	&	8.97	&	S14	&$	178.60	\pm	0.35	$&$	124.25	\pm	0.35	$&$	57.15	\pm	0.35	$&$	33.23	\pm	0.35	$&$	19.05	\pm	0.35	$&$	12.99	\pm	0.35	$&$	3.32	\pm	0.35	$&$	 -			$&$	2.67	\pm	0.05	$&$	 -			$&$	0.63	\pm	0.01	$&$	0.18	\pm	0.01	$&$	0.75	\pm	0.26	$&$	0.10	\pm	0.01	$&$	1.42	\pm	0.27	$&	0.56	\\	[0.01in]
1611+343	&	Q	&	1.401	&	9.57	&	W02	&$	77.40	\pm	0.42	$&$	53.84	\pm	0.42	$&$	24.77	\pm	0.42	$&$	14.40	\pm	0.42	$&$	8.26	\pm	0.42	$&$	5.63	\pm	0.42	$&$	1.44	\pm	0.42	$&$	 -			$&$	8.99	\pm	0.19	$&$	0.84	\pm	0.32	$&$	0.97	\pm	0.02	$&$	0.84	\pm	0.01	$&$	0.16	\pm	0.07	$&$	 -			$&$	1.24	\pm	0.20	$&	0.81	\\	[0.01in]
1633+382	&	Q	&	1.807	&	9.12	&	S14	&$	303.91	\pm	0.42	$&$	211.41	\pm	0.42	$&$	97.25	\pm	0.42	$&$	56.54	\pm	0.42	$&$	32.42	\pm	0.42	$&$	22.10	\pm	0.42	$&$	5.65	\pm	0.42	$&$	2.53	\pm	0.06	$&$	4.14	\pm	0.10	$&$	 -			$&$	1.31	\pm	0.05	$&$	0.29	\pm	0.01	$&$	0.31	\pm	0.22	$&$	 -			$&$	1.14	\pm	0.21	$&	0.86	\\	[0.01in]
1637+574	&	Q	&	0.751	&	9.18	&	W02	&$	121.05	\pm	0.37	$&$	84.21	\pm	0.37	$&$	38.73	\pm	0.37	$&$	22.52	\pm	0.37	$&$	12.91	\pm	0.37	$&$	8.80	\pm	0.37	$&$	2.25	\pm	0.37	$&$	6.79	\pm	0.17	$&$	3.63	\pm	0.09	$&$	 -			$&$	1.28	\pm	0.04	$&$	0.94	\pm	0.01	$&$	1.00	\pm	0.30	$&$	0.12	\pm	0.03	$&$	0.82	\pm	0.10	$&	0.95	\\	[0.01in]
1642+690	$^{\dagger}$&	G	&	0.751	&	7.76	&	W02	&$	71.41	\pm	0.37	$&$	49.68	\pm	0.37	$&$	22.85	\pm	0.37	$&$	13.29	\pm	0.37	$&$	7.62	\pm	0.37	$&$	5.19	\pm	0.37	$&$	1.33	\pm	0.37	$&$	5.07	\pm	0.16	$&$	1.47	\pm	0.05	$&$	 -			$&$	0.15	\pm	0.01	$&$	 -			$&$	1.12	\pm	0.29	$&$	0.12	\pm	0.07	$&$	1.93	\pm	0.30	$&	0.57	\\	[0.01in]
1730-130	&	Q	&	0.902	&	 --	&	 --	&$	326.14	\pm	0.39	$&$	226.88	\pm	0.39	$&$	104.36	\pm	0.39	$&$	60.68	\pm	0.39	$&$	34.79	\pm	0.39	$&$	23.72	\pm	0.39	$&$	6.07	\pm	0.39	$&$	 -			$&$	6.67	\pm	0.21	$&$	 -			$&$	0.58	\pm	0.01	$&$	0.62	\pm	0.01	$&$	0.54	\pm	0.20	$&$	 -			$&$	1.15	\pm	0.40	$&	0.86	\\	[0.01in]
1749+096	$^{\dagger}$&	Q	&	0.320	&	 --	&	 --	&$	65.26	\pm	0.23	$&$	45.40	\pm	0.23	$&$	20.88	\pm	0.23	$&$	12.14	\pm	0.23	$&$	6.96	\pm	0.23	$&$	4.75	\pm	0.23	$&$	1.21	\pm	0.23	$&$	 -			$&$	0.91	\pm	0.02	$&$	 -			$&$	0.78	\pm	0.02	$&$	 -			$&$	0.19	\pm	0.12	$&$	0.07	\pm	0.01	$&$	0.22	\pm	0.17	$&	0.90	\\	[0.01in]
1749+701	&	B	&	0.713	&	8.77	&	S14	&$	103.26	\pm	0.36	$&$	71.83	\pm	0.36	$&$	33.04	\pm	0.36	$&$	19.21	\pm	0.36	$&$	11.01	\pm	0.36	$&$	7.51	\pm	0.36	$&$	1.92	\pm	0.36	$&$	8.67	\pm	0.29	$&$	2.75	\pm	0.12	$&$	1.40	\pm	0.28	$&$	0.40	\pm	0.02	$&$	 -			$&$	 -			$&$	 -			$&$	1.79	\pm	0.29	$&	0.95	\\	[0.01in]
1803+784	&	Q	&	0.680	&	7.92	&	S14	&$	52.42	\pm	0.35	$&$	36.46	\pm	0.35	$&$	16.77	\pm	0.35	$&$	9.75	\pm	0.35	$&$	5.59	\pm	0.35	$&$	3.81	\pm	0.35	$&$	0.98	\pm	0.35	$&$	 -			$&$	3.42	\pm	0.06	$&$	0.71	\pm	0.27	$&$	0.68	\pm	0.01	$&$	0.21	\pm	0.01	$&$	0.66	\pm	0.27	$&$	0.18	\pm	0.07	$&$	1.47	\pm	0.19	$&	0.70	\\	[0.01in]
1807+698	&	B	&	0.050	&	10.1	&	S14	&$	2.83	\pm	0.05	$&$	1.97	\pm	0.05	$&$	0.91	\pm	0.05	$&$	0.53	\pm	0.05	$&$	0.30	\pm	0.05	$&$	0.21	\pm	0.05	$&$	0.05	\pm	0.05	$&$	0.83	\pm	0.02	$&$	0.23	\pm	0.01	$&$	0.39	\pm	0.13	$&$	0.06	\pm	0.01	$&$	 -			$&$	0.08	\pm	0.04	$&$	0.01	\pm	0.01	$&$	1.45	\pm	0.31	$&	0.66	\\	[0.01in]
1823+568	&	Q	&	0.653	&	7.94	&	S14	&$	84.55	\pm	0.35	$&$	58.82	\pm	0.35	$&$	27.06	\pm	0.35	$&$	15.73	\pm	0.35	$&$	9.02	\pm	0.35	$&$	6.15	\pm	0.35	$&$	1.57	\pm	0.35	$&$	5.30	\pm	0.14	$&$	1.35	\pm	0.04	$&$	1.22	\pm	0.24	$&$	0.52	\pm	0.01	$&$	0.52	\pm	0.01	$&$	0.69	\pm	0.28	$&$	0.11	\pm	0.03	$&$	0.87	\pm	0.17	$&	0.86	\\	[0.01in]
1845+797	$^{\dagger}$&	G	&	0.057	&	8.83	&	S14	&$	2.58	\pm	0.06	$&$	1.79	\pm	0.06	$&$	0.82	\pm	0.06	$&$	0.48	\pm	0.06	$&$	0.27	\pm	0.06	$&$	0.19	\pm	0.06	$&$	0.05	\pm	0.06	$&$	1.12	\pm	0.06	$&$	0.53	\pm	0.04	$&$	0.17	\pm	0.16	$&$	0.03	\pm	0.01	$&$	 -			$&$	 -			$&$	 -			$&$	2.09	\pm	0.21	$&	0.98	\\	[0.01in]
1928+738	&	Q	&	0.303	&	8.91	&	W02	&$	34.73	\pm	0.22	$&$	24.16	\pm	0.22	$&$	11.11	\pm	0.22	$&$	6.46	\pm	0.22	$&$	3.70	\pm	0.22	$&$	2.53	\pm	0.22	$&$	0.65	\pm	0.22	$&$	 -			$&$	 -			$&$	1.06	\pm	0.35	$&$	 -			$&$	0.44	\pm	0.01	$&$	0.50	\pm	0.16	$&$	0.09	\pm	0.03	$&$	0.86	\pm	0.18	$&	0.94	\\	[0.01in]
1936-155	$^{\dagger}$&	Q	&	1.657	&	 --	&	 --	&$	63.08	\pm	0.42	$&$	43.88	\pm	0.42	$&$	20.19	\pm	0.42	$&$	11.74	\pm	0.42	$&$	6.73	\pm	0.42	$&$	4.59	\pm	0.42	$&$	1.17	\pm	0.42	$&$	 -			$&$	4.62	\pm	0.15	$&$	0.42	\pm	0.32	$&$	0.90	\pm	0.03	$&$	0.27	\pm	0.01	$&$	 -			$&$	 -			$&$	1.53	\pm	0.21	$&	0.71	\\	[0.01in]
2121+053	&	Q	&	1.941	&	8.78	&	S14	&$	218.98	\pm	0.42	$&$	152.33	\pm	0.42	$&$	70.07	\pm	0.42	$&$	40.74	\pm	0.42	$&$	23.36	\pm	0.42	$&$	15.93	\pm	0.42	$&$	4.07	\pm	0.42	$&$	 -			$&$	4.87	\pm	0.09	$&$	0.91	\pm	0.46	$&$	1.16	\pm	0.05	$&$	0.36	\pm	0.01	$&$	 -			$&$	0.20	\pm	0.07	$&$	1.38	\pm	0.15	$&	0.86	\\	[0.01in]
2128-123	$^{\dagger}$&	Q	&	0.501	&	9.61	&	W02	&$	115.74	\pm	0.31	$&$	80.51	\pm	0.31	$&$	37.04	\pm	0.31	$&$	21.53	\pm	0.31	$&$	12.35	\pm	0.31	$&$	8.42	\pm	0.31	$&$	2.15	\pm	0.31	$&$	 -			$&$	8.82	\pm	0.29	$&$	 -			$&$	2.87	\pm	0.06	$&$	 -			$&$	1.05	\pm	0.16	$&$	0.48	\pm	0.03	$&$	0.82	\pm	0.09	$&	0.99	\\	[0.01in]
2145+067	&	Q	&	0.999	&	8.87	&	S14	&$	243.52	\pm	0.40	$&$	169.41	\pm	0.40	$&$	77.93	\pm	0.40	$&$	45.31	\pm	0.40	$&$	25.98	\pm	0.40	$&$	17.71	\pm	0.40	$&$	4.53	\pm	0.40	$&$	 -			$&$	5.03	\pm	0.11	$&$	3.58	\pm	0.55	$&$	0.96	\pm	0.02	$&$	0.77	\pm	0.01	$&$	1.27	\pm	0.21	$&$	 -			$&$	0.98	\pm	0.15	$&	0.73	\\	[0.01in]
2155-152	$^{\dagger}$&	Q	&	0.672	&	7.59	&	W02	&$	490.05	\pm	0.35	$&$	340.91	\pm	0.35	$&$	156.82	\pm	0.35	$&$	91.17	\pm	0.35	$&$	52.27	\pm	0.35	$&$	35.64	\pm	0.35	$&$	9.12	\pm	0.35	$&$	 -			$&$	4.11	\pm	0.08	$&$	 -			$&$	0.49	\pm	0.01	$&$	1.00	\pm	0.03	$&$	0.69	\pm	0.18	$&$	0.22	\pm	0.04	$&$	0.76	\pm	0.08	$&	1.00	\\	[0.01in]
2200+420	&	B	&	0.069	&	8.23	&	W02	&$	5.95	\pm	0.07	$&$	4.14	\pm	0.07	$&$	1.90	\pm	0.07	$&$	1.11	\pm	0.07	$&$	0.63	\pm	0.07	$&$	0.43	\pm	0.07	$&$	0.11	\pm	0.07	$&$	3.58	\pm	0.08	$&$	0.98	\pm	0.02	$&$	0.23	\pm	0.08	$&$	0.12	\pm	0.01	$&$	 -			$&$	0.08	\pm	0.03	$&$	0.01	\pm	0.01	$&$	0.73	\pm	0.37	$&	0.93	\\	[0.01in]
2201+315	$^{\dagger}$&	Q	&	0.298	&	8.87	&	W02	&$	114.34	\pm	0.22	$&$	79.54	\pm	0.22	$&$	36.59	\pm	0.22	$&$	21.27	\pm	0.22	$&$	12.20	\pm	0.22	$&$	8.32	\pm	0.22	$&$	2.13	\pm	0.22	$&$	 -			$&$	3.35	\pm	0.08	$&$	 -			$&$	0.24	\pm	0.01	$&$	0.77	\pm	0.01	$&$	0.62	\pm	0.12	$&$	0.13	\pm	0.04	$&$	0.73	\pm	0.57	$&	0.68	\\	[0.01in]
2209+236	$^{\dagger}$&	Q	&	1.125	&	8.46	&	S14	&$	17.17	\pm	0.41	$&$	11.95	\pm	0.41	$&$	5.49	\pm	0.41	$&$	3.19	\pm	0.41	$&$	1.83	\pm	0.41	$&$	1.25	\pm	0.41	$&$	0.32	\pm	0.41	$&$	 -			$&$	1.76	\pm	0.04	$&$	0.41	\pm	0.38	$&$	0.78	\pm	0.02	$&$	0.83	\pm	0.02	$&$	 -			$&$	 -			$&$	0.43	\pm	0.14	$&	0.18	\\	[0.01in]
2223-052	&	Q	&	1.404	&	 --	&	 --	&$	245.16	\pm	0.42	$&$	170.55	\pm	0.42	$&$	78.45	\pm	0.42	$&$	45.61	\pm	0.42	$&$	26.15	\pm	0.42	$&$	17.83	\pm	0.42	$&$	4.56	\pm	0.42	$&$	 -			$&$	2.79	\pm	0.11	$&$	 -			$&$	1.24	\pm	0.02	$&$	0.42	\pm	0.01	$&$	1.19	\pm	0.24	$&$	0.25	\pm	0.04	$&$	1.07	\pm	0.28	$&	0.77	\\	[0.01in]
2230+114	&	Q	&	1.037	&	8.93	&	S14	&$	450.16	\pm	0.40	$&$	313.16	\pm	0.40	$&$	144.05	\pm	0.40	$&$	83.75	\pm	0.40	$&$	48.02	\pm	0.40	$&$	32.74	\pm	0.40	$&$	8.38	\pm	0.40	$&$	 -			$&$	9.79	\pm	0.30	$&$	2.02	\pm	0.40	$&$	1.33	\pm	0.05	$&$	0.82	\pm	0.01	$&$	0.51	\pm	0.21	$&$	 -			$&$	0.86	\pm	0.09	$&	0.99	\\	[0.01in]
2243-123	$^{\dagger}$&	Q	&	0.630	&	8.81	&	S14	&$	70.10	\pm	0.34	$&$	48.77	\pm	0.34	$&$	22.43	\pm	0.34	$&$	13.04	\pm	0.34	$&$	7.48	\pm	0.34	$&$	5.10	\pm	0.34	$&$	1.30	\pm	0.34	$&$	 -			$&$	4.44	\pm	0.09	$&$	 -			$&$	1.27	\pm	0.03	$&$	0.30	\pm	0.01	$&$	0.71	\pm	0.18	$&$	 -			$&$	1.34	\pm	0.26	$&	0.79	\\	[0.01in]
2345-167	$^{\dagger}$&	Q	&	0.576	&	8.72	&	W02	&$	155.44	\pm	0.33	$&$	108.13	\pm	0.33	$&$	49.74	\pm	0.33	$&$	28.92	\pm	0.33	$&$	16.58	\pm	0.33	$&$	11.30	\pm	0.33	$&$	2.89	\pm	0.33	$&$	 -			$&$	4.13	\pm	0.13	$&$	 -			$&$	1.02	\pm	0.03	$&$	1.30	\pm	0.02	$&$	 -			$&$	0.94	\pm	0.23	$&$	0.56	\pm	0.23	$&	0.65	\\	[0.01in]
2351+456	$^{\dagger}$&	Q	&	1.986	&	9.29	&	S14	&$	292.78	\pm	0.42	$&$	203.67	\pm	0.42	$&$	93.69	\pm	0.42	$&$	54.47	\pm	0.42	$&$	31.23	\pm	0.42	$&$	21.29	\pm	0.42	$&$	5.45	\pm	0.42	$&$	 -			$&$	7.26	\pm	0.24	$&$	 -			$&$	0.34	\pm	0.01	$&$	0.62	\pm	0.01	$&$	0.98	\pm	0.38	$&$	 -			$&$	1.25	\pm	0.42	$&	0.55	\\	[0.01in]
\tableline
\multicolumn{21}{l}{$^\dagger$The different values for the deprojected $r_{\nu}$ in this source have been calculated using the approximate apparent speed $\beta_{app}$, rather than the viewing angle $\theta$.
}\\
\end{tabular}
}
\end{center}
\end{sidewaystable*}

Figure \ref{fits} (and corresponding Figure \ref{fitsAppendix} in Appendix) show plots of the transverse size versus core for these objects. In general, there is a clear trend of the core size to increase with distance, as expected if this effect arises from probing the upstream regions of the jet. In order to quantify this effect, we fitted for the core size assuming the form $R\propto r^{\epsilon}$. As our first a priori effort, we fitted all available data points for each source, regardless of any large variation or departures. We indicate these with a black straight line in Figure \ref{fits}. However, as we discussed in Sec. \ref{caveats}, it is possible that some of these  data points deviate due to several factors which, albeit related with the innermost jet physics of the source, may not follow the overall jet shape trend. We performed an alternative fit on these sources considering that some of the points may be outliers based both on an inspection by eye and a random sample consensus (RANSAC) model, although we consider the latter not reliable due to the small amount of data. Such fits are shown by a dashed red line in Figure \ref{fits}.

Examples of sources which may show such outliers are 0003-066, 1156+295, 1308+326, 2200+420 or 2230+114, where we observe that the outmost points show a larger size than expected by extrapolation of the innermost data. In cases such as 0133+476, 0149+218, 0716+714, 0748+126, 0804+499 or 2121+053, it is the innermost data point which shows a noticeable larger size and, although it may also indicate a geometry break, we did not include it in the alternative fit due to the lack of statistics. Other sources such as 0106+013, 0851+202 seem to show large deviations in the extremal data points. We will discuss these aspects below.

\begin{figure}
\includegraphics[scale=0.45,trim=0cm 2.5cm 0cm 2.5cm, clip=true]{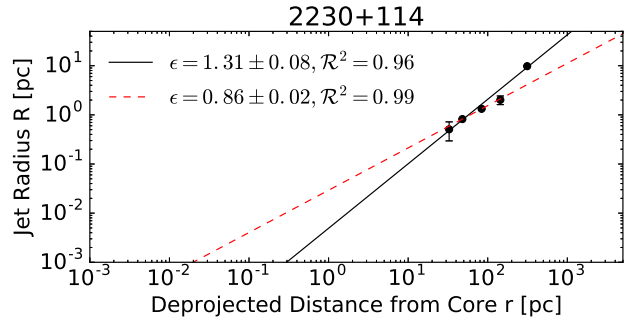}
\caption{Sample plot of the core size as a function of distance from the central engine. Dots indicate the compiled data from the literature for these objects where values for 4 or mode independent frequencies were found. The black straight line is a fit of the form $R\propto r^\epsilon$ using all the available data points, whereas the red dotted line shows the fit including only selected points (see text). The top left corner shows the fitted value for $\epsilon$ and its error (quadratic sum of uncertainties due to $k_r$ not included), together with the goodness of the fit. Plots for all sources studied here are shown in the appendix.}
\label{fits}
\end{figure}

If we consider the alternative fitting, free of outliers, and hence less affected by possible biases such as possible geometry breaks and spurious data due to jet blending or time variability effects, as discussed above, a good number of sources (32 of 56, or $\sim$57\%) lie within the lines $1/2<\epsilon<1$ within $1\sigma$, in agreement with a range of geometries including parabolical or conical streamline. There are also several objects with clearly $\epsilon<1$ (5 out of 56, or 9\%), indicating a quasi--parabolical streamline\footnote{Note the case of 2251+158, which showed $k_r=0.6-0.8$. The fitted value will be contaminated by that factor and should correspond to $\epsilon/k_r$, which would imply a more proper estimate for the streamline morphology, assuming the fit were reliable.}. However, a significant portion of the sources studied here (19 out of 56, or 34\%) show $\epsilon>1$,  which would correspond to a hyperbolic jet. 

A considerable number of sources show a large scatter in their data, with  $\mathcal{R}^2<0.85$. We find it interesting that a fraction of these also show $\epsilon>1$ (see e.g., 0716+714, with $\epsilon=1.44, \mathcal{R}^2=0.79$; 1803+698, with $\epsilon=1.45, \mathcal{R}^2=0.66$; 1936--155, with $\epsilon=1.53, \mathcal{R}^2=0.71$ or 2351+436, with $\epsilon=1.25, \mathcal{R}^2=0.55$). We consider the possibility that these large values for $\epsilon$, corresponding to a hyperbolic geometry may actually be spurious, and due to the lack of significant data for proper statistics and a reliable fit. In order to construct a  criteria to check for the data scatter, we consider these fits with $\mathcal{R}^2\geq0.85$. Under this criteria, 11\%,  60\% and 29\% of the sources have a small ($\epsilon<1/2$), intermediate ($1/2<\epsilon<1$)  and large ($\epsilon>1$) geometry value respectively.

\begin{figure}
\includegraphics[angle=0,scale=0.45,trim=0cm 0cm 0cm 0cm,clip=true]{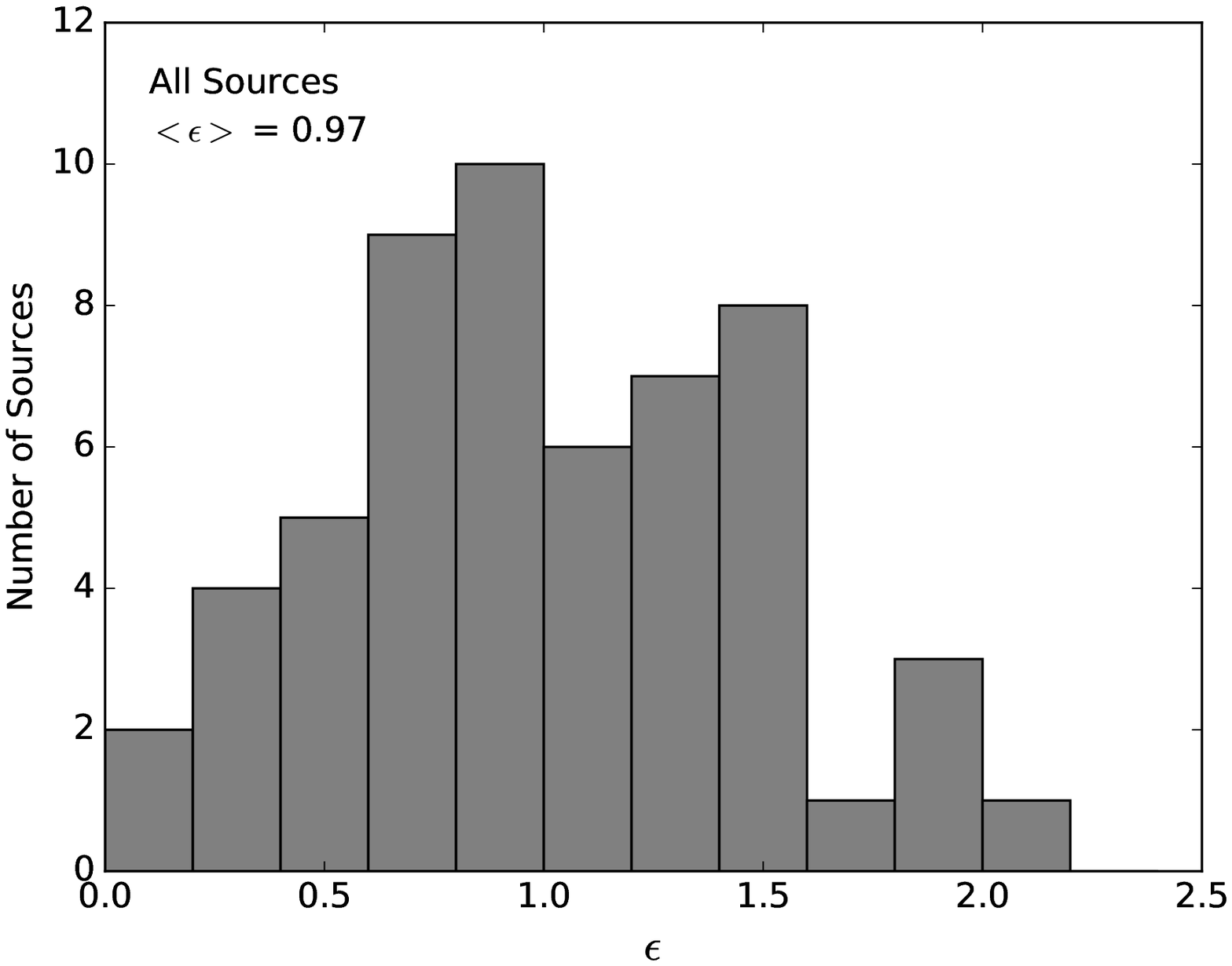}\\
\includegraphics[angle=0,scale=0.45,trim=0cm 0cm 0cm 0cm,clip=true]{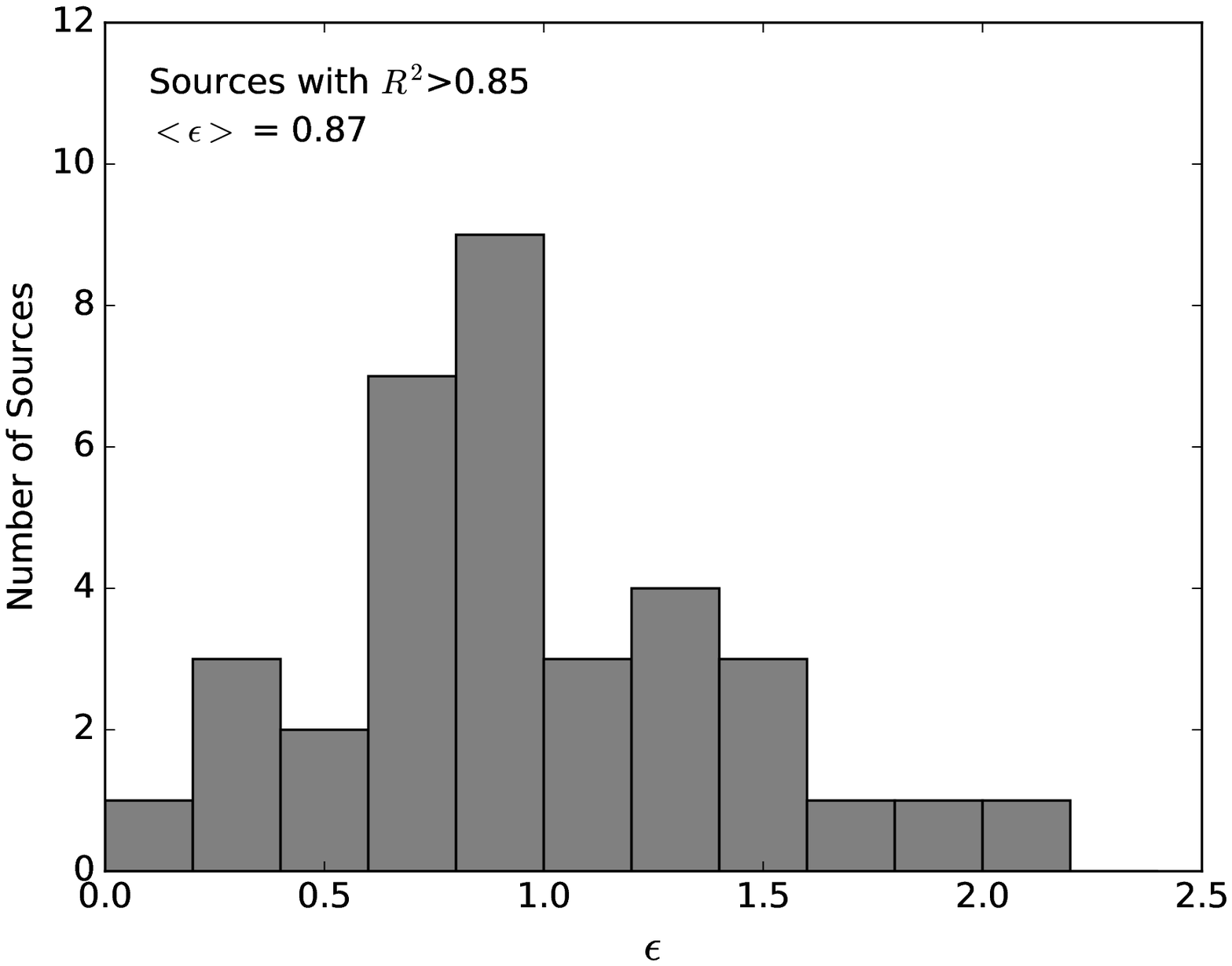}
\caption{Distribution of the $\epsilon$ values for all the sources. Top: all sources; Bottom: only these sources for which $R^2>0.85$ is found}
\label{histogram}
\end{figure}

In Figure \ref{histogram} we show a histogram of the values found for $\epsilon$. The median value is $\left< \epsilon \right>=0.97$. A Kolmogorov--Smirnov test indicates the distribution to be different from a gaussian with a significance of 90\%. Similar median values are found for $R\leq0.75$, whereas the median significantly decreases for $R\geq0.75$. If we restrict ourselves to only these sources with $\mathcal{R}^2\geq0.85$, then $\left< \epsilon \right>=0.85$ and the distribution is different from a gaussian with a significance of 70\%. We do not attempt to make a comparative statistics between various kinds of objects (quasars, blazars or radio galaxies) given that the small quantity of objects other than quasars in our sample would provide poor statistics and this analysis would be not robust.

\section{Discussion}

\subsection{Jet Geometry -- External Medium Connection}

We have found that a good number, but not all, of the sources can be described with a conical to parabolical geometry. In the following paragraphs we consider various models discussing the physical properties in (magneto)--hydrodynamic jets and their connection with ambient pressure of the external medium which seem to be required in order to lead to the various geometries seen above.

\subsubsection{Conical Jet}

In an adiabatic conical jet ($\epsilon = 1$), the hydrodynamic pressure decreases as $r^{-2 \gamma}$ (with $\gamma$ the ratio of specific heats). Thus, the ambient pressure in the interstellar matter (ISM) will be scaled as $p_{\rm ism} \propto r^{-b}, b > 2.5$ if the jet pressure is balanced with the ISM pressure inside the SGI. This is steeper than the case of the Bondi accretion flows. On the other hand, in the case of a magnetized jet with purely toroidal field component $B_{\phi}$, which is good approximation far downstream of SMBHs,  $b=4$ is required in the self-similar solution of the steady jet \citep{Zakamska08}. For a general (non-self-similar) case, $b > 2$ is allowed in analytical and numerical results in order to have an asymptotically conical streamline \citep{Tchekhovskoy08,Lyubarsky09,Komissarov09}. Or, if the jet is highly over-pressured compared with the ISM pressure, $p_{\rm jet} > p_{\rm ism}$, then the jet would be subject to a quasi-conical expansion. Though, in the case of a slowly decreasing ISM pressure (compared with the jet pressure), re-collimation shocks (oblique shock and Mach disk system) can be formed in the downstream \citep[e.g.][]{Sanders83}, where $p_{\rm jet} < p_{\rm ism}$. 

\subsubsection{Parabolic Jet}

A self-similar solution has been found for the hydrodynamic (non-magnetized) case with the purely parabolic ($\epsilon = 1/2$) streamline where the ISM pressure is decreasing with a power-law index of $b=2$ \citep{Zakamska08}. So far, no analytical or numerical solutions have been found in the non-magnetized cases except $\epsilon$ = 0, 1/2, and 1. On the other hand, the magnetized jets (with dominant $B_{\phi}$) can be purely parabolic ($\epsilon = 1/2$) in analytical and numerical solutions where the ISM is decreasing with a power-law index of $b=2$ \citep{Tchekhovskoy08,Lyubarsky09}. \cite{Komissarov09} suggest a quasi-parabolic streamline $1/2 < \epsilon < 1$ can be also obtained with $b=2$. The steady-state force-free magnetosphere around a Kerr black hole is examined along parabolic  streamlines ($\epsilon = 1/2$) \citep{Blandford76,BZ77}. \cite{McKinneyNarayan07} examined that the fiducial general relativistic MHD (GRMHD) simulation jet \citep[e.g.][]{McKinneyGammie04} can be reasonably consistent with quasi-parabolic structures ($1/2 < \epsilon < 1$, which are organized by a force-free field with a steeper radial distribution of the toroidal current on the equatorial plane than the purely parabolic case \citep[$\epsilon = 1/2$]{BZ77}. The agreement between the GRMHD numerical models and quasi-parabolic force-free model is found to be good out as far as $\sim 10^3$ times larger than the black hole scale \citep{McKinneyNarayan07}.

\subsubsection{Other Geometries}


An ISM decrease of the form $b > 4$ may give a hyperboloidal shape with conical asymptotes \citep{Komissarov09}; although how this steep pressure gradient can be realized in a realistic ISM is not clear. In the same sense, we find some objects with $\epsilon<1/2$, which can be more collimated case than the purely parabolic flow. In magnetized jets (with dominant $B_{\phi}$), this configuration is established when $b <  2$ \citep{Komissarov09}. We also find some objects with $\epsilon \sim 0$, which would be asymptotically cylindrical configuration. This category may be identified as that well-collimated flows ($\epsilon < 1/2$) converge asymptotically to cylinders at the final phase of the magnetic acceleration far distant from the black hole. In self-similar solutions, lateral structure of a cylindrical jet in both non-magnetized and magnetized (with dominant $B_{\phi}$) can be obtained when the ambient pressure scaling is uniform along the jet \citep{Zakamska08}.

\subsection{Jet Collimation Break; Quasi--Parabolic to Conical Transition}

An analysis of jet geometries based on model--fitting jet components by \cite{Pushkarev12b} provided median values of  $\epsilon\sim1.2$  at 2.3~GHz and  $\epsilon\sim0.8$ at 8.6~GHz, respectively. They interpreted that the jet regions probed at 8.6~GHz were accelerating and collimating, whereas these at 2.3~GHz were freely expanding and decelerating\footnote{We note however that, as we have mentioned, it is not straightforward to interpret the physical properties of their sources at 2.3~GHz, showing mean $\epsilon>1$ values.}. Similarly, \cite{Pushkarev14} found a median $\epsilon\sim0.94$ for a sample at 15~GHz, suggesting that these sources with lower $\epsilon$ are closer or sustain larger viewing angles, thus resolving smaller linear distances from the super-massive black hole (SMBH). If this is the case, our values should be closer to, and possibly lower, than theirs, as we are probing regions that are in principle much closer the to SMBH in comparison. Our value $\left<\epsilon \right>=0.85$ when we consider the criteria $\mathcal{R}^2\geq0.85$ (see Figure \ref{histogram}) seems to pinpoint in this direction. 

Based on this, it seems that jets, classically thought to have a conical geometry on parsec scales, show a tendency towards a more quasi--parabolical geometry upstream. This suggests the existence of a jet collimation break. Indeed, inspection of Figure \ref{fits} indicates that 2200+420 might be a possible candidate where such break is observed at around 10~pc. Several works on M87  \cite{Asada12,Nakamura13} and NGC~6251 \citep{Tseng16} found such break occurring  on scales of also tens of parsec, near the Bondi radius.  Similarly, various works \citep{Homan15,Lee16} suggest that  jet features show a positive acceleration near the base of the jet, at around 10~pc, slowing down or even decelerating downstream the jet, suggesting that a jet acceleration break is also present.  \cite{Asada12,Nakamura13} indicate that in M87, this jet acceleration/collimation break occurs near the Bondi radius. Our results may suggest that such acceleration/collimation break may be the case as well in other sources.

As we have seen, some sources seem to show a larger size in their the outmost regions compared to what is expected by extrapolation of the innermost data. One possibility for this to happen can be attributed to blending due to opacity or resolution effects, where the obtained size may be a combination of the real core size plus a contribution of the innermost jet or new component, or time variability. On the other hand, based on the previous discussion, we cannot discard the possibility that a jet collimation break occurs and we are probing different jet geometries, although this is very unclear except possibly for the case of 2200+420, due to the small amount of data.

\subsection{Caveats}\label{caveats}
As we discussed above, in this paper we interpret that the compact VLBI core is associated with the throat of a diverging jet. It is only in this context that we can treat the VLBI core widths as equivalent to the innermost jet widths. This assumption seems to be valid in M87 \citep[see e.g.][]{Hada11,Nakamura13} or NGC~6251 \citep[see e.g.][]{Tseng16}, where it has been seen that the jet widths obtained in such way smoothly connects with the outer  jet profile and follows a similar radial dependence down to several tens of Schwarzschild radii.

In this context, one would expect the jet transverse size to increase as we observe it further away from the central engine (i.e, alternatively, and owing to the core shift effect, when we observe at lower frequencies). One may however think about other causes leading to a decrease of the core size with increasing frequency that may not be related with the upstream intrinsic jet size. The most obvious one is that, at higher frequencies, we are able to obtain much larger resolutions and thus we are fitting for much more smaller regions of the core, unresolved at lower frequencies. The best way to study this possibility would be to have simultaneous data at the same frequency with a another array with different baseline lengths and configuration, and hence, different resolution. 

Whilst unfortunately this is not currently possible, we can however examine this issue with our current data. First, we note that the compiled data was observed with various arrays. For example, VSOP at 5~GHz provides better resolution ($\sim0.2$ arcsec) than VLBA at 8.6 and 15~GHz ($\sim0.8$ and $\sim0.4$ arcsec respectively), yet no evidence for systematic lower core size at 5~GHz with VSOP data is found in the data. On the contrary, core sizes at 5~GHz seem to be systematically larger, in agreement with our assumptions. 
Second, we were able to find in the literature core sizes using different arrays for few sources: the data compiled by \cite{Jiang98} with ground arrays seem to indicate similar core sizes at 5~GHz, with the only exception of 1308+326, whose value differ $\sim50$\%. 

Additionally, as noted by \cite{Hada13}, the fitted values could be overestimating the core size in the direction of the jet due to some blending with the optically thin part of the jet. Detailed discussion on this can be found in \cite{Kovalev08} and \cite{Lee16}, where they find this effect to be of the order of few $\mu as$, much smaller than the values discussed here. In the transverse direction, contribution from extended or `sheath' emission not resolved by VLBI to the apparent core size could also exist, although we aim to measure the width of the jet.

It is possible that some scatter in the core size may arise due to temporal variation. For example, \cite{Hada13} found a variation of up to 23\% in the size of the M87 core semi--major axis at 15~GHz on time scales of 28 months. In this context, even when measurements are taken directly from the jet width, a large scatter in the data is expected \citep[see e.g., Figure 22 in][]{Jorstad05}. If this is the case, the results presented here may be contaminated by this scatter and variations, and may not reflect the actual jet shape. In order to avoid this problem, multi-epoch data analyses should be necessary. In this sense, we treat the results discussed here with caution and understand them as a preemptive effort towards future analysis on currently ongoing or future multi--frequency simultaneous surveys.

Even in the case of simultaneous data, it may still be dangerous to fit for a power-law index on individual sources when data spans only one or two orders of magnitude along the jet axis, especially when the available data is quite sparse (i.e, restricted to just few, $\lesssim10$ data points, rather than a more or less fair sampling of the region under investigation), and scatter may play an important role. The presence of  unresolved blobs, shocks and/or jet components, for example, may contaminate the inferred jet size, if not properly constrained with enough data. Furthermore, possible jet geometry breaks, if present, may not be properly identified in the case of sparse data and a inaccurate geometry may be derived instead.

The core shifts in \cite{Pushkarev12} were obtained under the assumption of $k_r\sim1$.  Although this seems to be a typical value in many sources \citep{OSullivan09,Sokolovsky11,Hada11,Algaba12}, it may not be the case in all of them. For example, it would be possible to find larger values of $k_r$ in regions with steep pressure gradients or, in the case that external absorption determines the core position, if similar density gradients on the external medium are found \citep{Lobanov98}. Such values have been found in J0241--0815 \citep{Kadler04} or Cyg~A \citep{Bach08}. Values of $k_r<1$ can also be found in the case of large gradient departures from equipartition or decelerating flows. For example, \cite{Kutkin14} found $k_r<0.8$ for J2251+158. If $k_r$ is significantly different from unity, the core shifts studied in \cite{Pushkarev12} may be under- or overestimated, propagating this effect onto our calculations. In general, and based on the literature, we will consider that $k_r\sim1$ is a good a priori approximation unless otherwise stated. Based on typical deviations observed for $k_r$ \citep[e.g.][]{Sokolovsky11}, we consider that an error of the order of $\sim$10\% on the estimated $\epsilon$ can be assumed.

Alternatively, another reason for the core size to decrease with frequency could be scintillation. In this case, angular broadening is manifested by an scaling of the angular size approximately as $\epsilon\sim2$ \citep[see e.g.][]{Lazio08}. This is much larger than the values discussed through this paper and is only found for one source, 1845+797, which is not known to be an scintillating object. We thus suggest that scintillation broadening is not the case here as we would expect a much steeper slope in general.

\subsection{A Global View}

A way to further study the global behaviour of all sources, to search for a common structural trend free of most of the caveats mentioned above, is to combine all the individual sources data. We note that, even in this case, if our sample consist on a mixture of various geometries, such  as semi--parabolical and conical, or a geometry break occurs, which we cannot know a priori, a fit may still not be relevant to discuss about a global geometry. It is thus an interesting exercise to consider all the data without the biases or unknowns related to the fitting and consider a different approach to examine the data.

\begin{figure}
\includegraphics[scale=0.58]{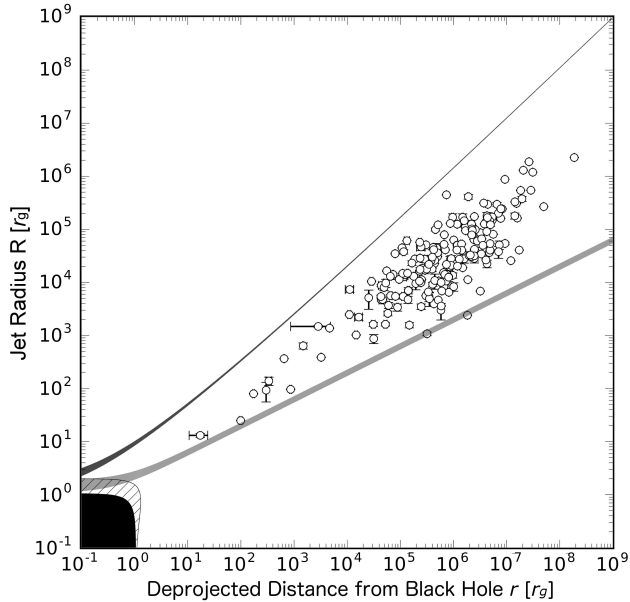}
\caption{Compilation of the core sizes as a function of distance from the central engine for all sources in units of gravitational radii ($r_{\mathrm g}=GM_{\mathrm BH}/c^2$) considering different models for the streamlines for a spinning black hole. Filled black region denotes the black hole (inside the event horizon), while the hatched area represents the ergosphere for the black hole spin parameter a = 0.998. Light gray area denotes the genuine parabolic streamline ($R \propto r^{1/2}$ at $R \gg r_{\mathrm g}$) of the force-free steady jet solution \citep{BZ77}, while the dark gray area denotes the quasi-conical streamline ($R \propto r^{0.97}$ at $R \gg r_{\mathrm g}$) of the force-free steady jet solution \citep{Narayan07,Tchekhovskoy08}. In both streamlines, a variation from $a = 0.5$ (upper boundary) to $a=0.998$ (lower boundary) is considered as a shaded area. Note that all streamlines are anchored at the even horizon $r_{\mathrm H}=r_{\rm g}(1+\sqrt{1-a^2})$ with the maximum angle $\theta=\pi/2$ in polar $(r, \theta)$ coordinates in the Boyer-Lindquist frame.}
\label{BHshape}
\end{figure}

Data points are converted into the units of the gravitational radius $r_{\rm g}=G M_{\rm BH}/c^2$ by using the black hole mass $M_{\rm BH}$ tabulated in \cite{WooUrry02} and \cite{Zamaninasab14} (see Table \ref{data}). Figure \ref{BHshape} shows the jet radius as a function of the jet axial distance in the units of $r_{\rm g}$ for the 43 objects for which this data was available.  Analytical streamlines of the force-free jet steady state solutions are overlaid for comparison; the genuine parabolic streamline ($\epsilon = 0.5$) of \cite{BZ77} and the quasi-conical streamline ($\epsilon = 0.97$) as a representative of \cite{BK79} by using \cite{Narayan07} and \cite{Tchekhovskoy08}. In this examination, we consider the outermost streamlines, which are anchored at the horizon radius with a polar angle of $\pi/2$ with different black hole spin parameters $a=0.5 - 0.998$, as the maximally allowable angle to be penetrated; i.e., the magnetic filed lines will touch the event horizon with this critical
angle on the equatorial plane inside the ergosphere. We can see that data points are well guided between two shaded areas.

Inspection of Figure \ref{BHshape} indicates that the collected data may not support the classical picture of a jet starting from the vicinity of the black hole with neither i) a conical geometry ($\epsilon=1$) nor ii) a genuine parabolic geometry ($\epsilon=1/2$). Instead, the data fits in an intermediate region where semi-parabolic streamlines ($0.5 < \epsilon < 1$) would exist. The fact that most of the objects in our study show $\epsilon > 1/2$ may support an idea that quasi parabolic structures are common in the upstream inside the SGI. The caveat is whether a quasi-parabolic streamline can be generally formed in radiatively efficient accretion flows. This can be  by utilizing general relativistic radiation magnetohydrodynamics (GRRMHD) simulations \citep[e.g][]{Sadowski15,McKinney15} in coming years. Our results may also support an idea that the magnetically organized jet on parsec scale may be norm by inferring the non--conical jet geometry, in agreement with \cite{Zamaninasab14}, who argues that dynamically important magnetic fields exist in the AGN jets on parsec scale. 

However, as discussed above, we would speculate that semi-parabolic streamlines, if they are originated in the vicinity of the SMBHs, may not extend beyond the scale of the SGI $\sim 10^5 - 10^6\, r_{\rm g}$. Instead, a jet collimation break may take place in the form of a transition from parabolic to conical geometry, as discovered in nearby radio galaxies \citep[e.g.][]{Asada12,Tseng16}. A structural transition may correspond to the observed feature of the jet bulk acceleration break on the scale of $\sim 10$ pc in MOJAVE samples  \citep{Lister09,Lee08}. If such mixture of semi--parabolic and conical geometries are existing on the sample, this would support the idea that a simple fit to the stacked data would not be useful to extract physical parameters. Alternatively, we cannot rule out the existence of a hidden or invisible jet starting from a comparatively more distant scale from the SMBH.

\begin{figure}
\includegraphics[scale=0.58]{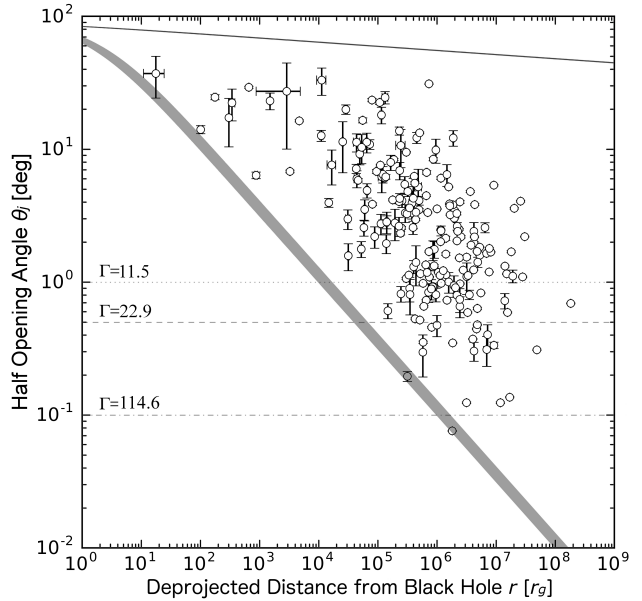}
\caption{Compilation of jet half opening angles as a function of distance from the central engine in units of gravitational radii. Grey lines indicate the limiting cases for the opening angle derived from the parabolic and quasi-conical jet. Dotted, dashed and dotted--dashed lines show Lorentz factor for 1, 0.5 and 0.1 degrees respectively}
\label{openingangle}
\end{figure}

We derived the jet intrinsic half opening angle $\theta_j$ from the values in Table \ref{data}. In Figure \ref{openingangle} we plot these as a function of distance from the central engine, in units of gravitational radii, following a similar manner of Figure \ref{BHshape}. Here the grey lines indicate the limiting cases for the opening angle derived from the parabolic and quasi-conical jet as before. It is clear that all data lie within these two limiting cases, as expected from inspection of Figure \ref{BHshape}. In addition, it is clear that, for small radii, the opening angle is quite large, suggesting that a quasi--conical expansion is unlikely in such regime (otherwise jets would be unrealistically wide even near to the jet base). This altogether also supports our view where quasi parabolic structures  are most common inside the SGI.

In the jet production standard model, the jet is expected to be causally connected with its symmetry axis, implying $\Gamma \theta_j<1$, where $\Gamma$ is the Lorentz factor \citep[e.g.][]{Tchekhovskoy09,Komissarov09}. An analysis on a sample of MOJAVE sources performed by \cite{ClausenBrown13} suggests $\Gamma\theta_j\sim0.2$. Assuming this factor for the population we investigate here, we can derive an estimate of the Lorentz factor based on the observed jet half opening angles. In Figure \ref{openingangle} we show the derived values for $\Gamma$ for $\theta_j=1, 0.5$ and $0.1^{\circ}$ respectively for guidance. The values of  $ \Gamma \sim10 - 20$ we find are in agreement with values $\Gamma\sim15$ from observations  by e.g. \cite{Jorstad05} and \cite{Hovatta09}, which indicates that the derivation of half opening angles from core size analyses is reasonable.

\section{Conclusions}
We have obtained from the literature core sizes of a sample of AGNs at various frequencies. Based on the assumption that these cores represent the upstream regions of the jet, and taking into account the core shift effect to locate the relative position of these regions, we have studied the jet morphology of the upstream unresolved regions for the AGNs in our sample. Although measuring the jet geometry based on the method discussed here provides only a limited number of data points distributed over just two or three orders of magnitude, leading to much larger associated uncertainties compared with direct jet widths measurements, we can probe on jet scales that are unaccessible otherwise.

When we consider the fitted data with criteria to avoid outliers which may arise due to jet collimation breaks, blending or resolution effects or time variations, and we account for the goodness of the fit, 60\% of the sources show quasi--parabolic structure, with $1/2<\epsilon<1$, and the median geometry value is $\left< \epsilon \right>=0.85$. This is in agreement with previous values found on small jet scales and seems to suggest that a semi--parabolic jet shape may be more common near the innermost few parsecs of the jet, in contrast with from the conical shapes typically found on deca--parsec scales or further.

The combined data fits in a region between a genuine parabolic and conical geometries. This supports the idea that near the vicinity of the central engine, a semi--parabolic streamline geometry would exist. The large jet half opening angles ($\theta_j>10^\circ$) derived from the data near the SMBH ($r\lesssim10^4~r_g$) seem to support this idea. The derived Lorentz factors obtained considering causality arguments are in agreement with observed ones and provide an alternative consistency check to our results. We speculate that the quasi parabolic streamlines discussed here may not extend beyond the sphere of gravitational influence of the SMBH, and a transition from parabolic to conical geometry may occur.

Studies of  transverse profiles and streamline geometries in conjunction with core shift measurements on other sources, such as the one performed on M87, would be desirable in order to make a more robust comparison with our method, although to the knowledge of the authors, such studies are very limited. Future sub-mm VLBI will be crucial to test the parabolic jet hypothesis within the black hole sphere of influence, $10^3 - 10^5~r_g$.\\

\acknowledgments

\footnotesize{This research has made use of data from the MOJAVE database that is maintained by the MOJAVE team \citep{Lister09}. This research has made use of the NASA/IPAC Extragalactic Database (NED) which is operated by the Jet Propulsion Laboratory, California Institute of Technology, under contract with the National Aeronautics and Space Administration. The National Radio Astronomy Observatory is operated by Associated Universities, Inc., under contract with the National Science Foundation. SL was supported by the National Research Foundation of Korea (NRF) grant funded by the Korea government (MSIP) (No. NRF-2016R1C1B2006697). We thank the referee for several helpful suggestions.}

\appendix

\section{Plots of core size versus distance from the SMBH}

\begin{figure*}
\centering
\subfigure{\includegraphics[trim=0cm 0cm 0cm 2.5cm, clip=true,width=0.24\textwidth]{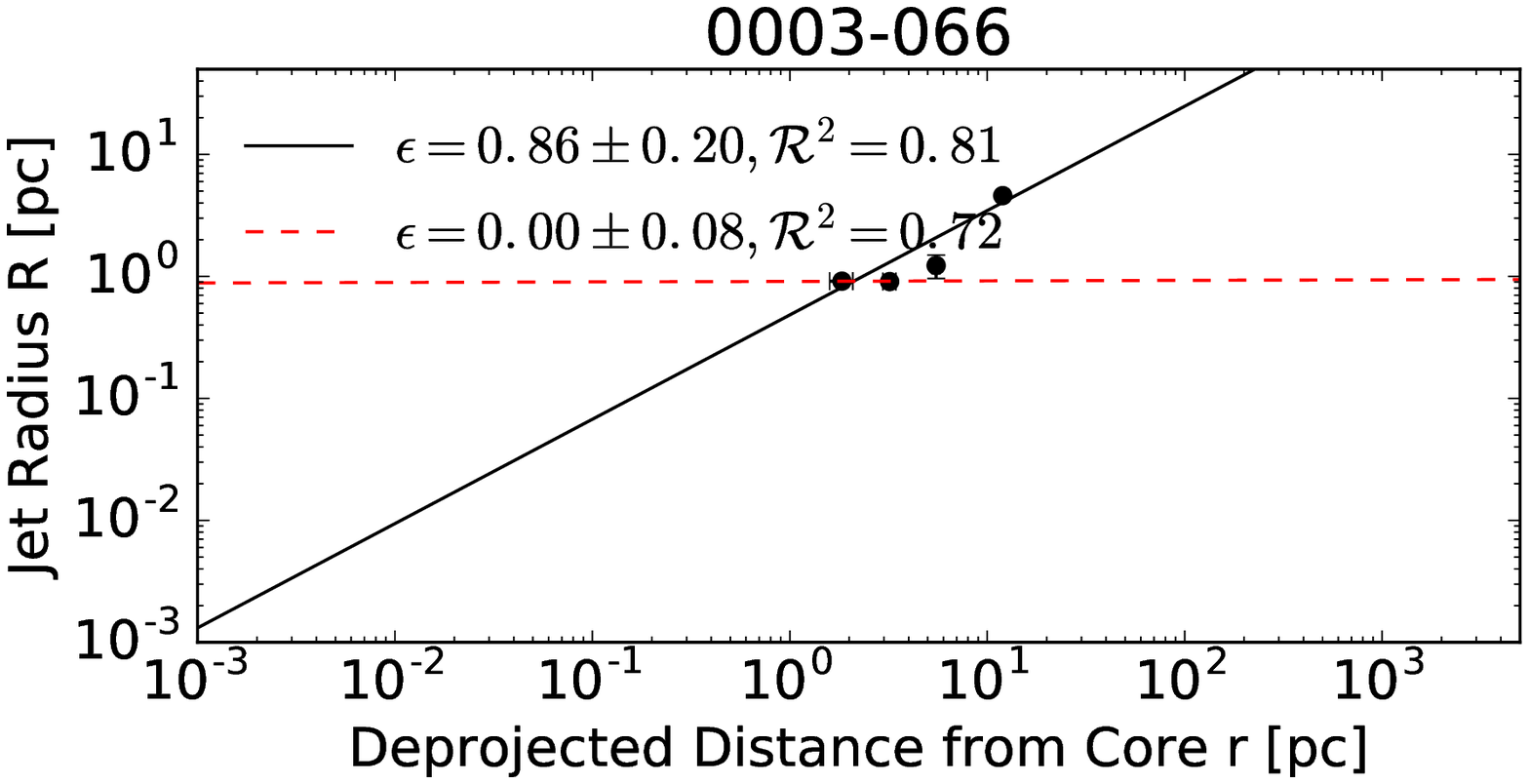}}\
\subfigure{\includegraphics[trim=0cm 0cm 0cm 2.5cm, clip=true,width=0.24\textwidth]{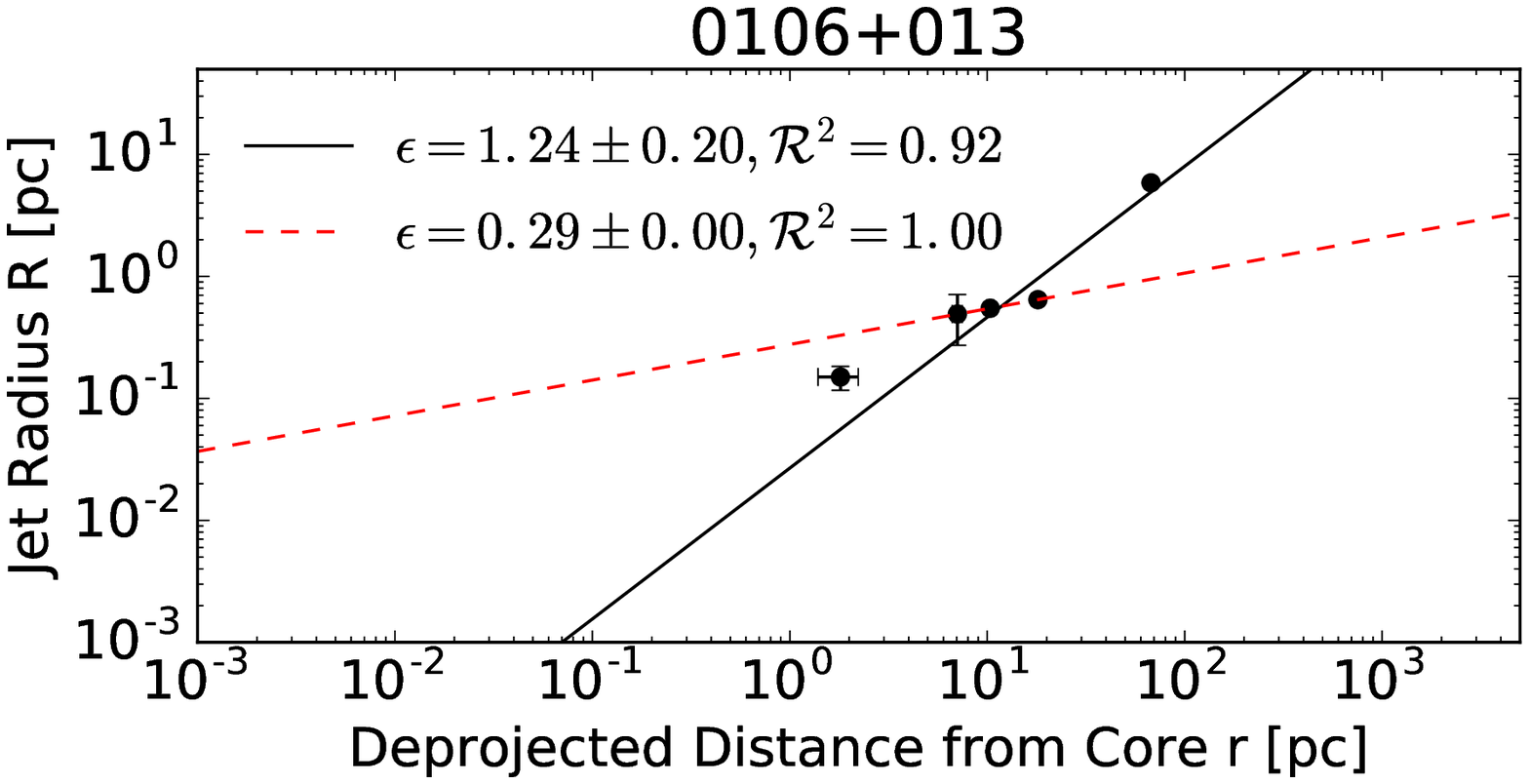}}\
\subfigure{\includegraphics[trim=0cm 0cm 0cm 2.5cm, clip=true,width=0.24\textwidth]{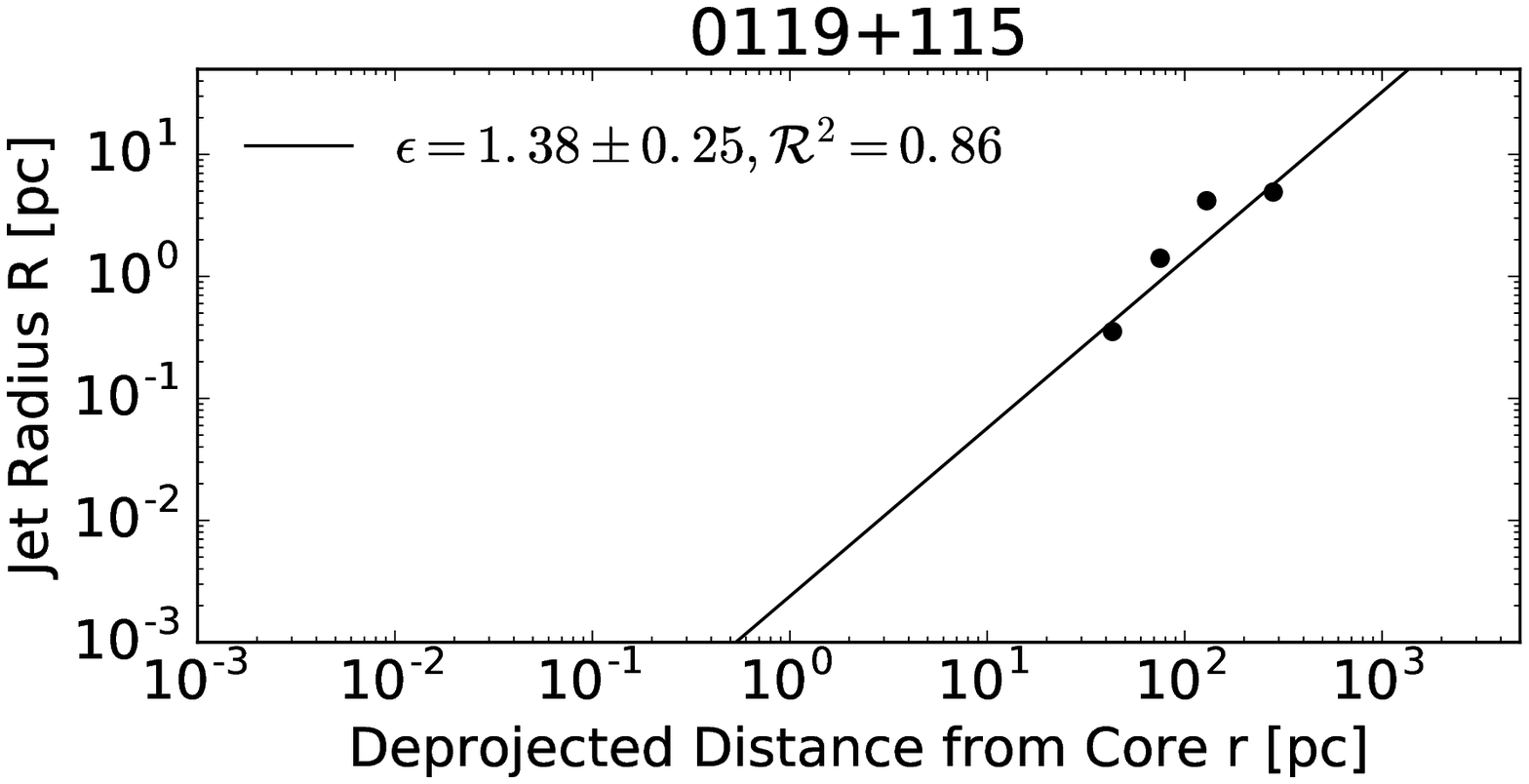}}\
\subfigure{\includegraphics[trim=0cm 0cm 0cm 2.5cm, clip=true,width=0.24\textwidth]{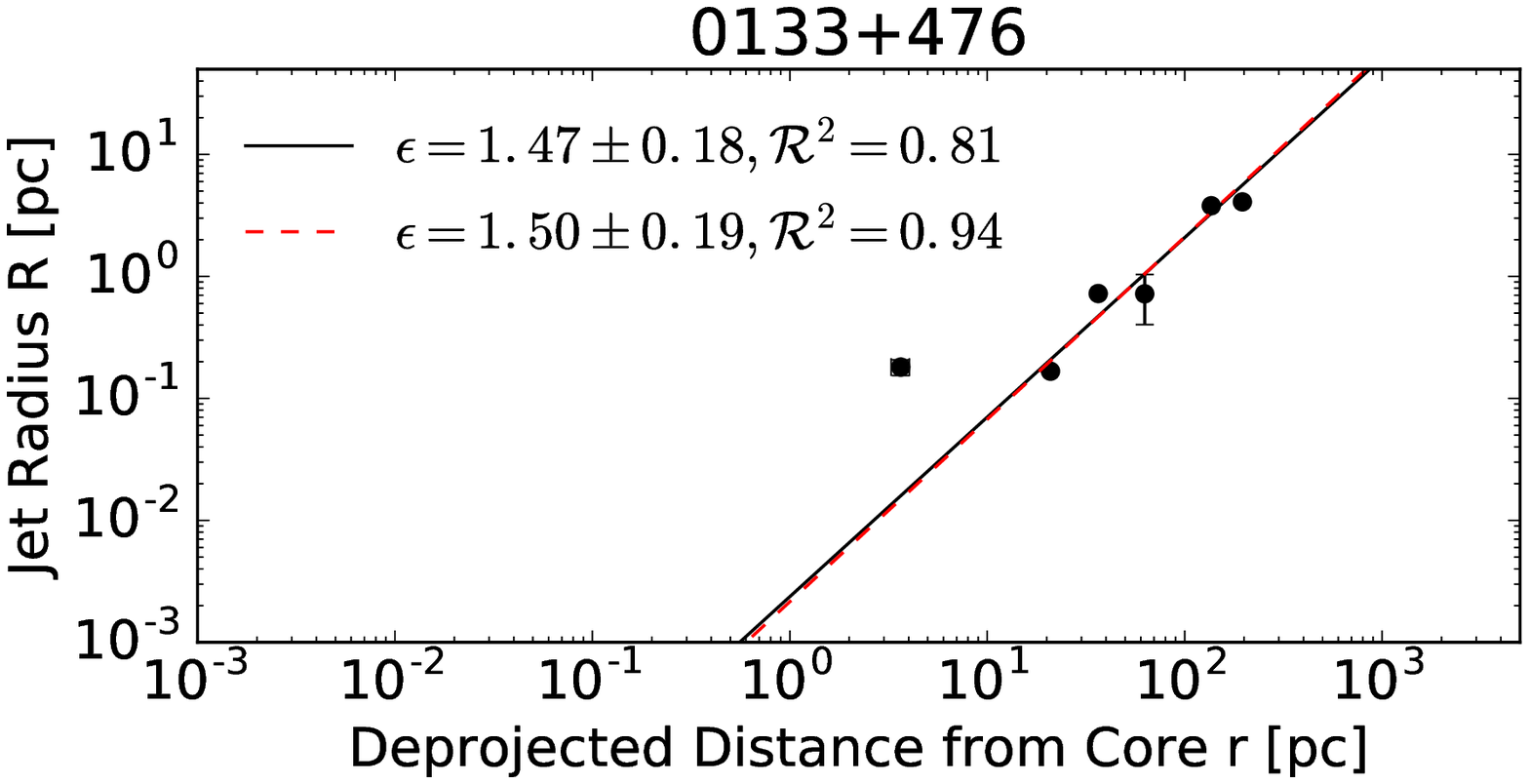}}\
\\
\vspace{-0.8cm} 
\subfigure{\includegraphics[trim=0cm 0cm 0cm 2.5cm, clip=true,width=0.24\textwidth]{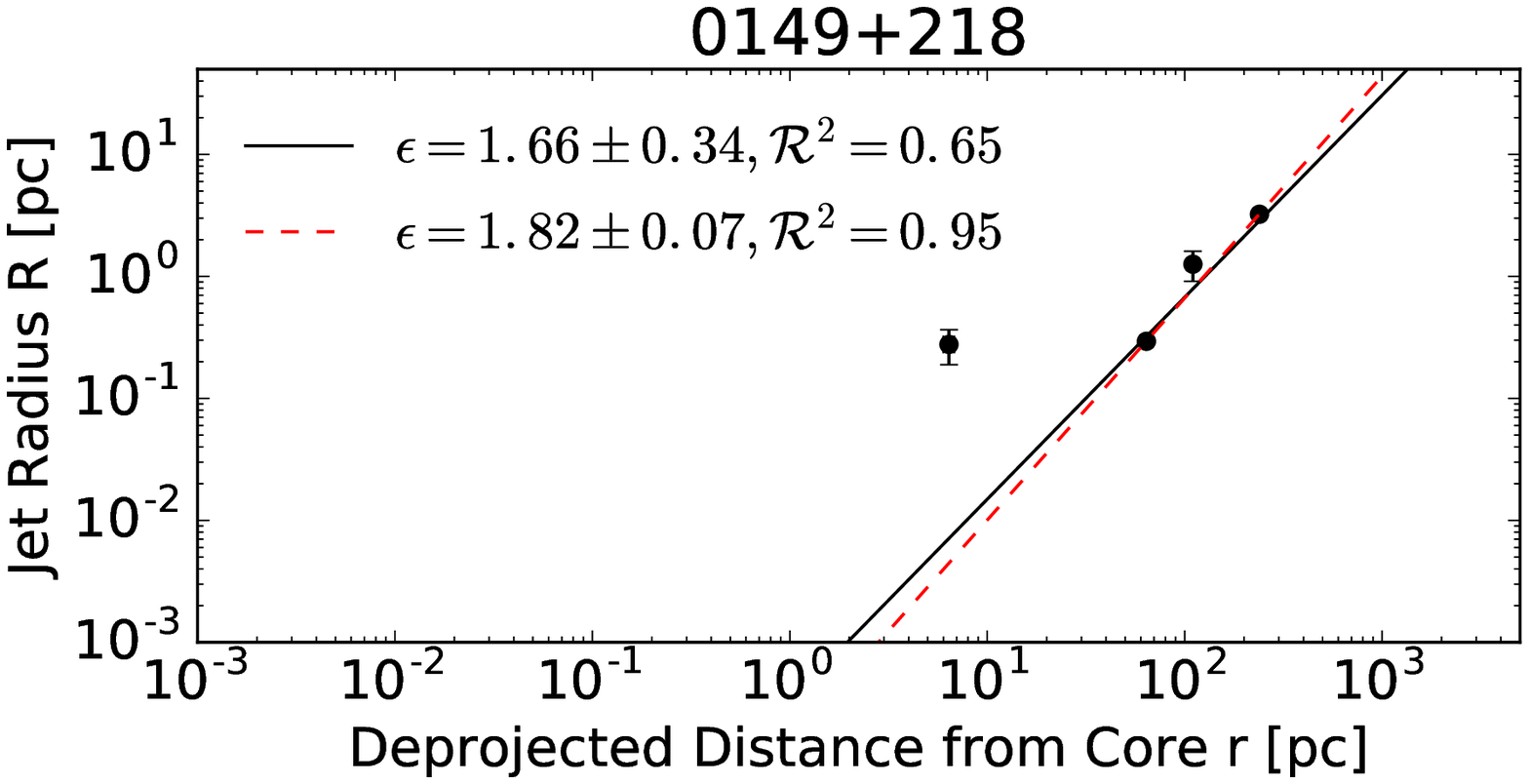}}\
\subfigure{\includegraphics[trim=0cm 0cm 0cm 2.5cm, clip=true,width=0.24\textwidth]{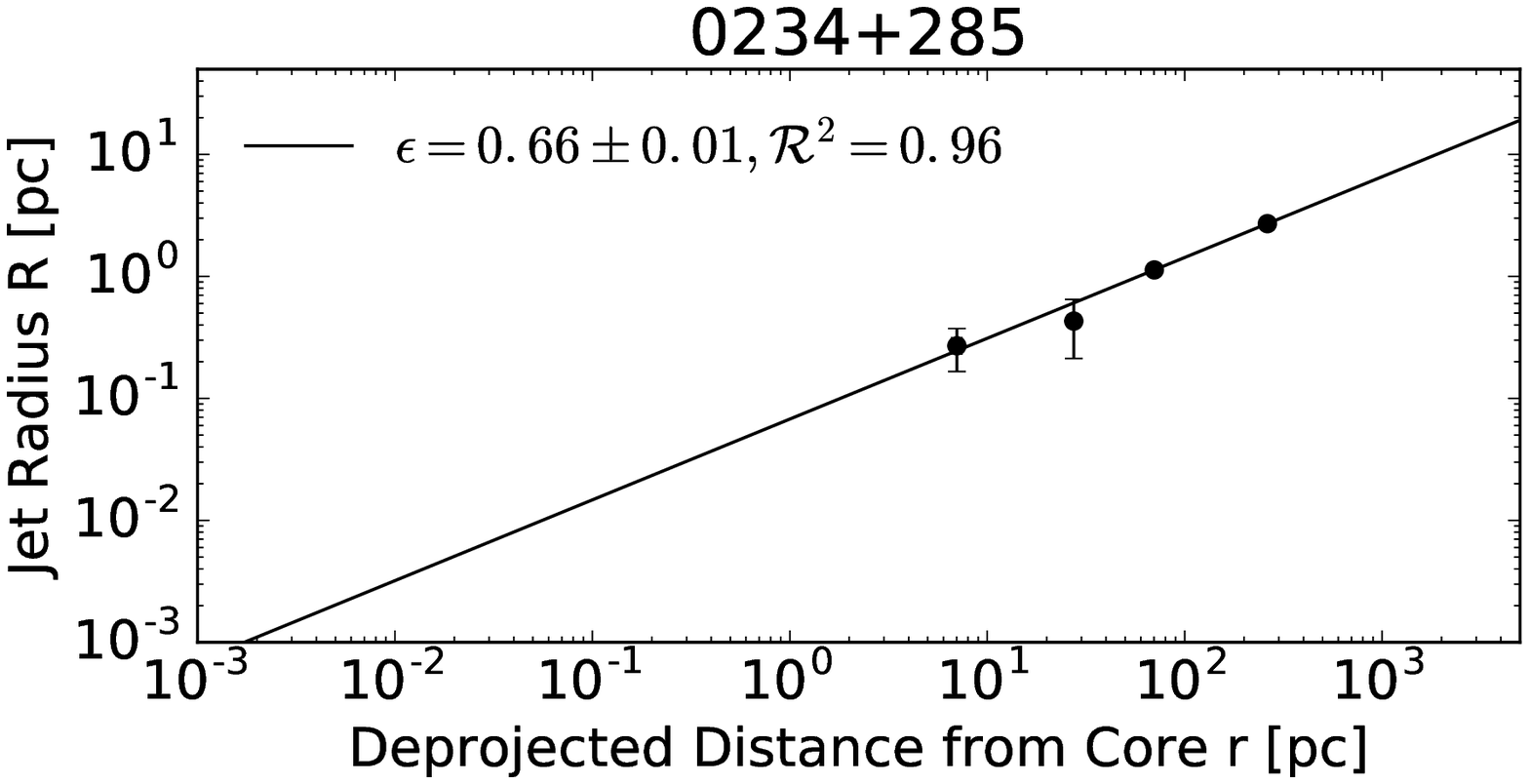}}\
\subfigure{\includegraphics[trim=0cm 0cm 0cm 2.5cm, clip=true,width=0.24\textwidth]{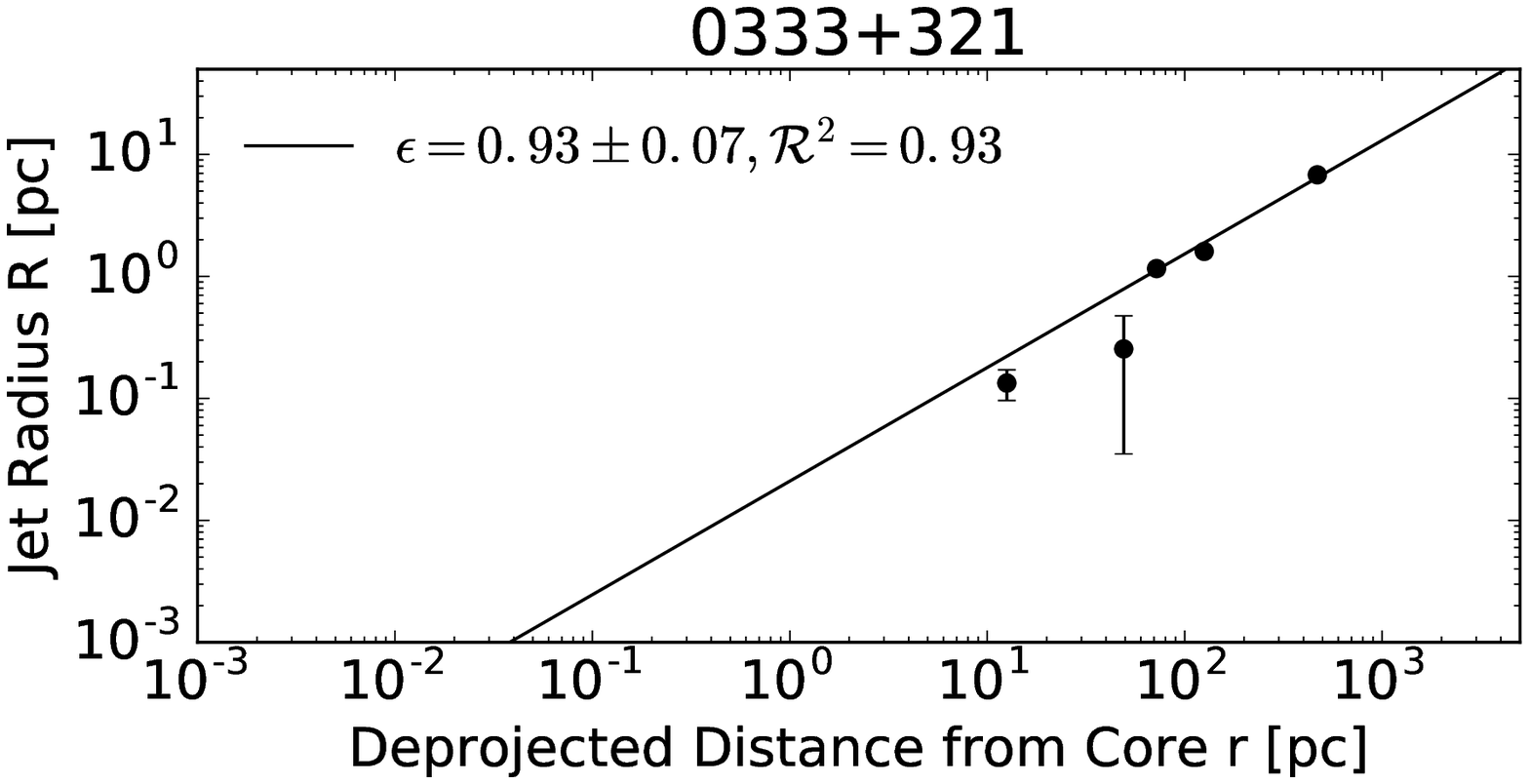}}\
\subfigure{\includegraphics[trim=0cm 0cm 0cm 2.5cm, clip=true,width=0.24\textwidth]{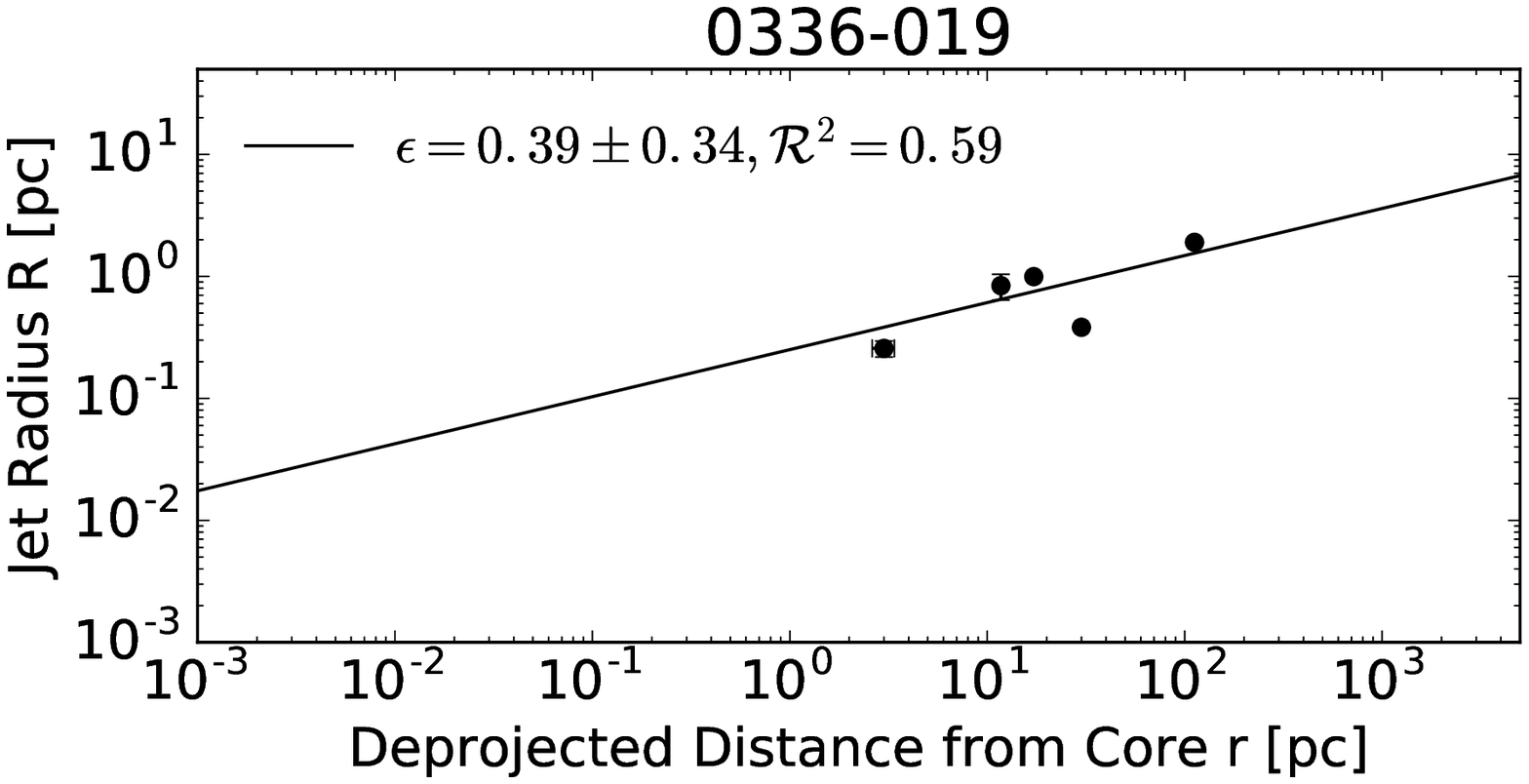}}\
\\
\vspace{-0.8cm} 
\subfigure{\includegraphics[trim=0cm 0cm 0cm 2.5cm, clip=true,width=0.24\textwidth]{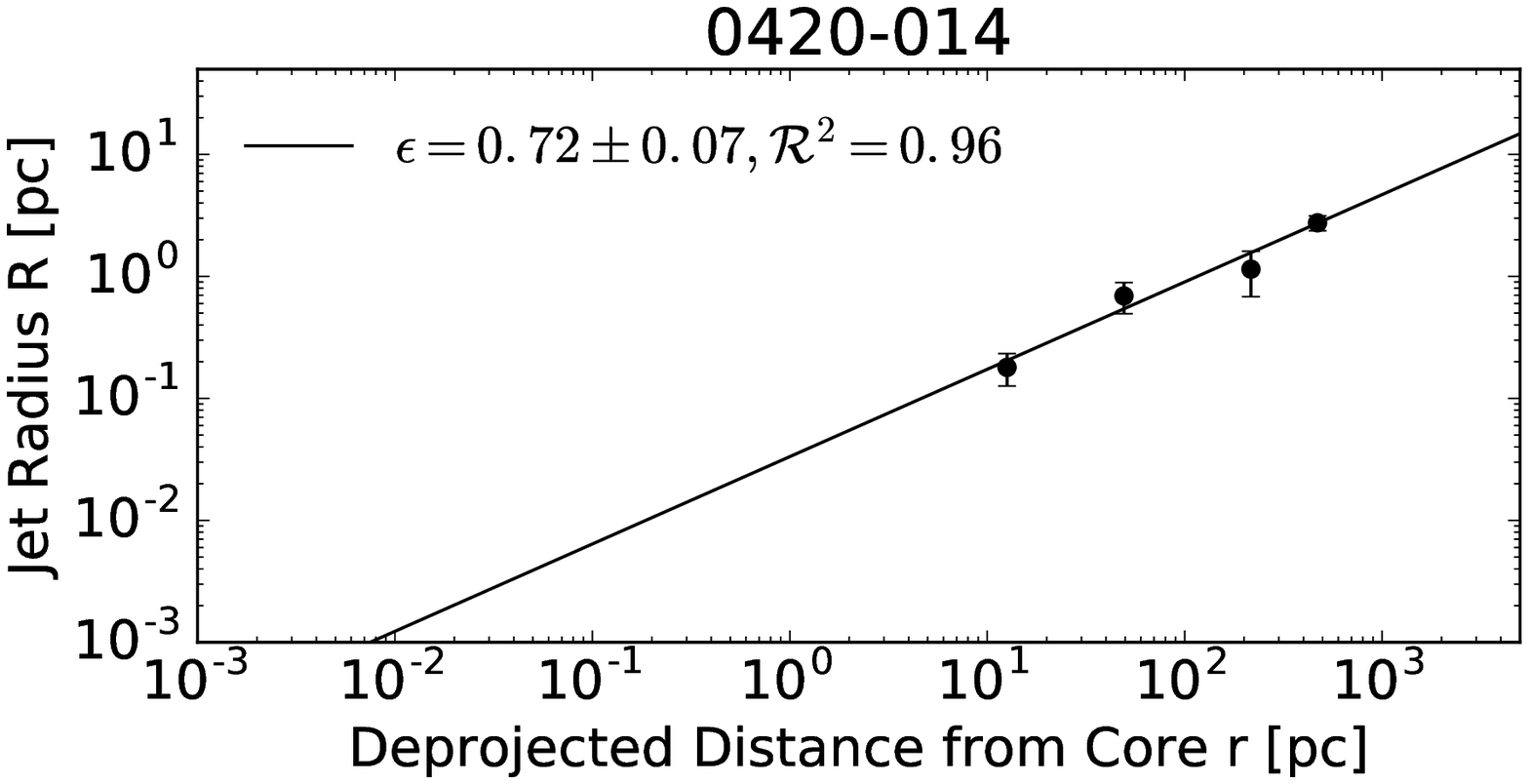}}\
\subfigure{\includegraphics[trim=0cm 0cm 0cm 2.5cm, clip=true,width=0.24\textwidth]{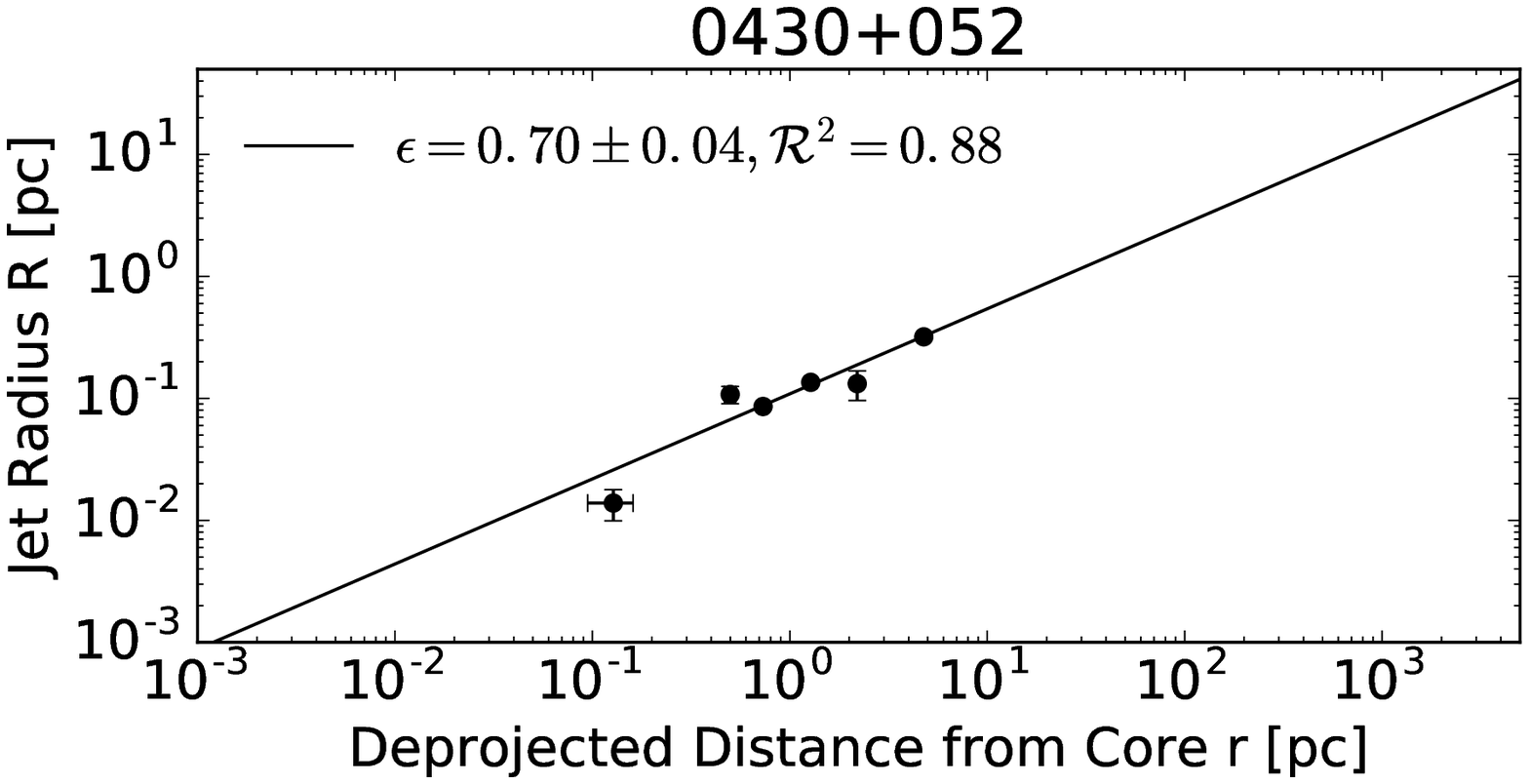}}\
\subfigure{\includegraphics[trim=0cm 0cm 0cm 2.5cm, clip=true,width=0.24\textwidth]{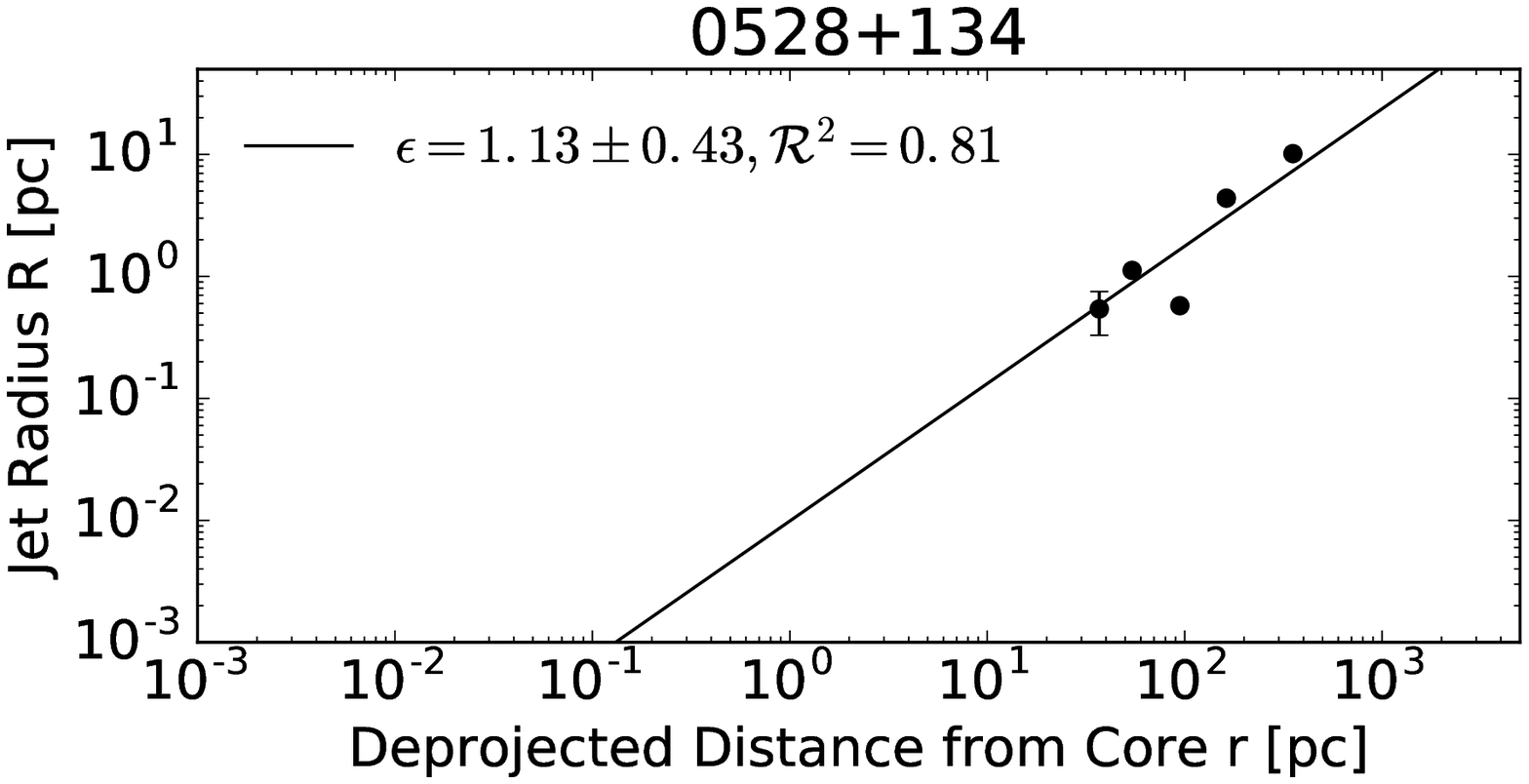}}\
\subfigure{\includegraphics[trim=0cm 0cm 0cm 2.5cm, clip=true,width=0.24\textwidth]{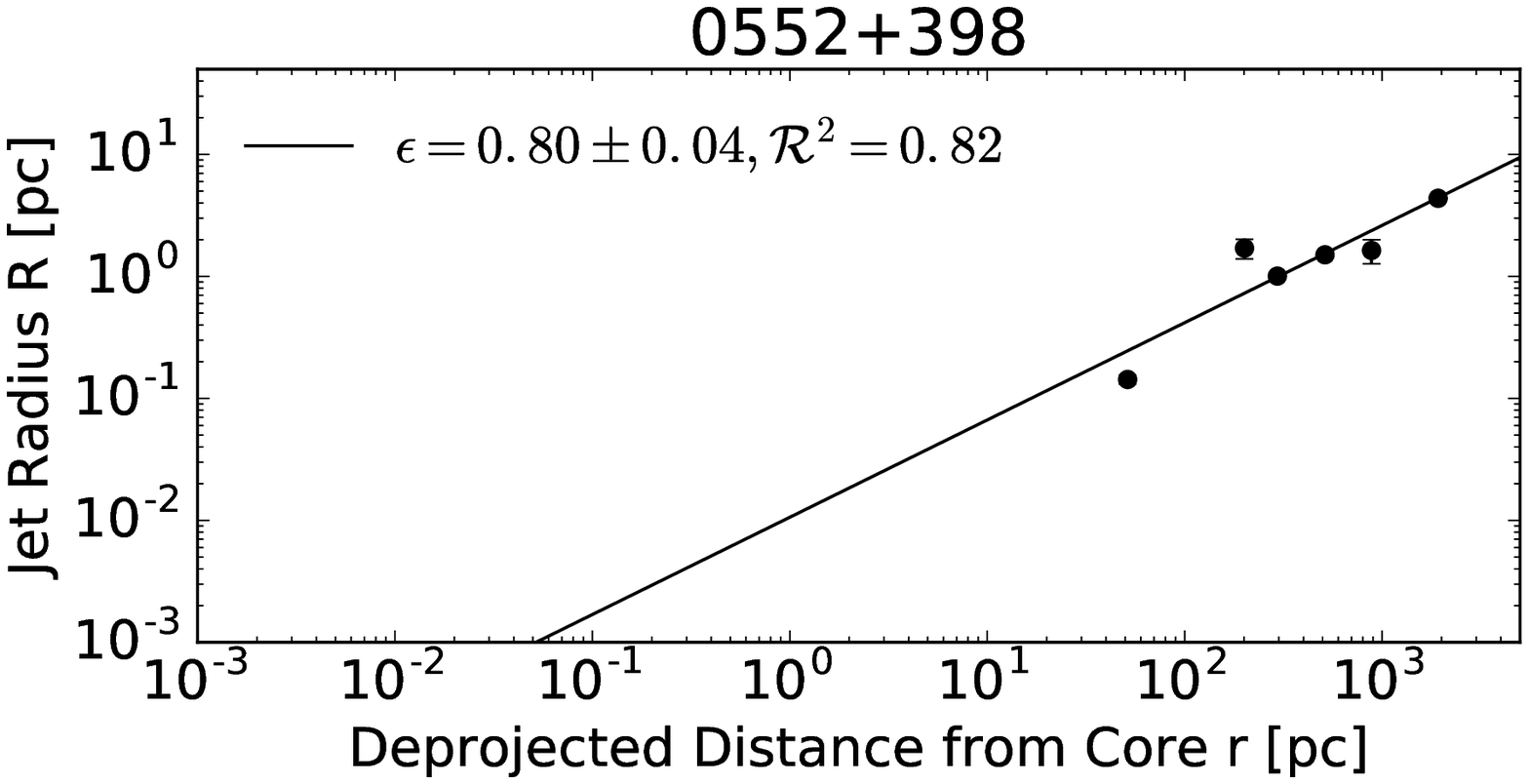}}\
\\
\vspace{-0.8cm} 
\subfigure{\includegraphics[trim=0cm 0cm 0cm 2.5cm, clip=true,width=0.24\textwidth]{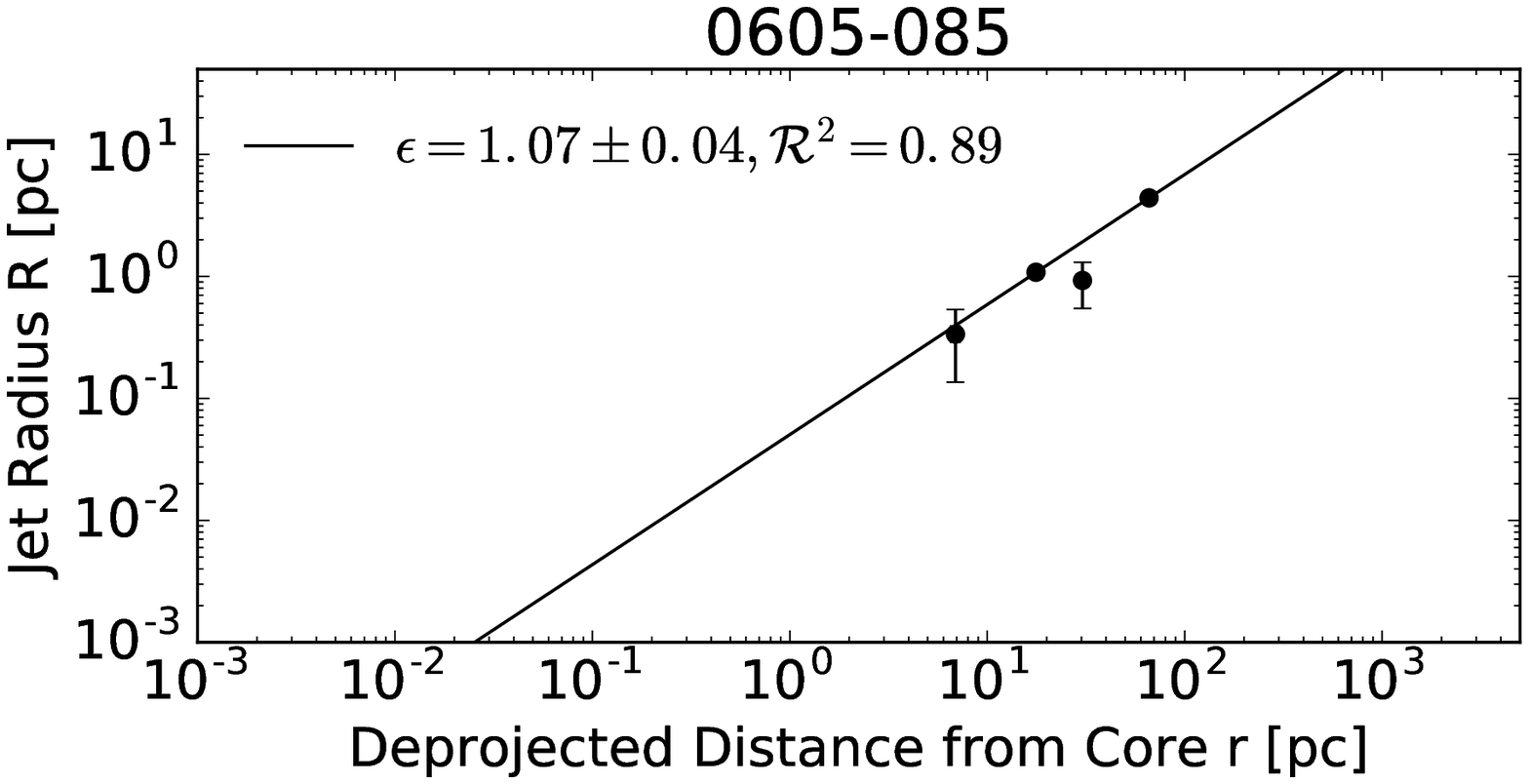}}\
\subfigure{\includegraphics[trim=0cm 0cm 0cm 2.5cm, clip=true,width=0.24\textwidth]{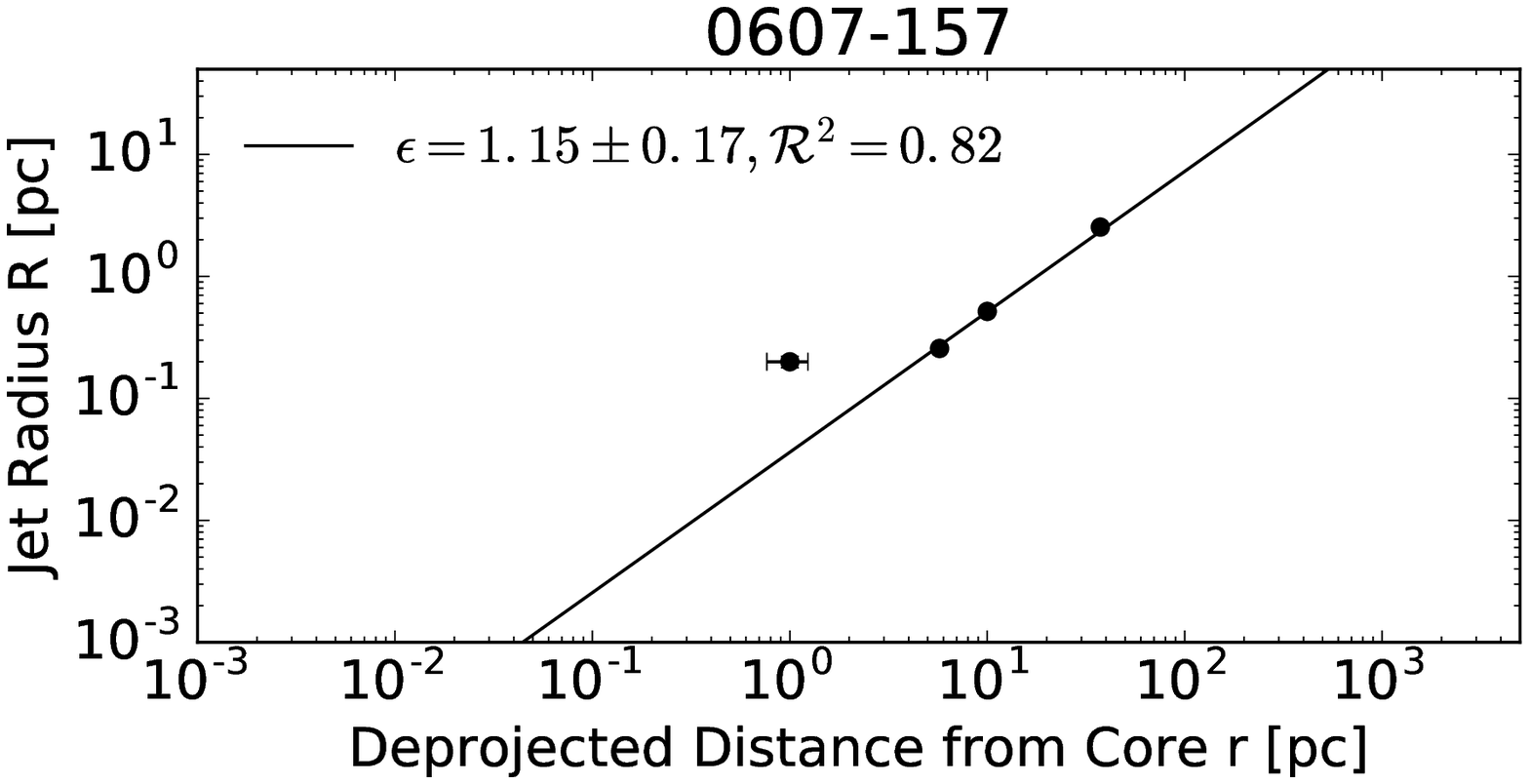}}\
\subfigure{\includegraphics[trim=0cm 0cm 0cm 2.5cm, clip=true,width=0.24\textwidth]{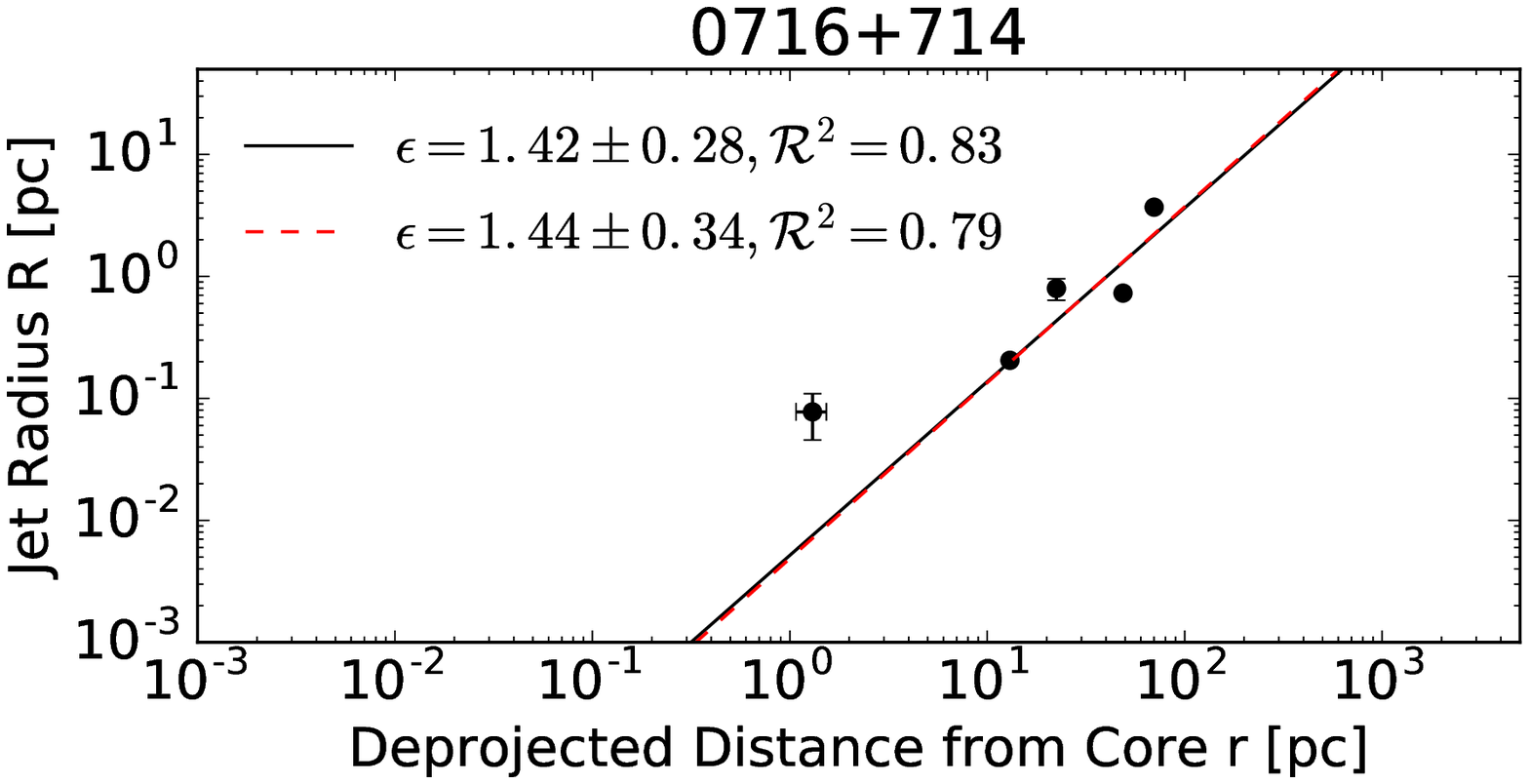}}\
\subfigure{\includegraphics[trim=0cm 0cm 0cm 2.5cm, clip=true,width=0.24\textwidth]{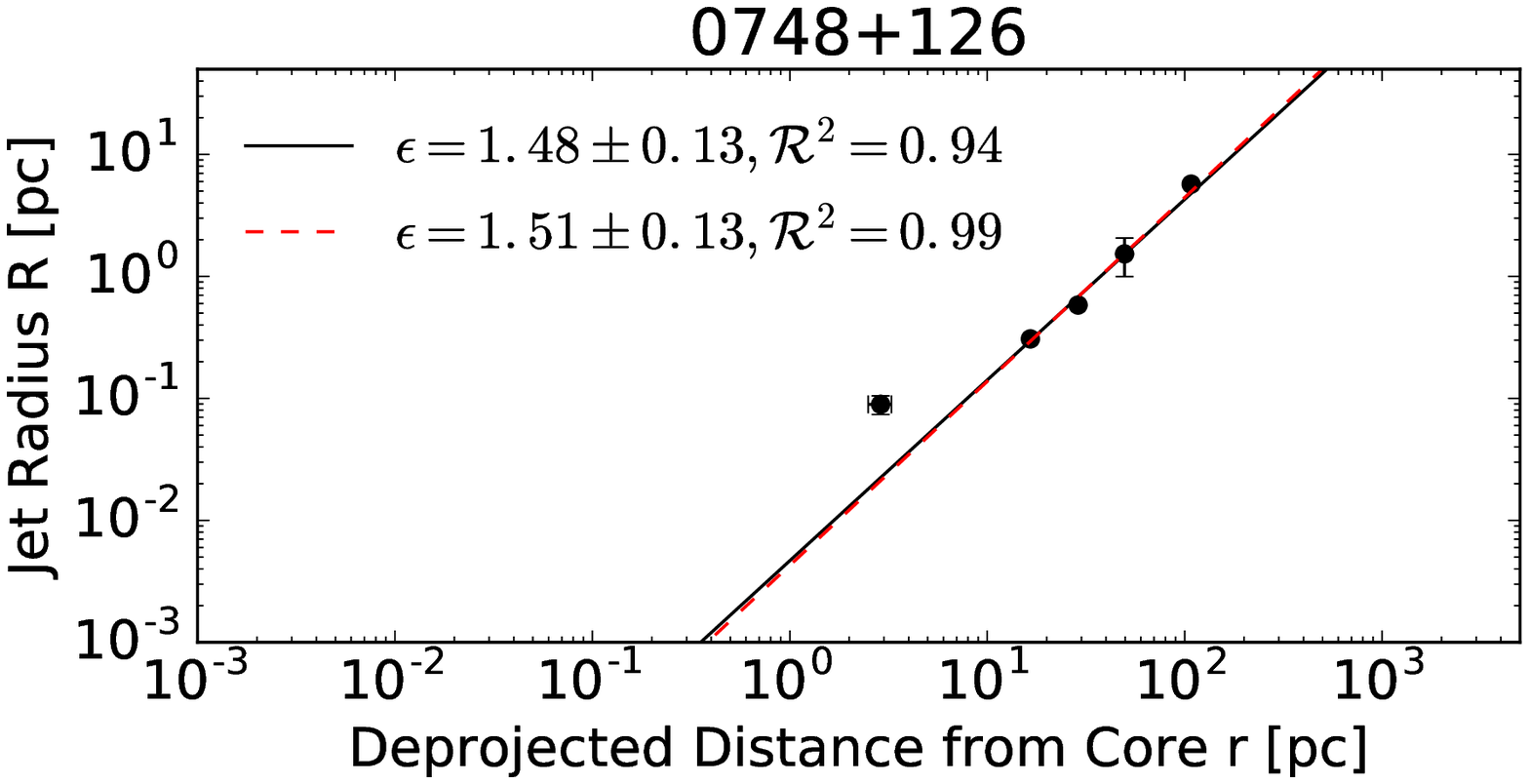}}\
\\
\vspace{-0.8cm} 
\subfigure{\includegraphics[trim=0cm 0cm 0cm 2.5cm, clip=true,width=0.24\textwidth]{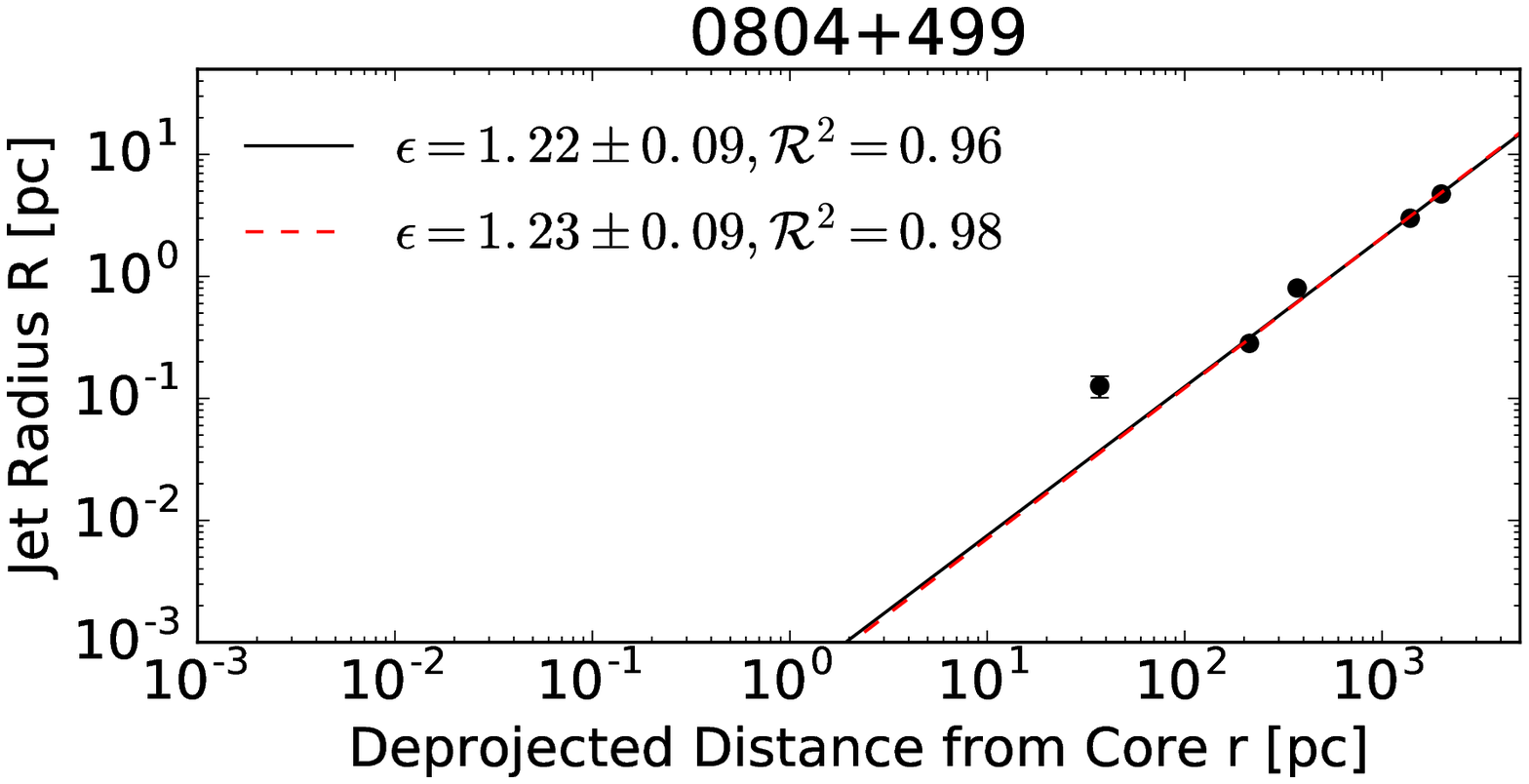}}\
\subfigure{\includegraphics[trim=0cm 0cm 0cm 2.5cm, clip=true,width=0.24\textwidth]{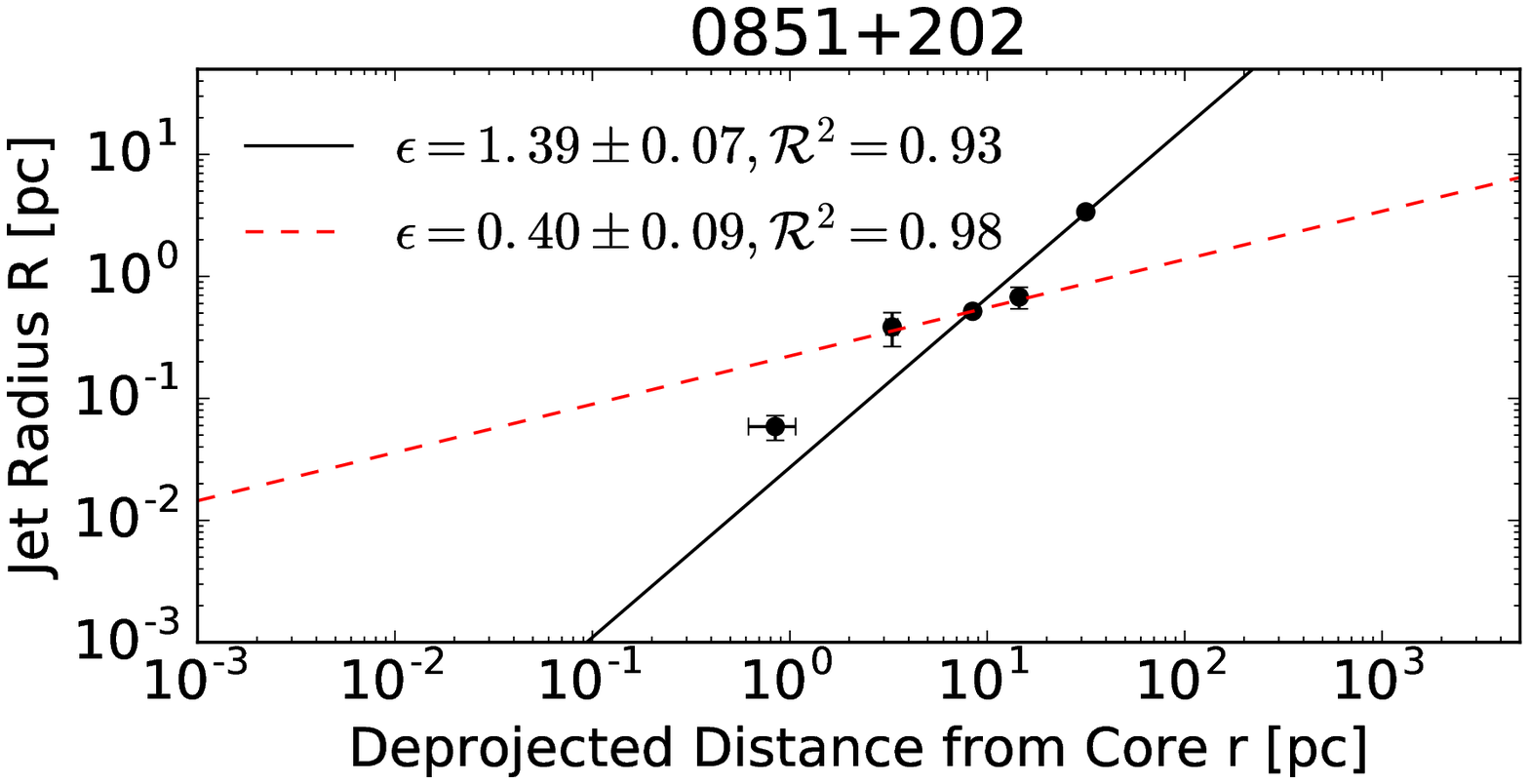}}\
\subfigure{\includegraphics[trim=0cm 0cm 0cm 2.5cm, clip=true,width=0.24\textwidth]{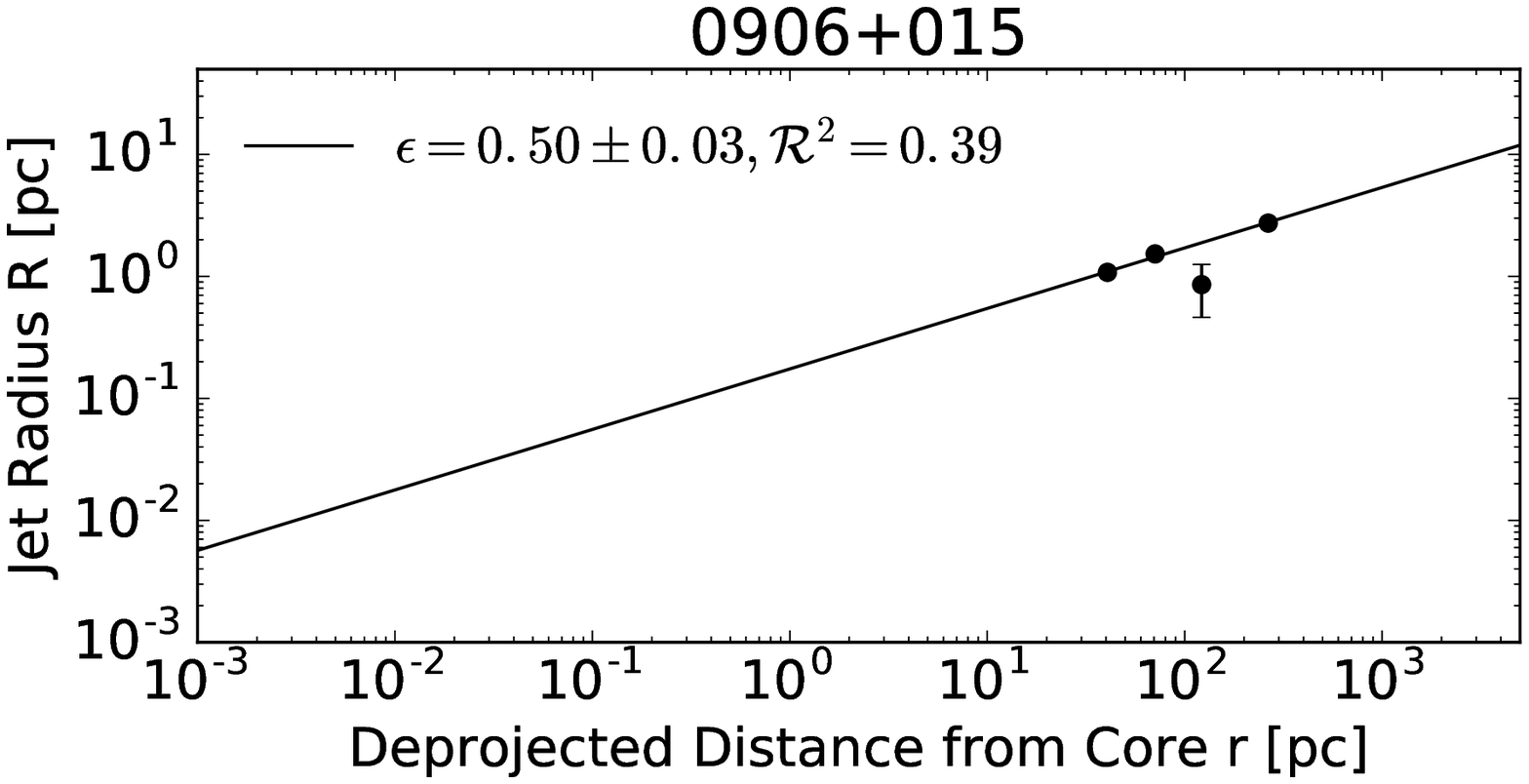}}\
\subfigure{\includegraphics[trim=0cm 0cm 0cm 2.5cm, clip=true,width=0.24\textwidth]{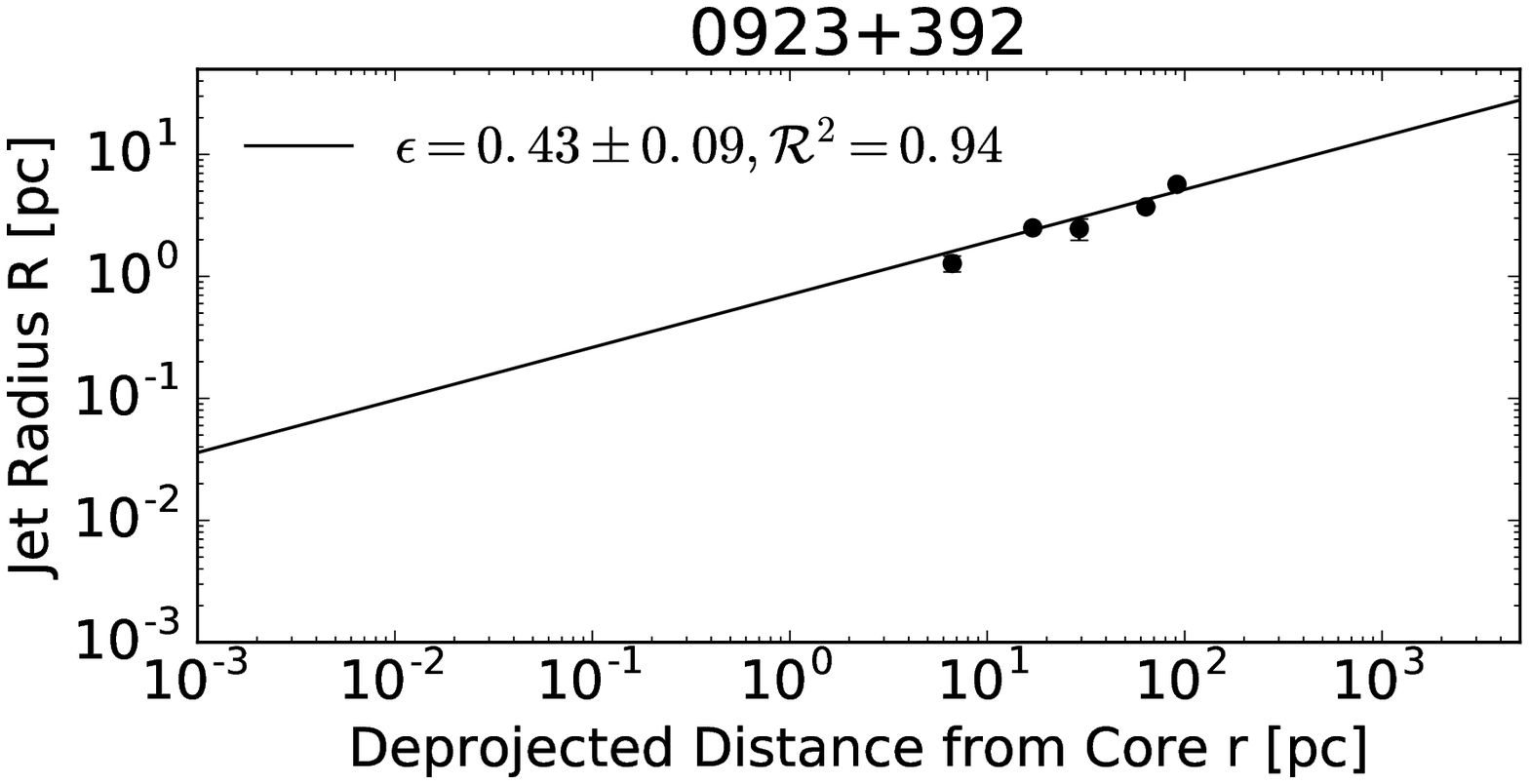}}\
\\
\vspace{-0.8cm} 
\subfigure{\includegraphics[trim=0cm 0cm 0cm 2.5cm, clip=true,width=0.24\textwidth]{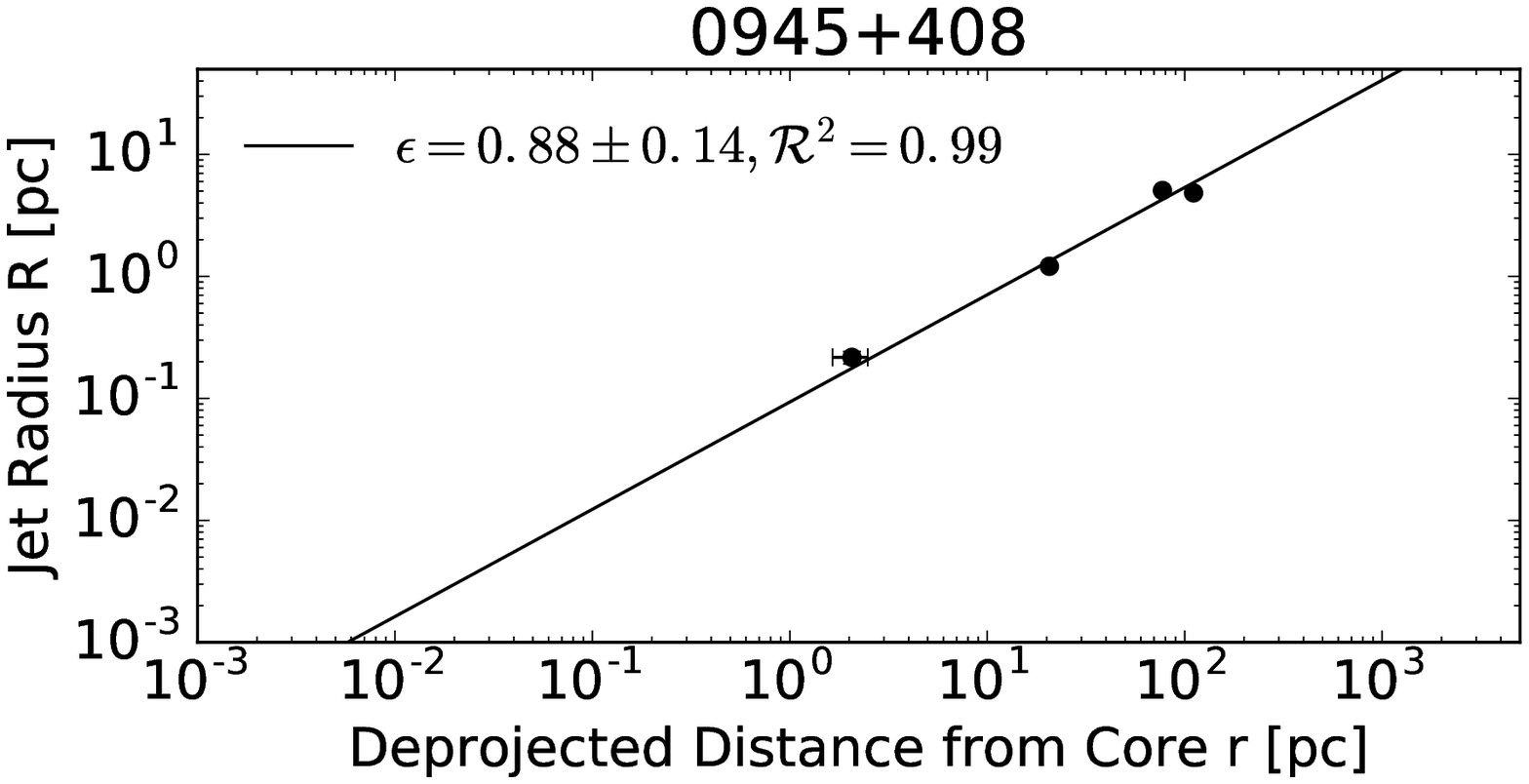}}\
\subfigure{\includegraphics[trim=0cm 0cm 0cm 2.5cm, clip=true,width=0.24\textwidth]{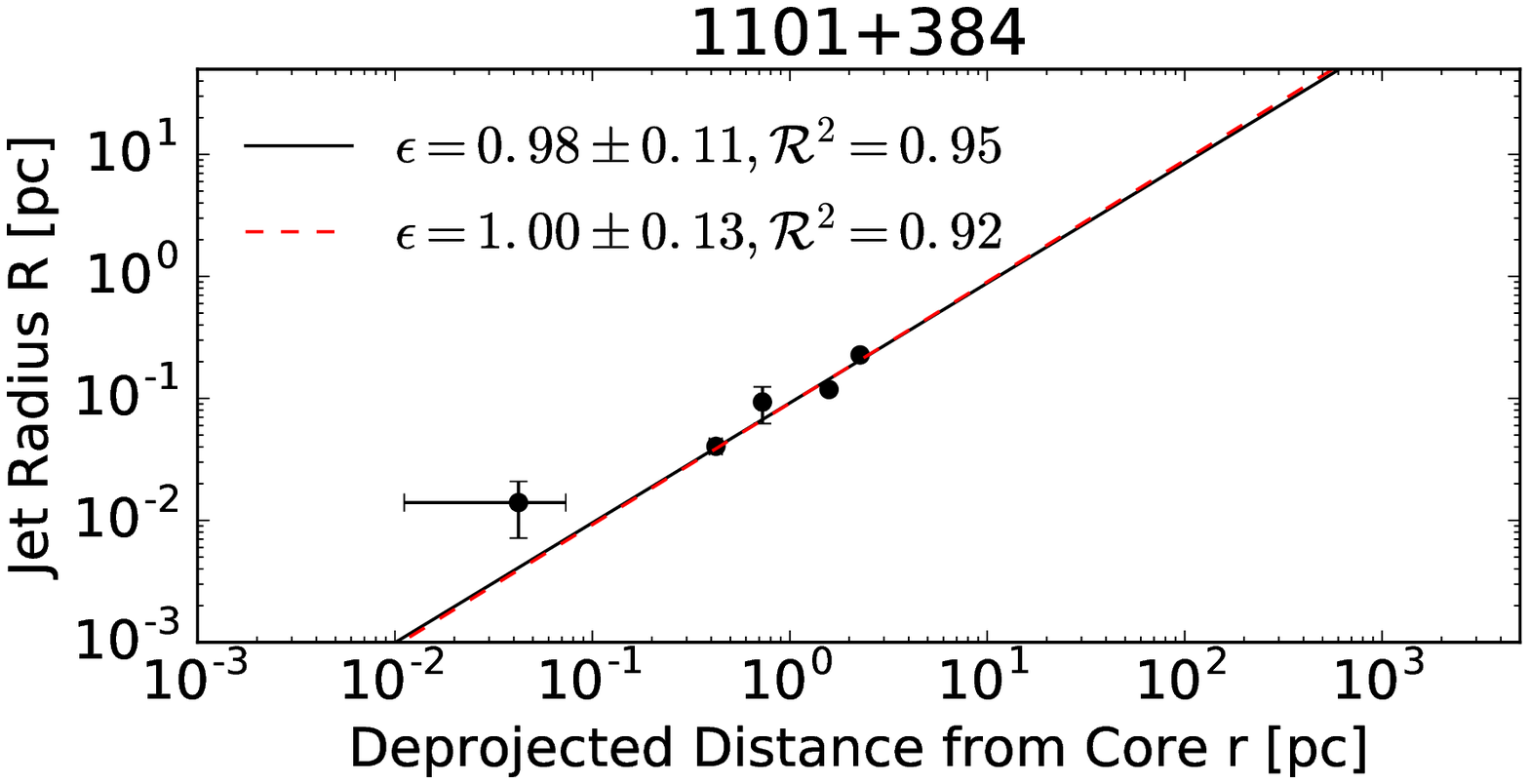}}\
\subfigure{\includegraphics[trim=0cm 0cm 0cm 2.5cm, clip=true,width=0.24\textwidth]{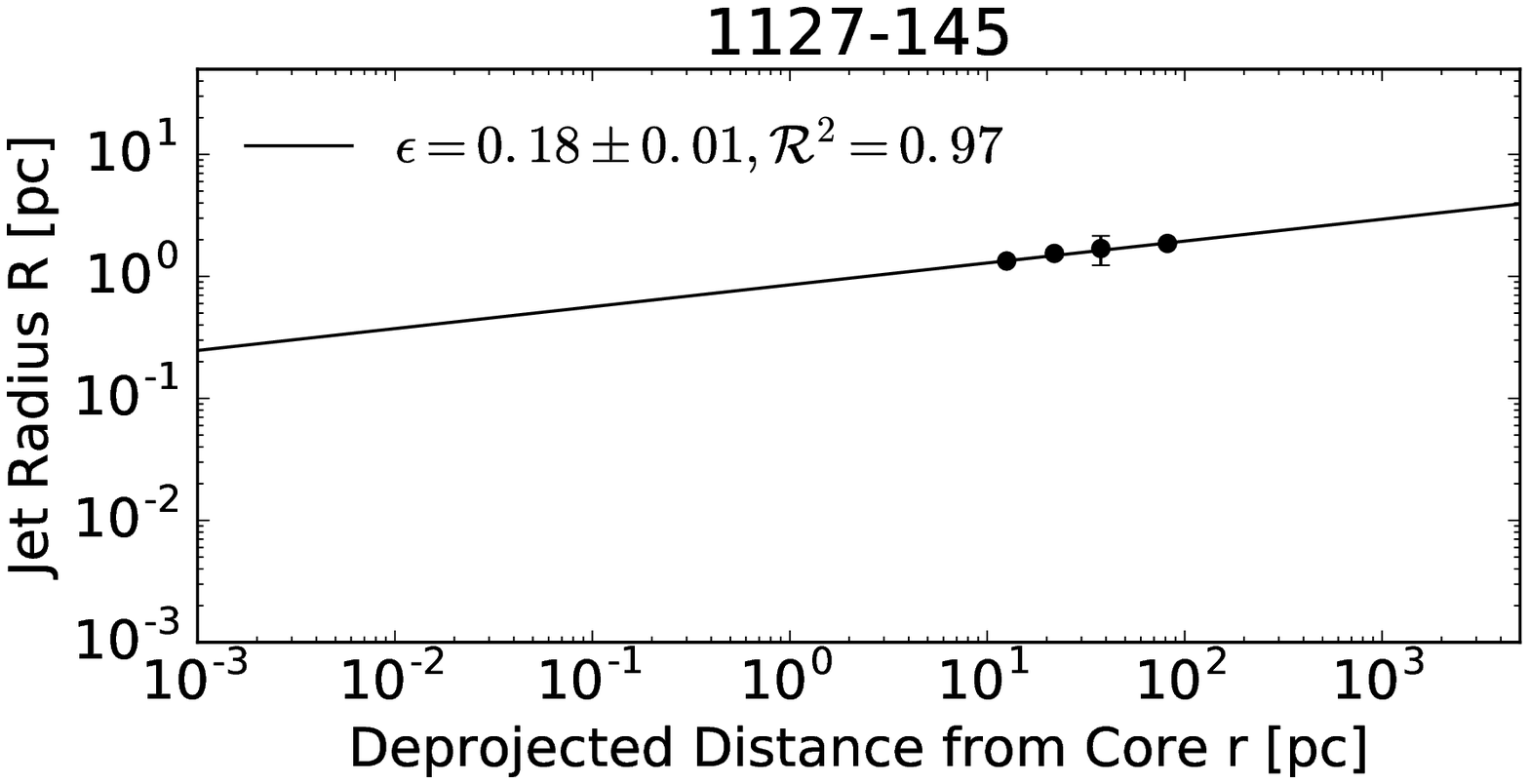}}\
\subfigure{\includegraphics[trim=0cm 0cm 0cm 2.5cm, clip=true,width=0.24\textwidth]{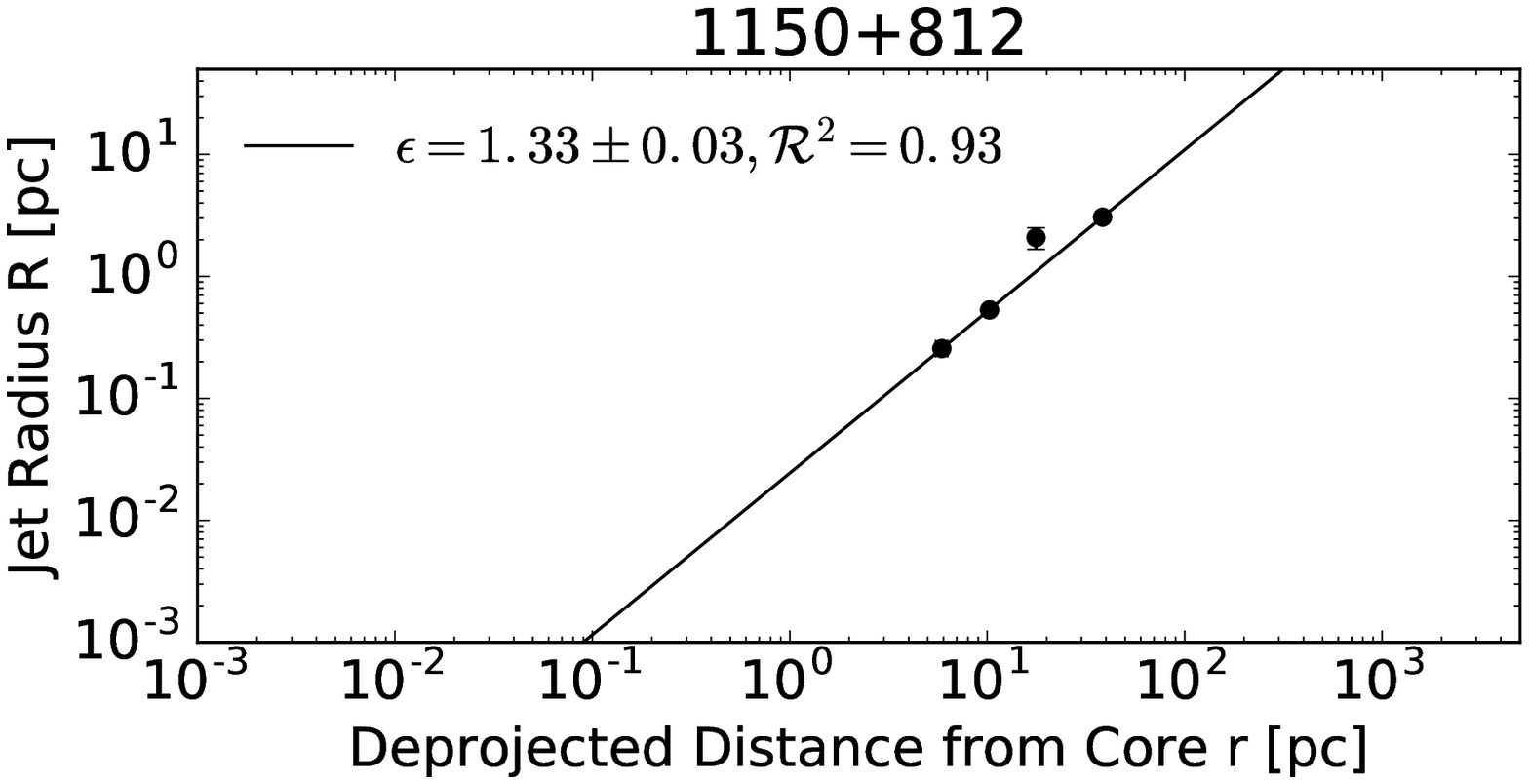}}\
\\
\vspace{-0.8cm} 
\subfigure{\includegraphics[trim=0cm 0cm 0cm 2.5cm, clip=true,width=0.24\textwidth]{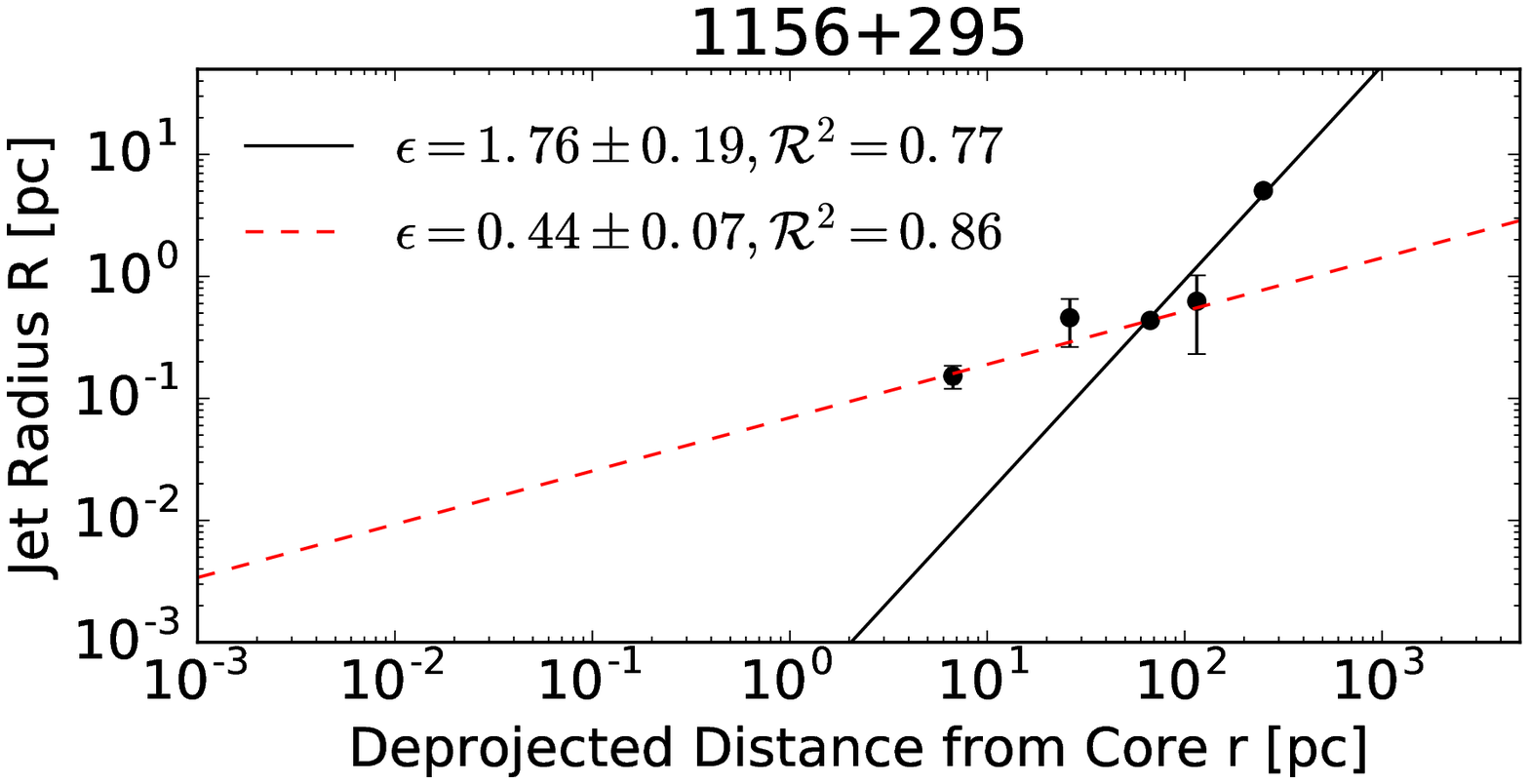}}\
\subfigure{\includegraphics[trim=0cm 0cm 0cm 2.5cm, clip=true,width=0.24\textwidth]{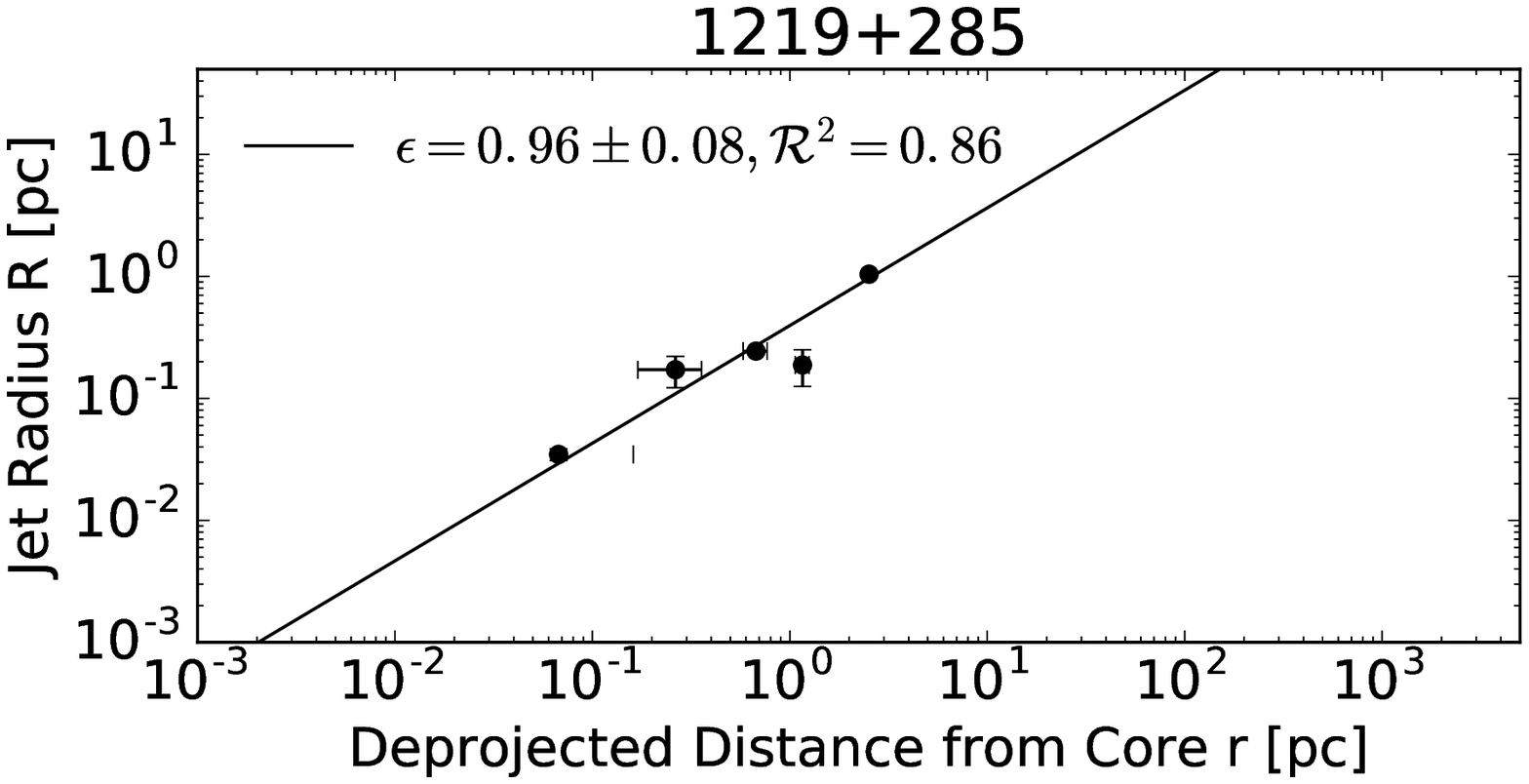}}\
\subfigure{\includegraphics[trim=0cm 0cm 0cm 2.5cm, clip=true,width=0.24\textwidth]{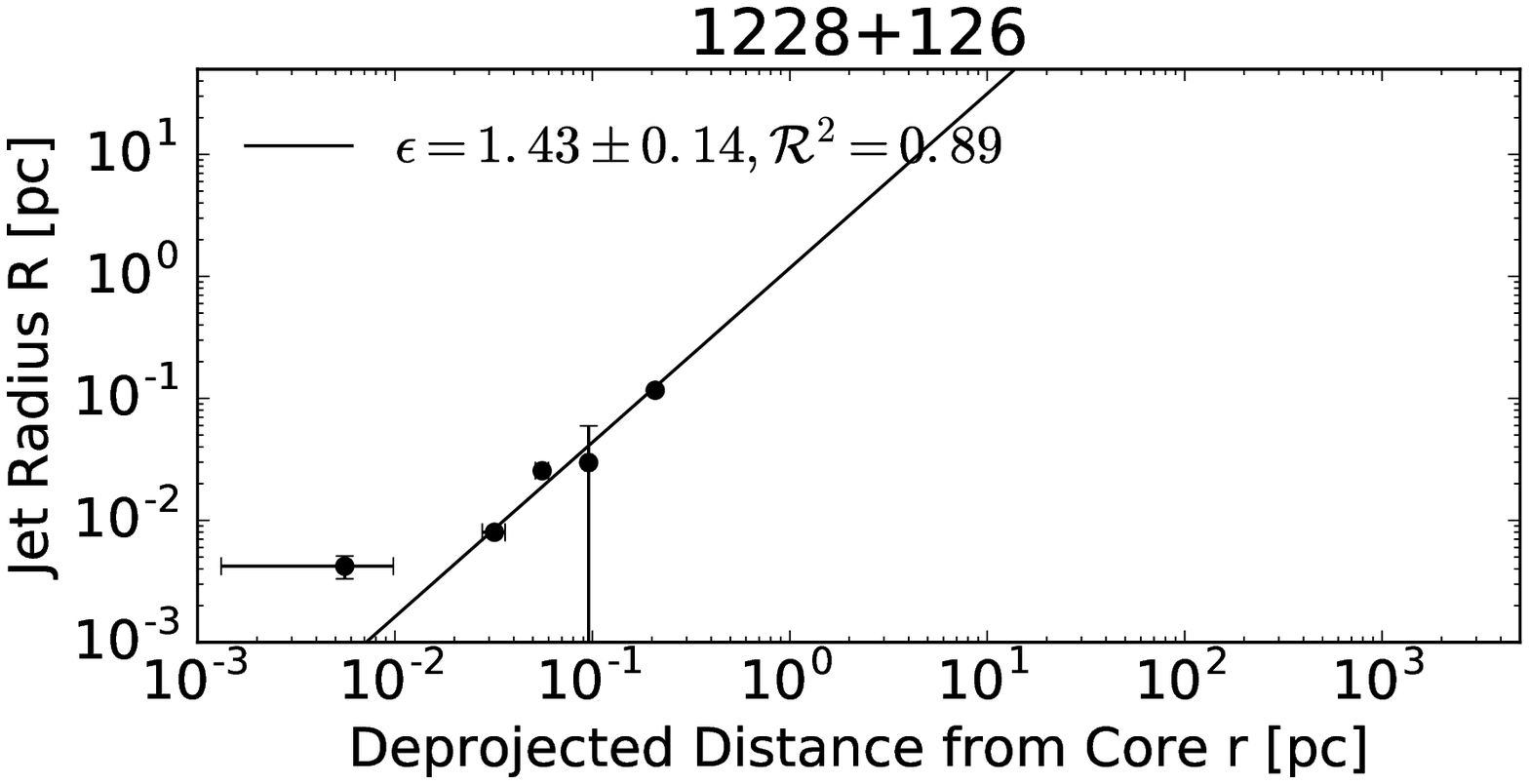}}\
\subfigure{\includegraphics[trim=0cm 0cm 0cm 2.5cm, clip=true,width=0.24\textwidth]{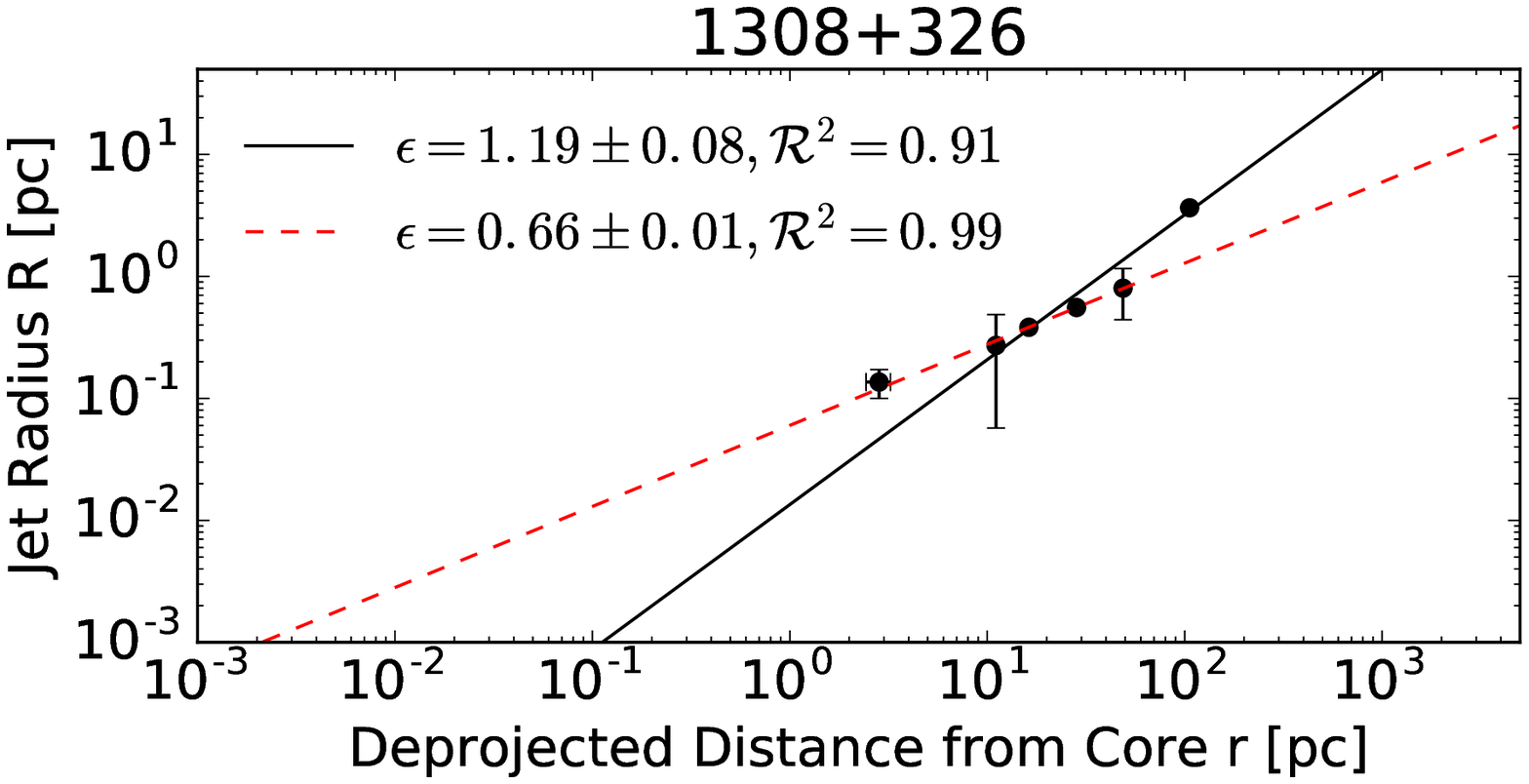}}\
\\
\vspace{-0.8cm} 
\subfigure{\includegraphics[trim=0cm 0cm 0cm 2.5cm, clip=true,width=0.24\textwidth]{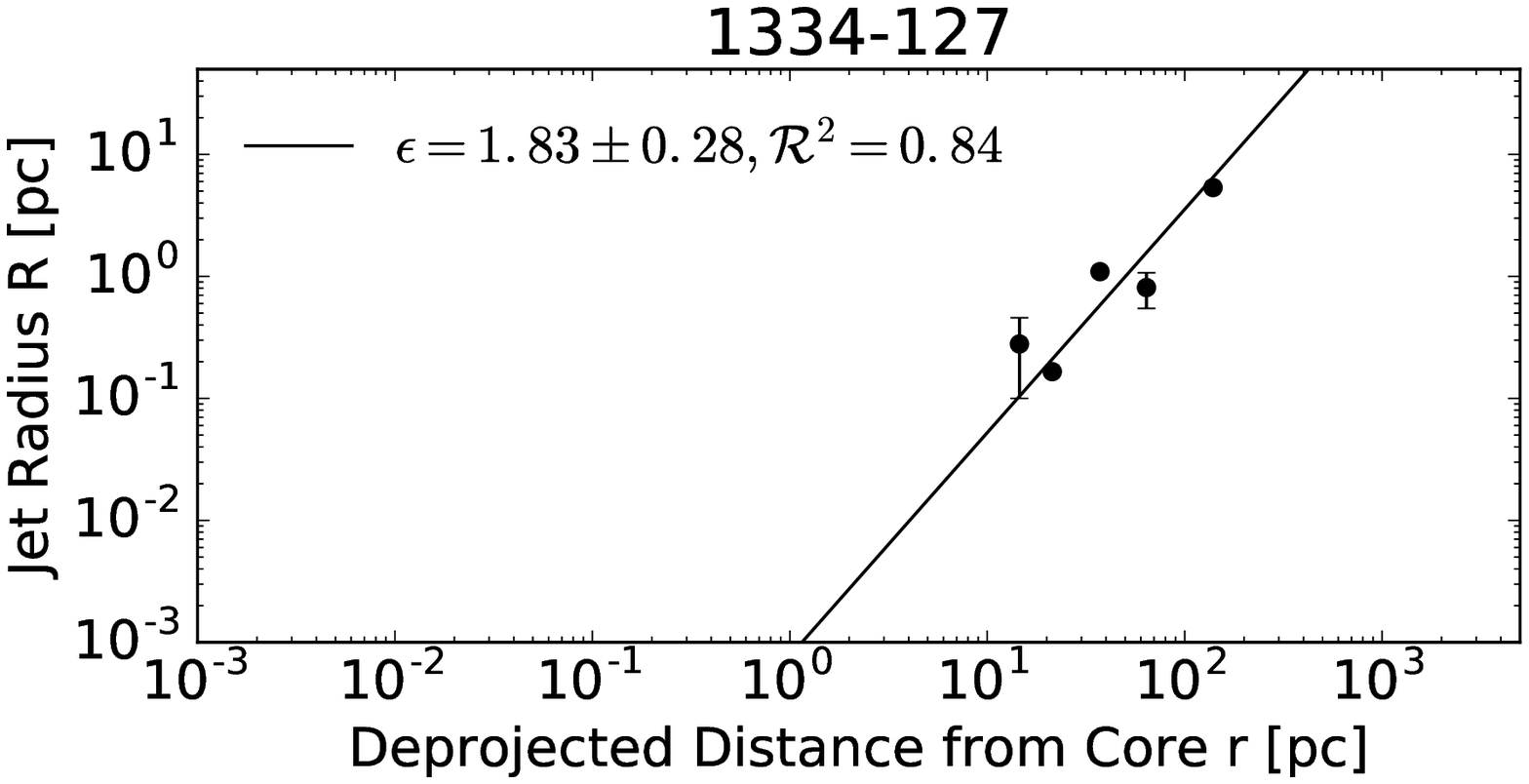}}\
\subfigure{\includegraphics[trim=0cm 0cm 0cm 2.5cm, clip=true,width=0.24\textwidth]{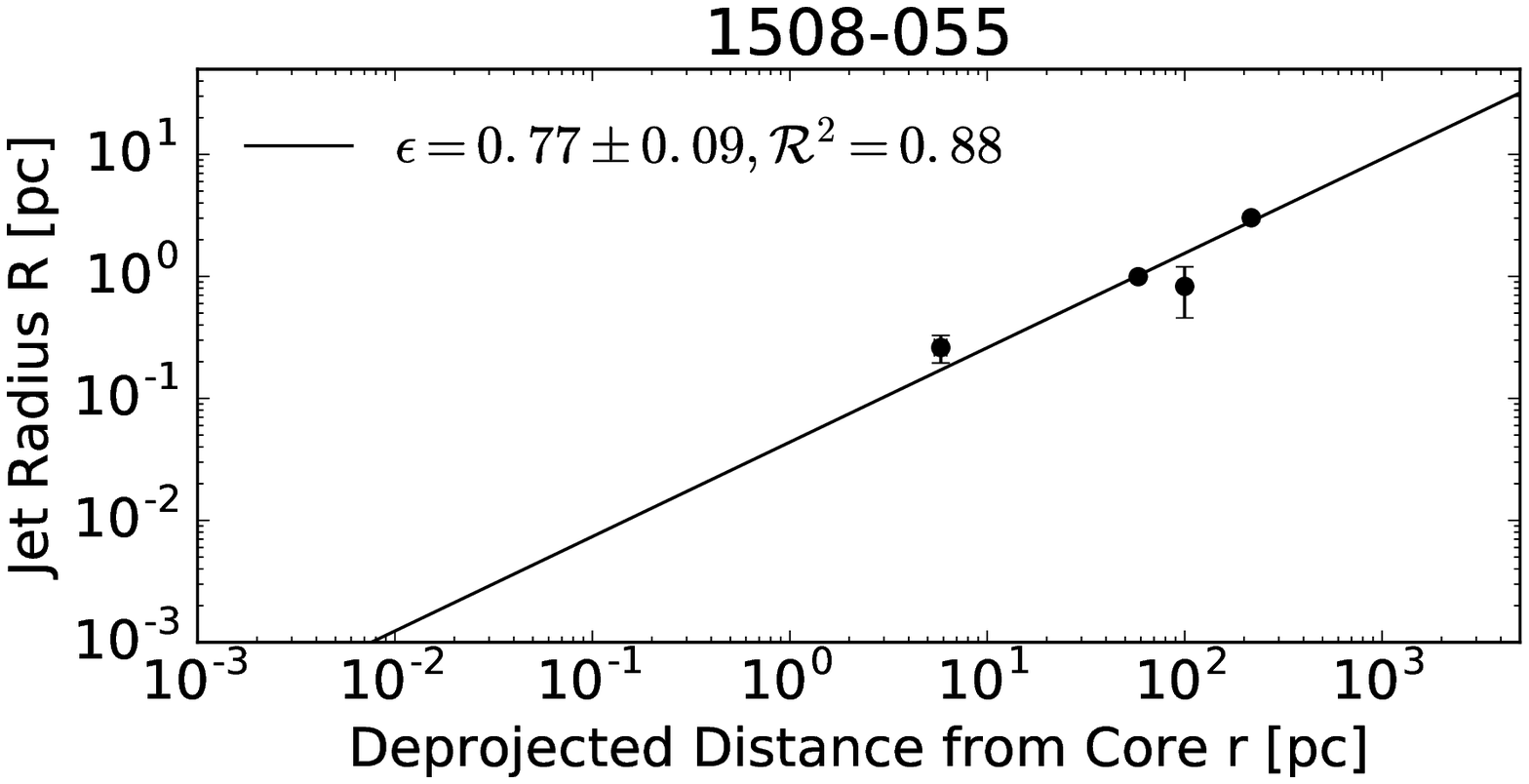}}\
\subfigure{\includegraphics[trim=0cm 0cm 0cm 2.5cm, clip=true,width=0.24\textwidth]{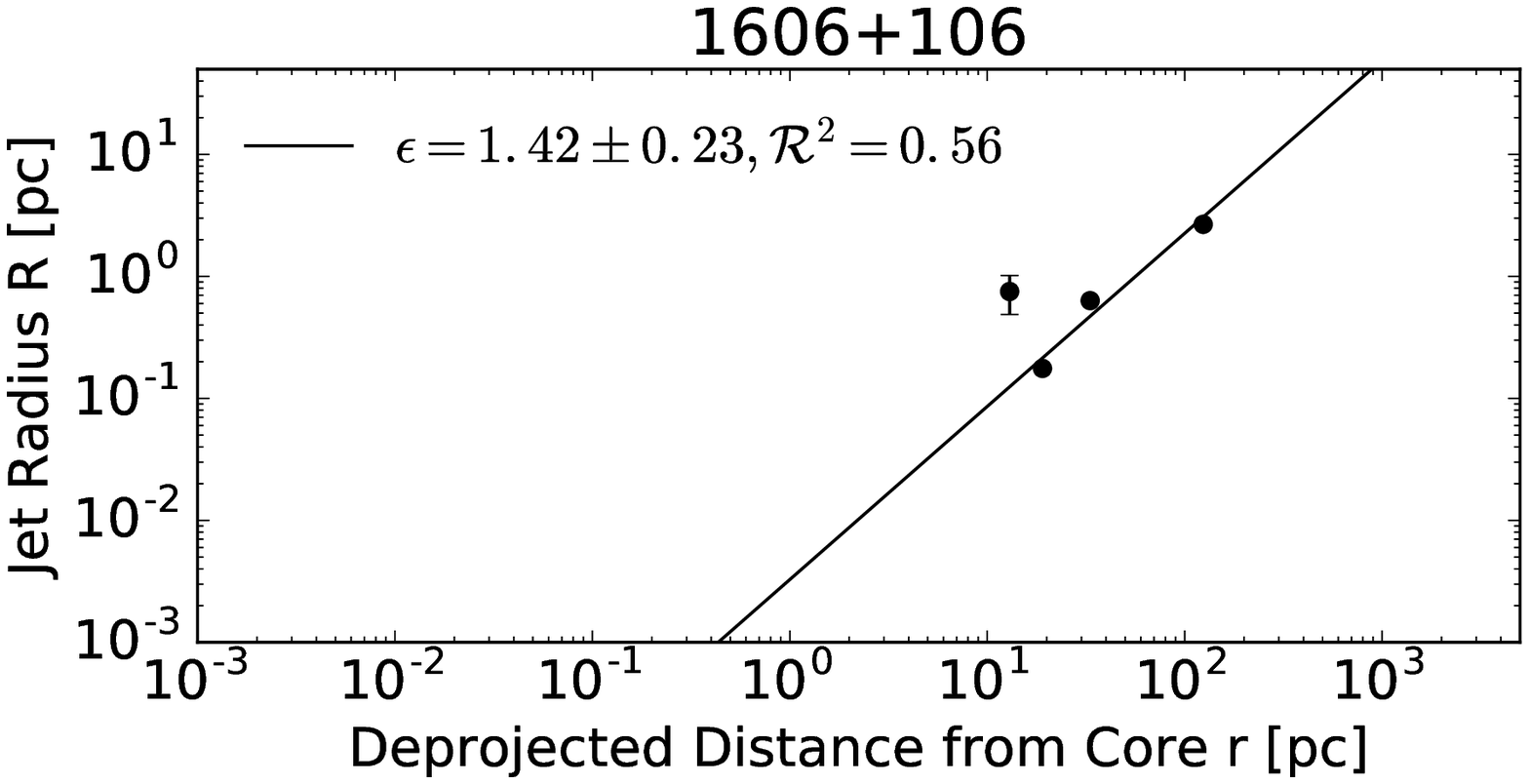}}\
\subfigure{\includegraphics[trim=0cm 0cm 0cm 2.5cm, clip=true,width=0.24\textwidth]{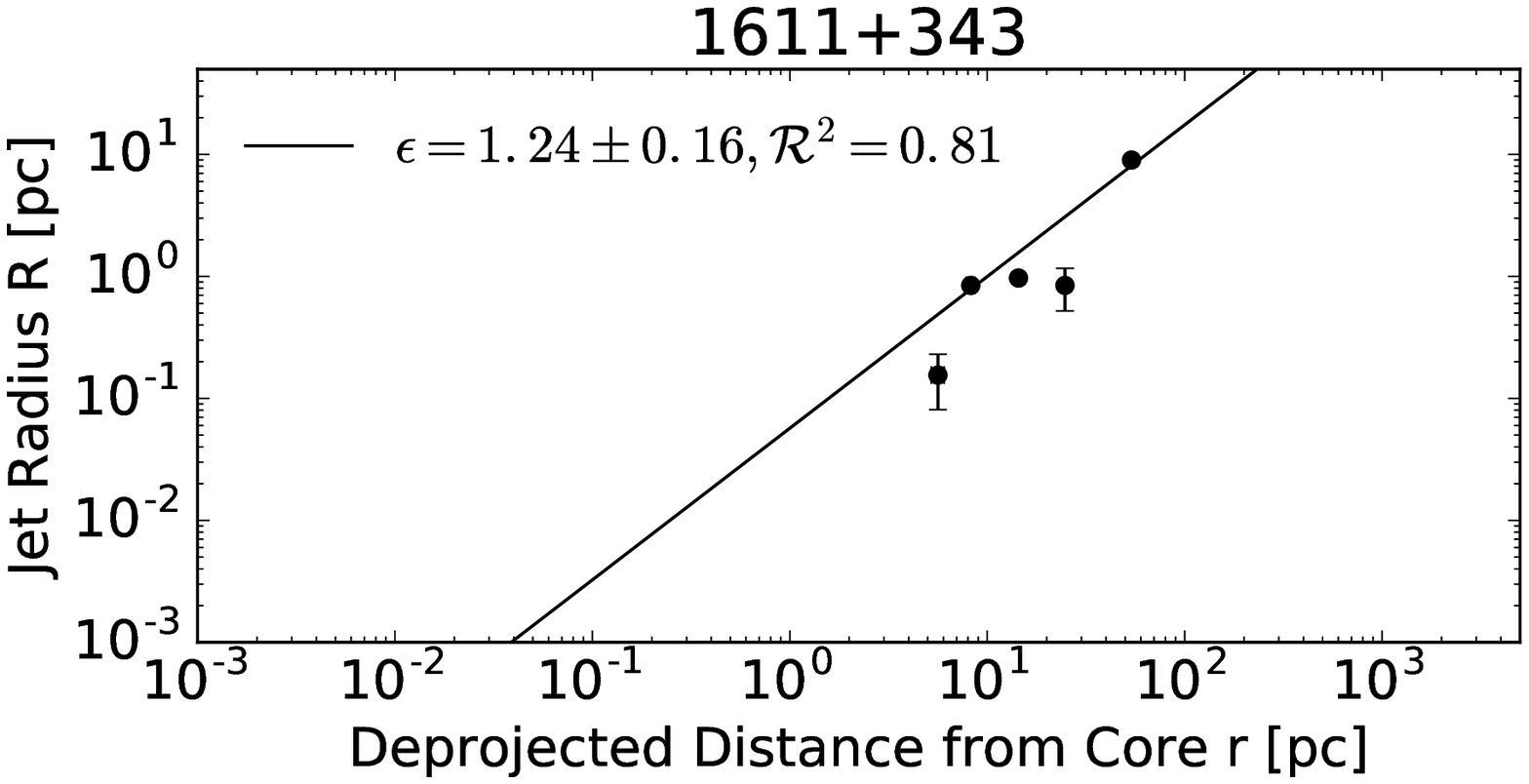}}\
\\
\vspace{-0.8cm} 
\subfigure{\includegraphics[trim=0cm 0cm 0cm 2.5cm, clip=true,width=0.24\textwidth]{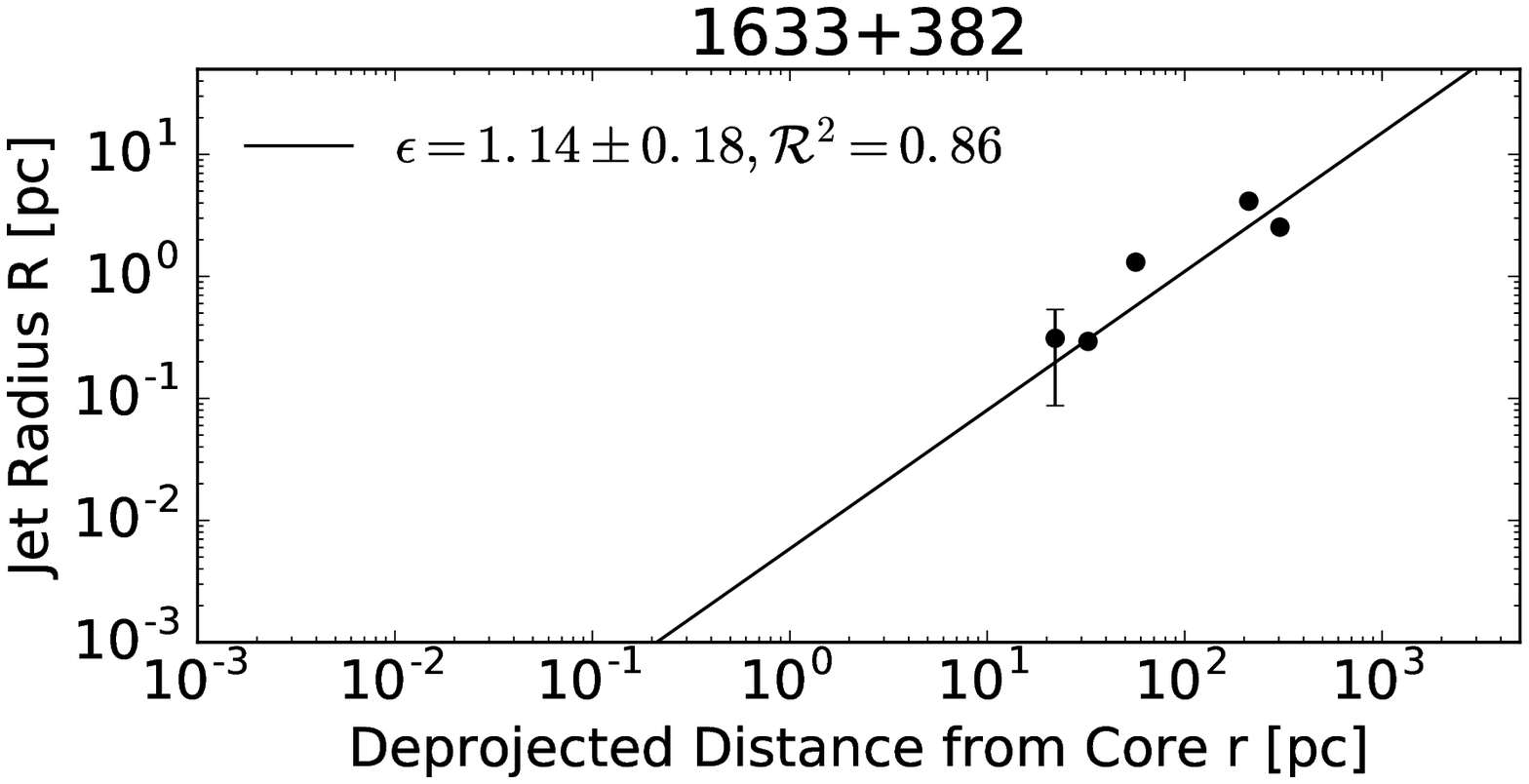}}\
\subfigure{\includegraphics[trim=0cm 0cm 0cm 2.5cm, clip=true,width=0.24\textwidth]{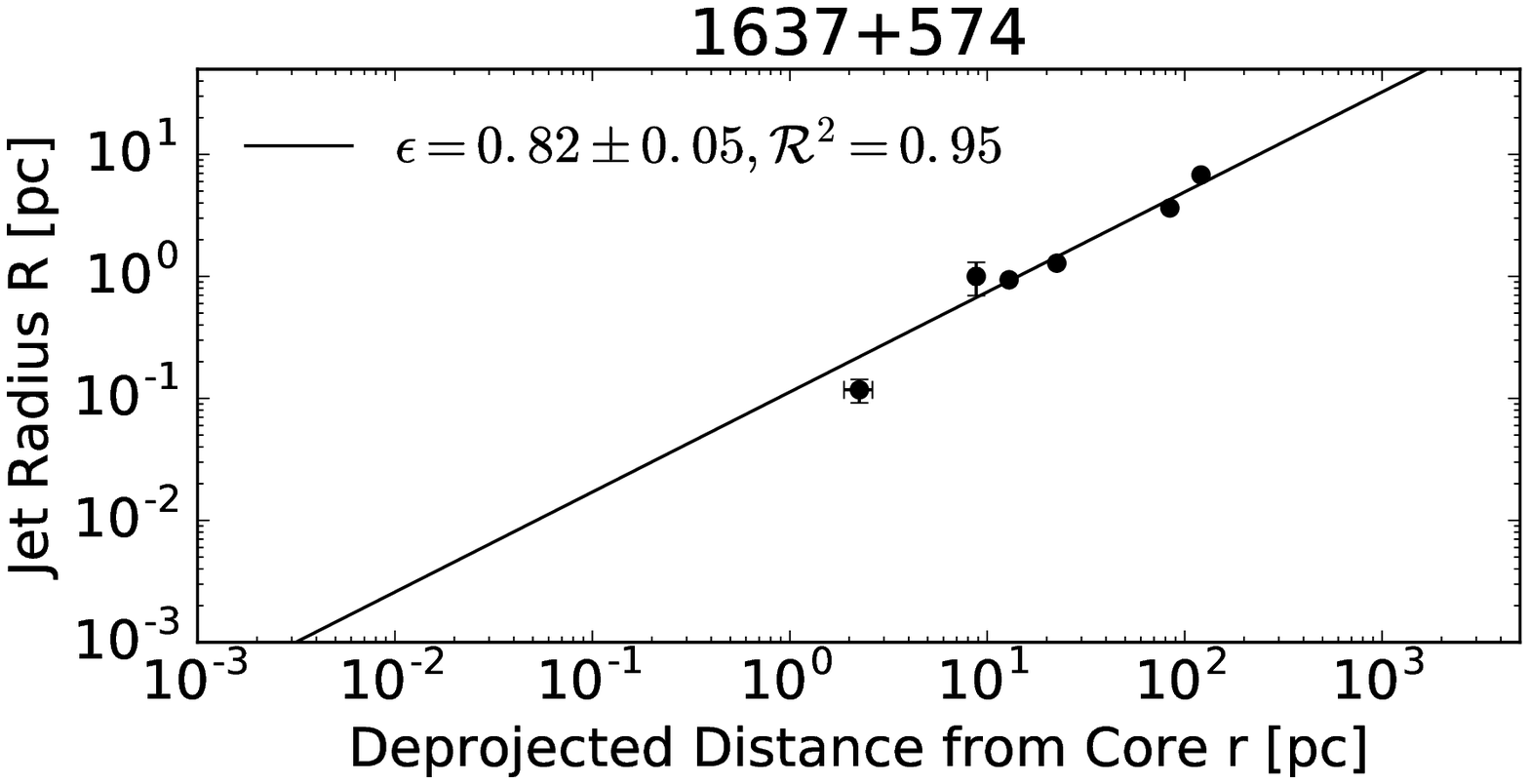}}\
\subfigure{\includegraphics[trim=0cm 0cm 0cm 2.5cm, clip=true,width=0.24\textwidth]{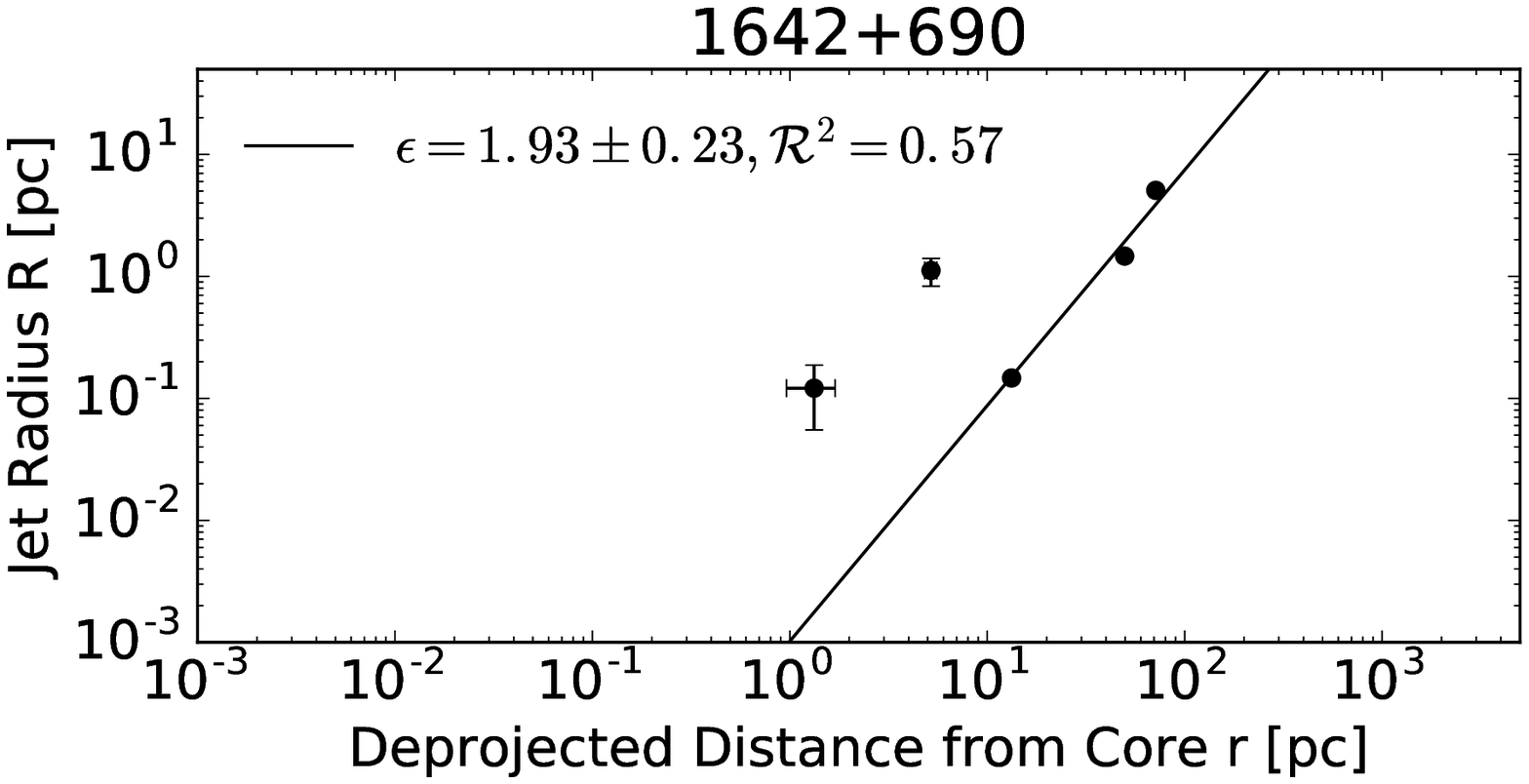}}\
\subfigure{\includegraphics[trim=0cm 0cm 0cm 2.5cm, clip=true,width=0.24\textwidth]{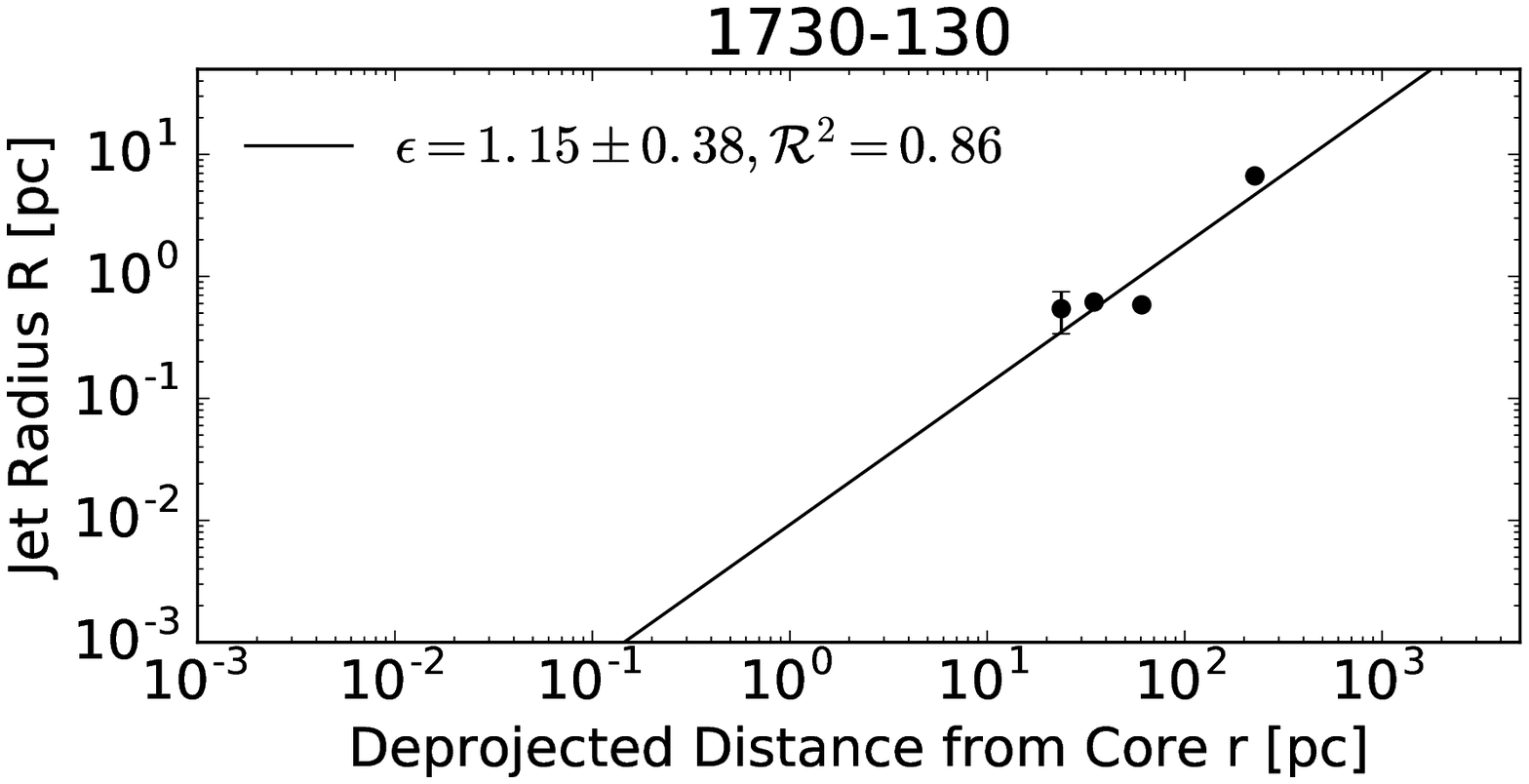}}\
\\
\vspace{-0.8cm} 
\subfigure{\includegraphics[trim=0cm 0cm 0cm 2.5cm, clip=true,width=0.24\textwidth]{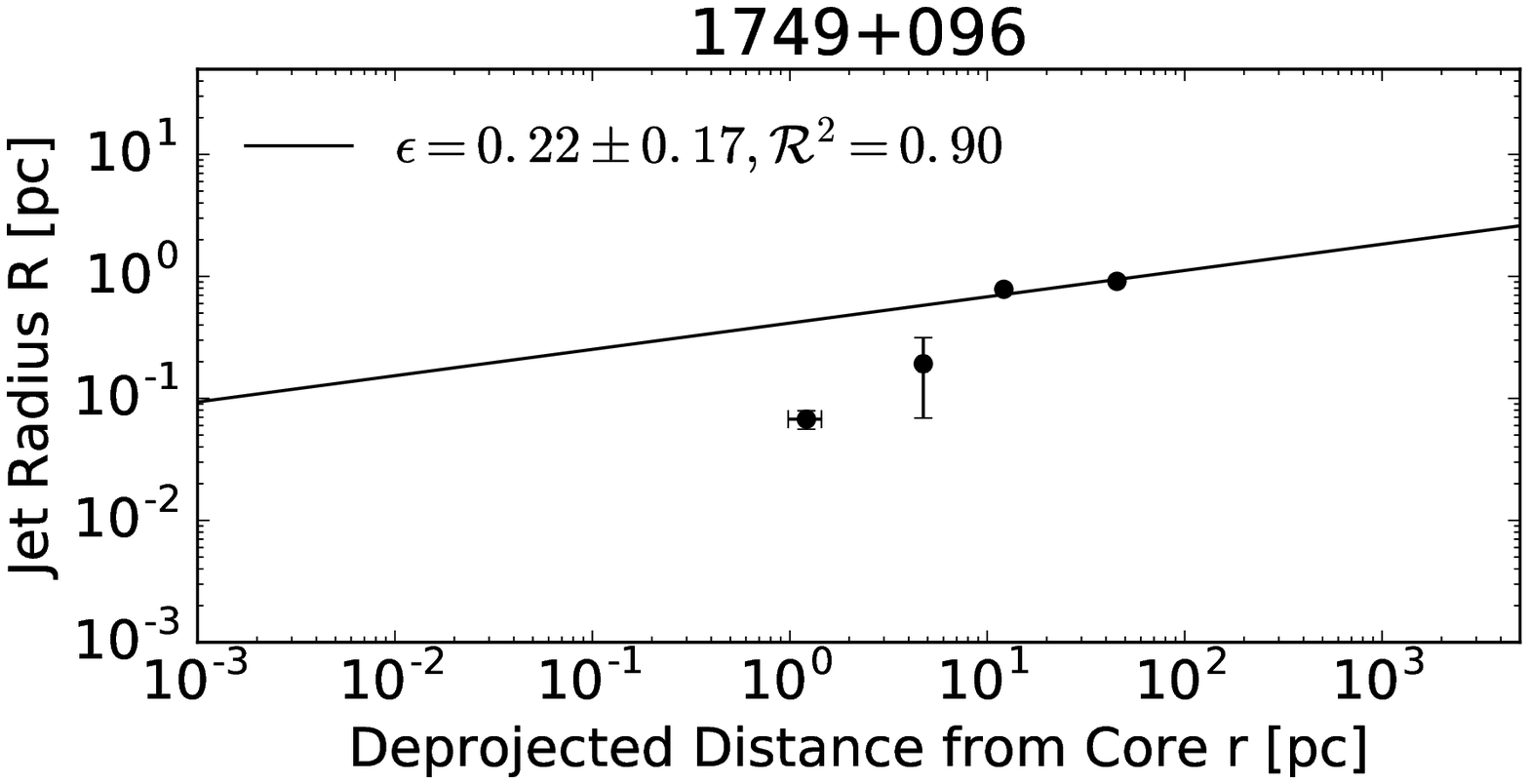}}\
\subfigure{\includegraphics[trim=0cm 0cm 0cm 2.5cm, clip=true,width=0.24\textwidth]{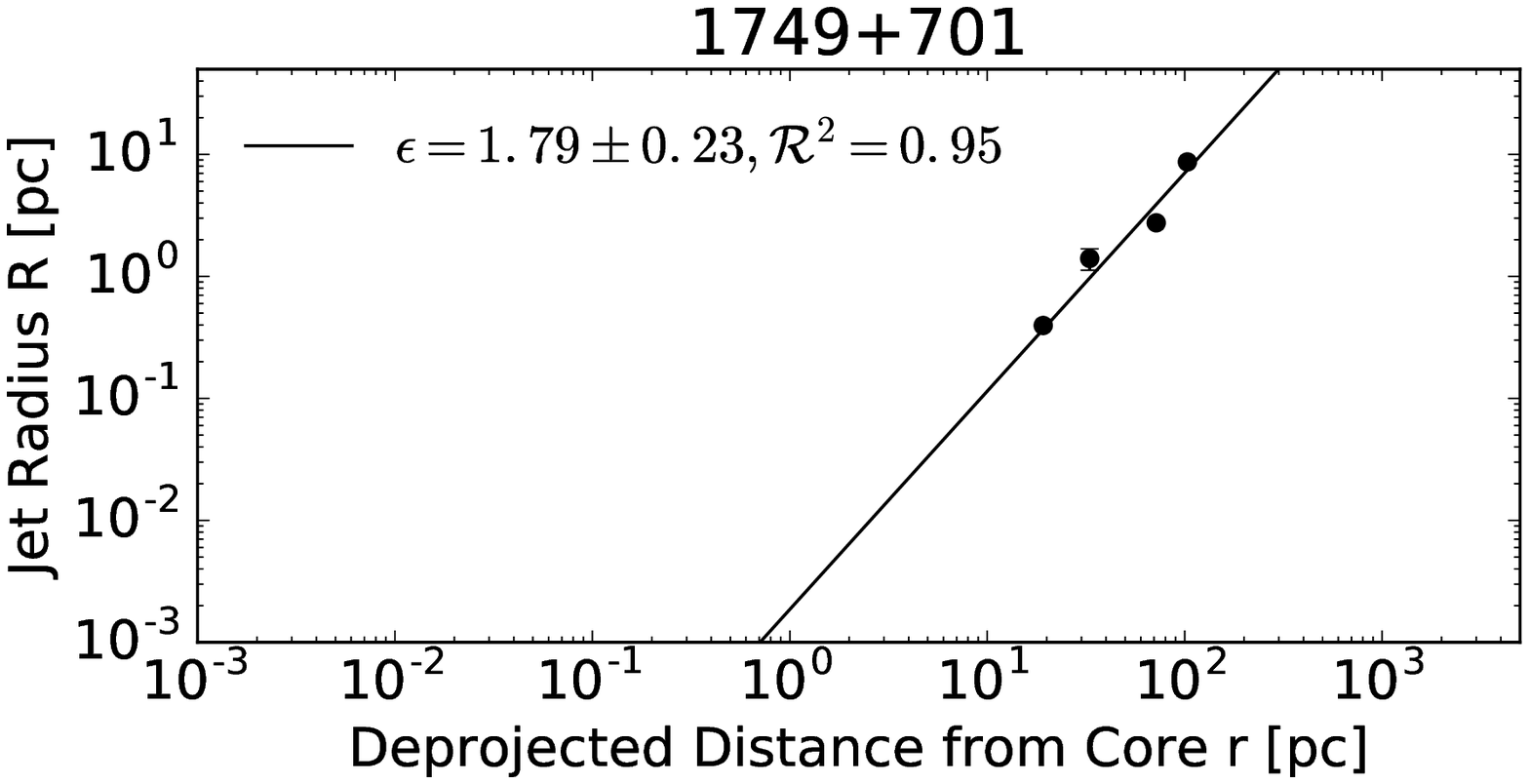}}\
\subfigure{\includegraphics[trim=0cm 0cm 0cm 2.5cm, clip=true,width=0.24\textwidth]{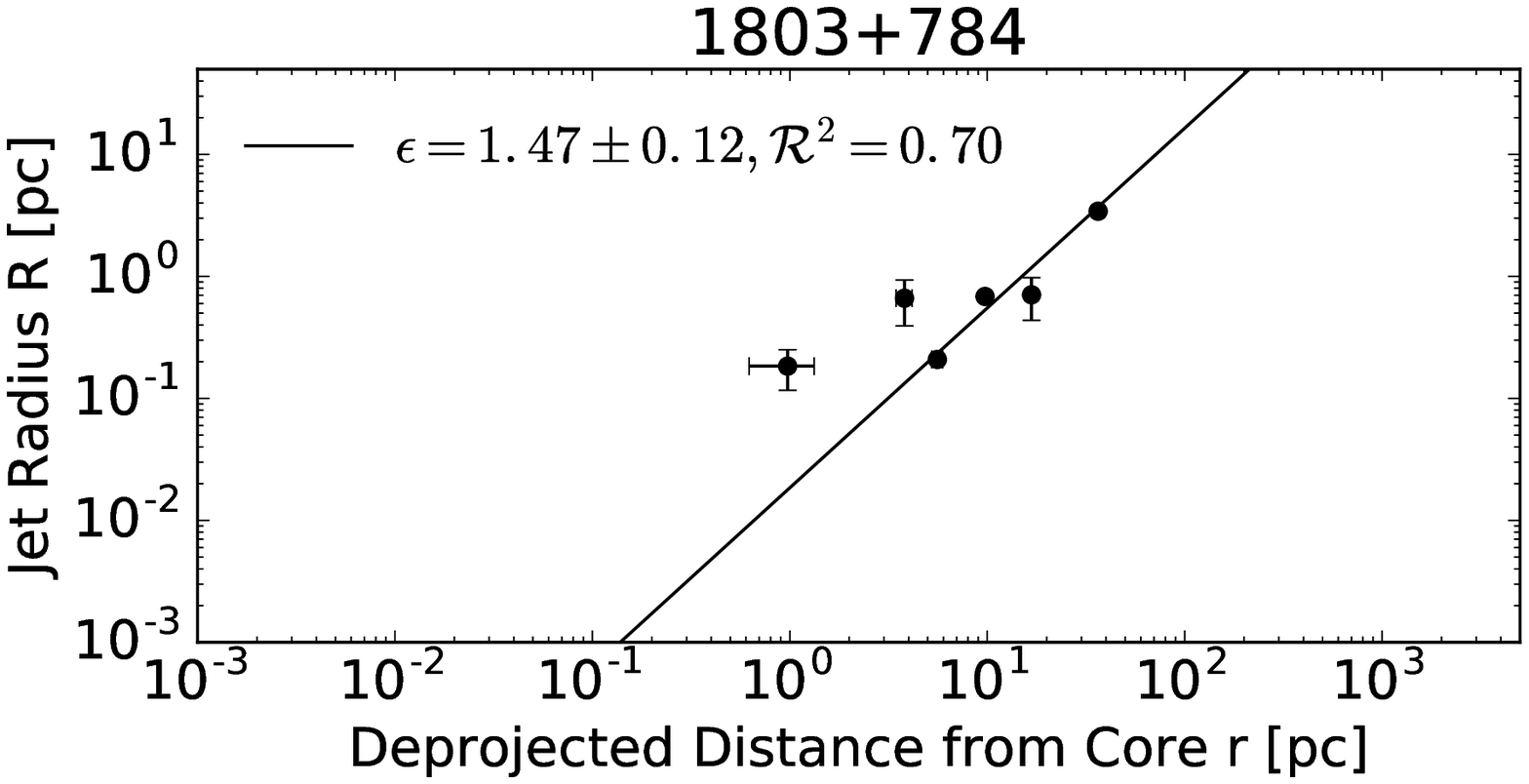}}\
\subfigure{\includegraphics[trim=0cm 0cm 0cm 2.5cm, clip=true,width=0.24\textwidth]{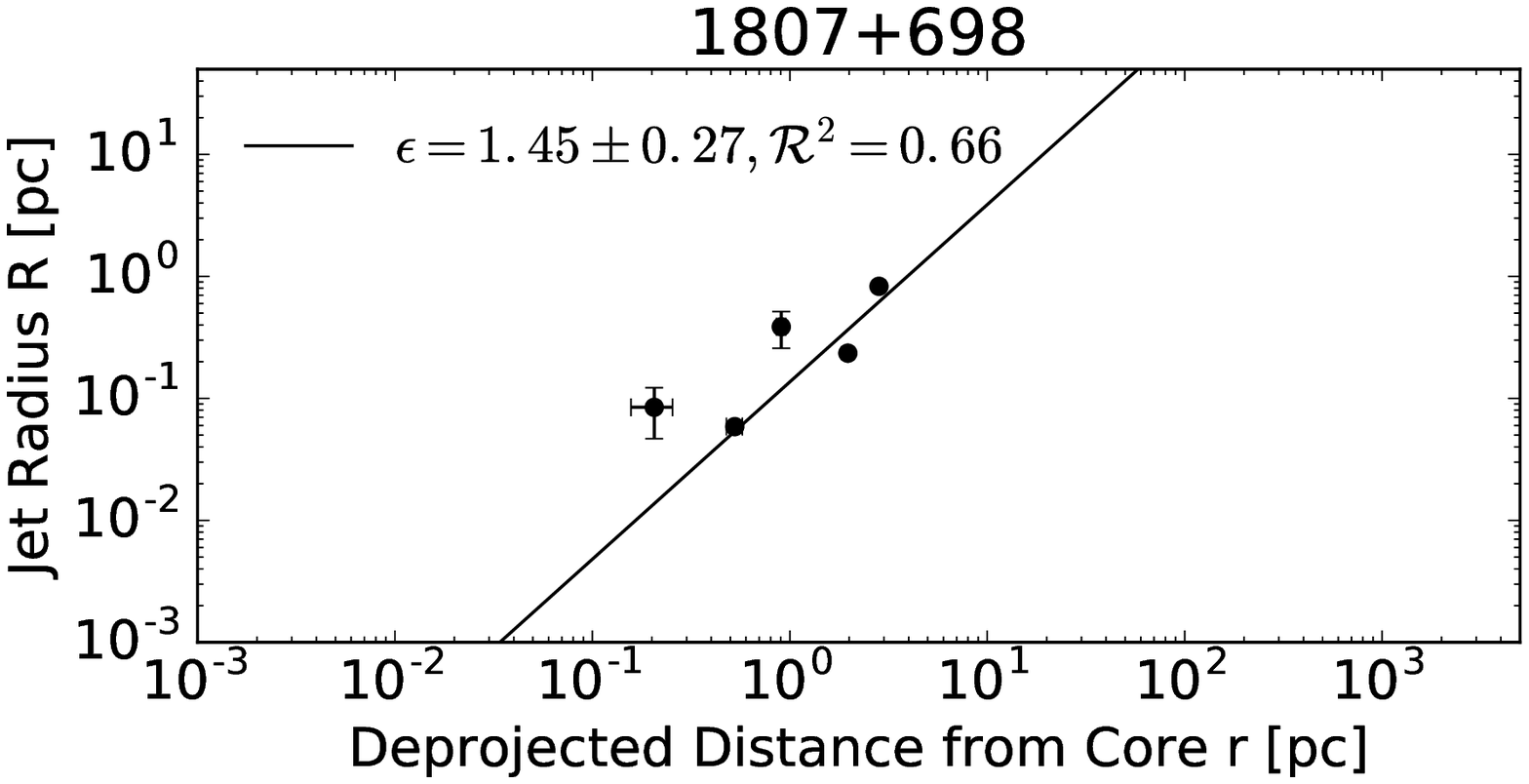}}\

\caption{Core size as a function of distance from the central engine for all sources discussed here. Dots indicate the compiled data from the literature for these objects where values for 4 or mode independent frequencies were found. The black straight line is a fit of the form $R\propto r^\epsilon$ using all the available data points, whereas the red dotted line shows the fit including only selected points (see text). On top of each figure, the source name is shown. The top left corner shows the fitted value for $\epsilon$ and its error (quadratic sum of uncertainties due to $k_r$ not included), together with the goodness of the fit.}
\label{fitsAppendix}
\end{figure*}

\begin{figure*}
\centering
\subfigure{\includegraphics[trim=0cm 0cm 0cm 2.5cm, clip=true,width=0.24\textwidth]{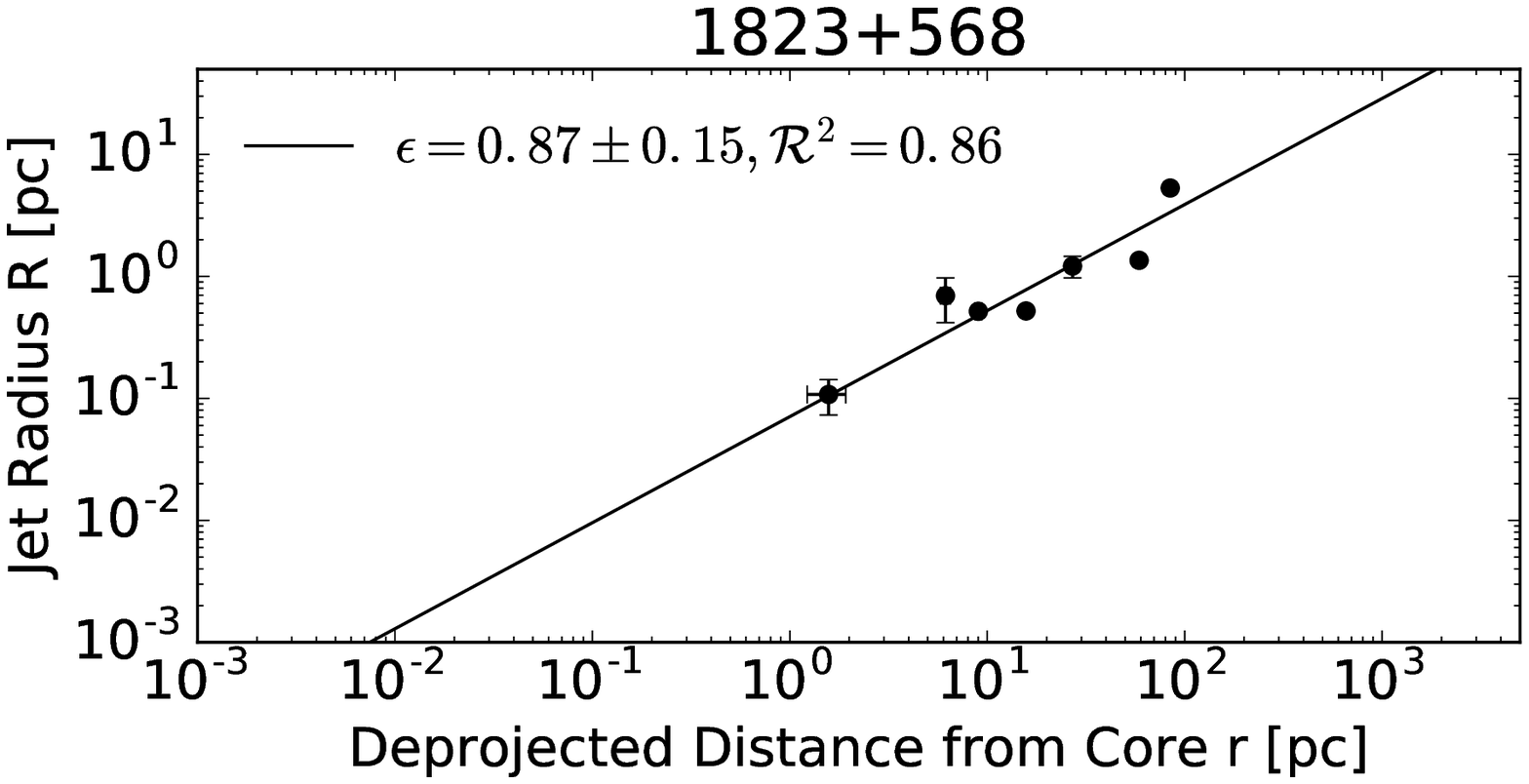}}\
\subfigure{\includegraphics[trim=0cm 0cm 0cm 2.5cm, clip=true,width=0.24\textwidth]{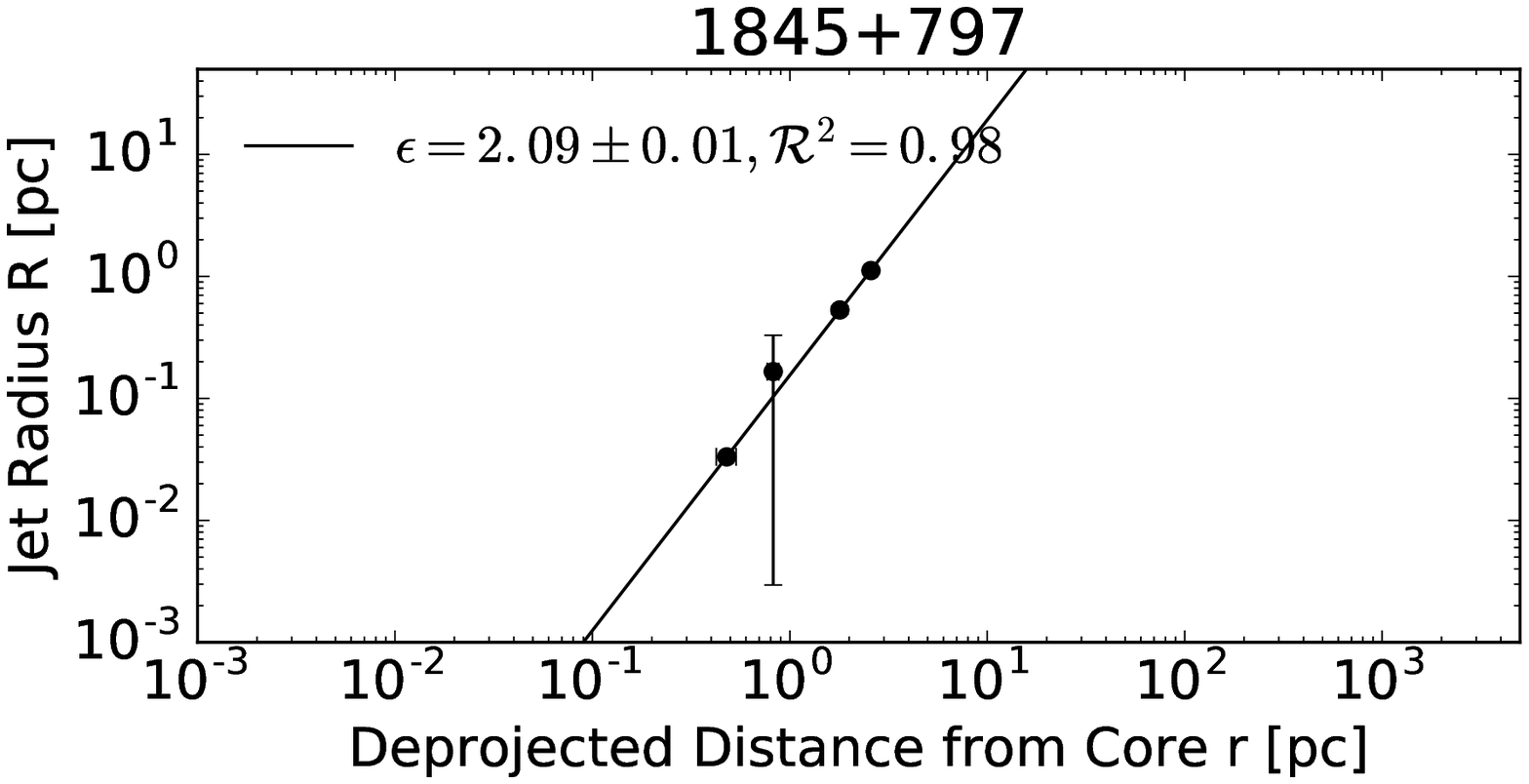}}\
\subfigure{\includegraphics[trim=0cm 0cm 0cm 2.5cm, clip=true,width=0.24\textwidth]{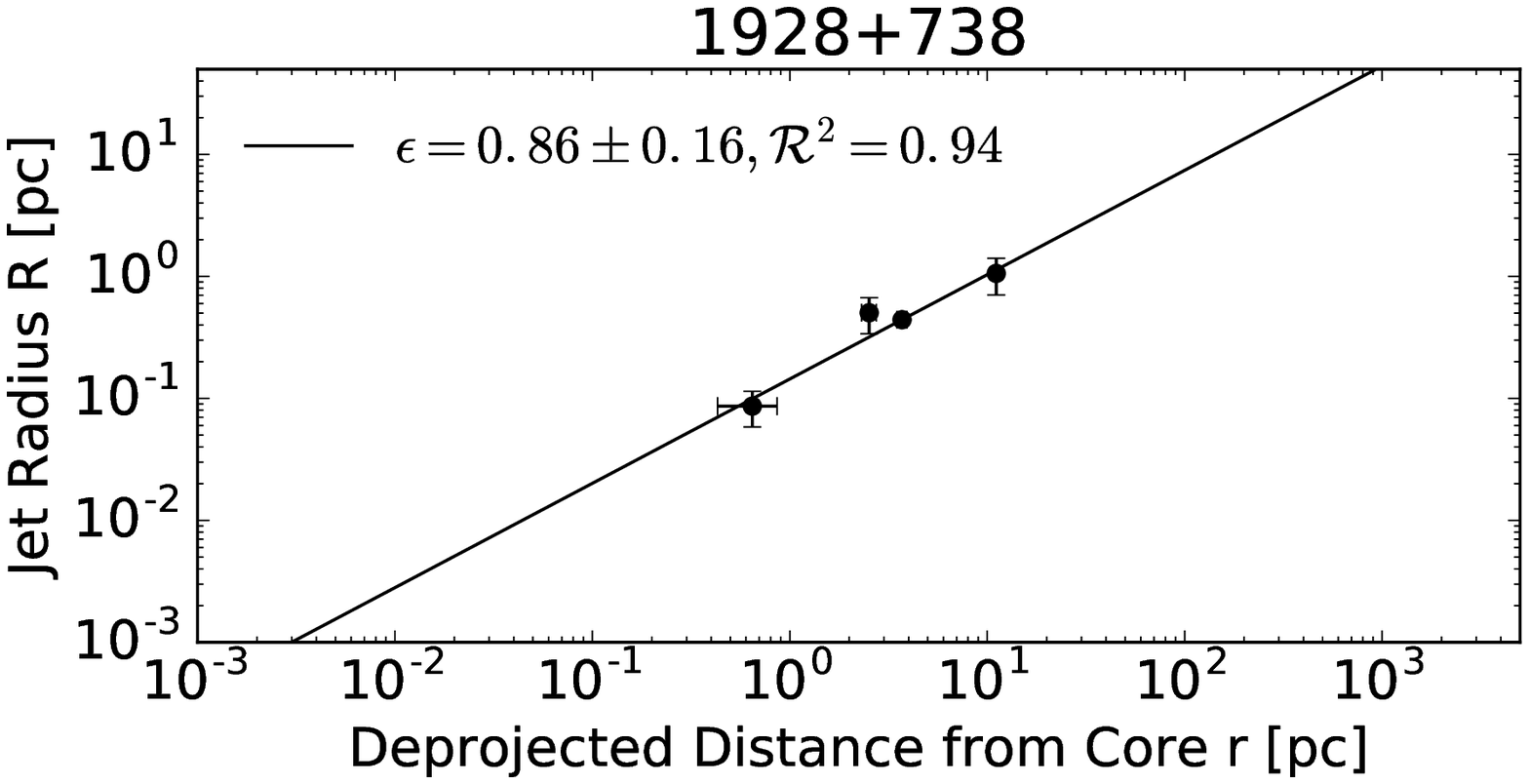}}\
\subfigure{\includegraphics[trim=0cm 0cm 0cm 2.5cm, clip=true,width=0.24\textwidth]{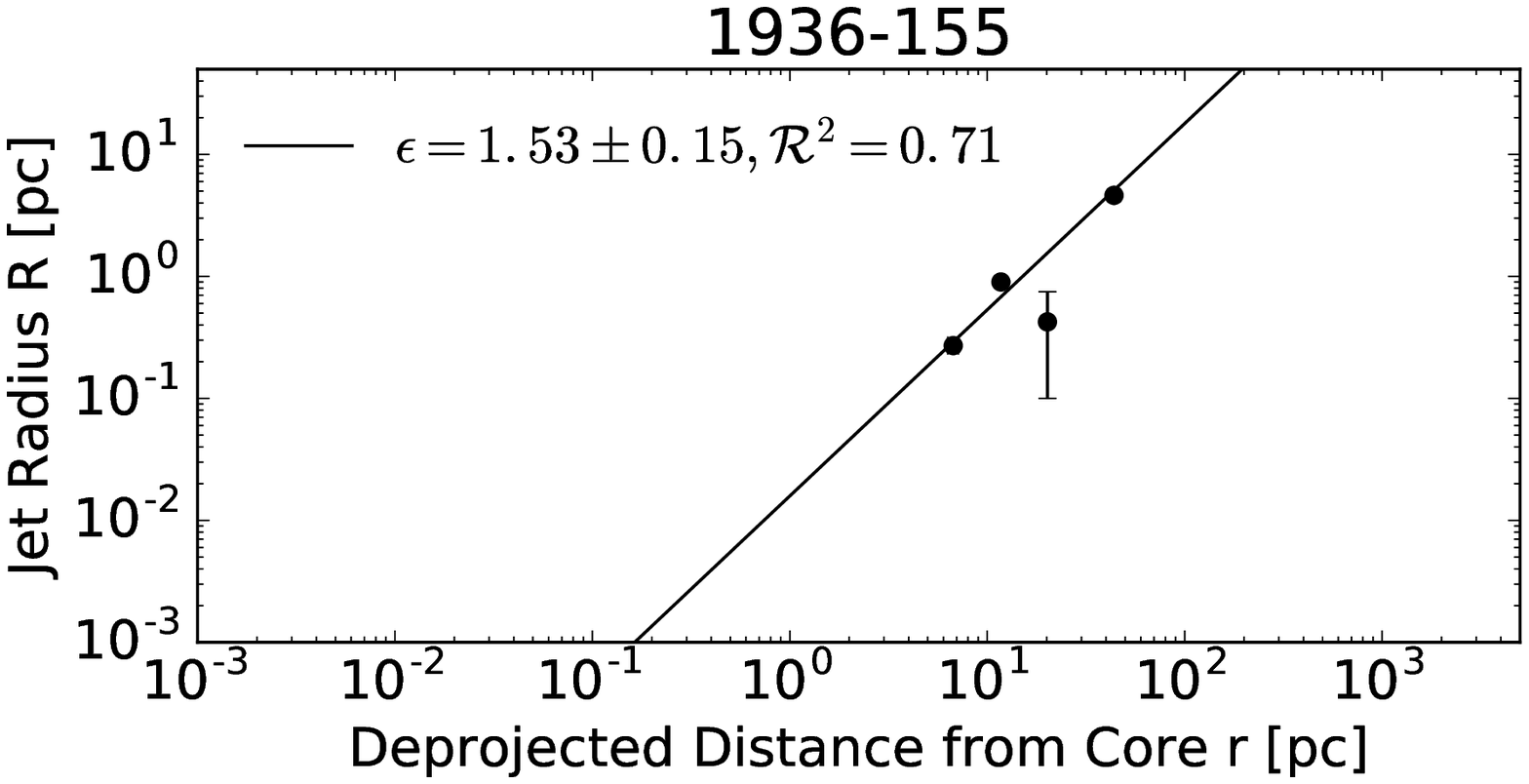}}\
\\
\vspace{-0.8cm} 
\subfigure{\includegraphics[trim=0cm 0cm 0cm 2.5cm, clip=true,width=0.24\textwidth]{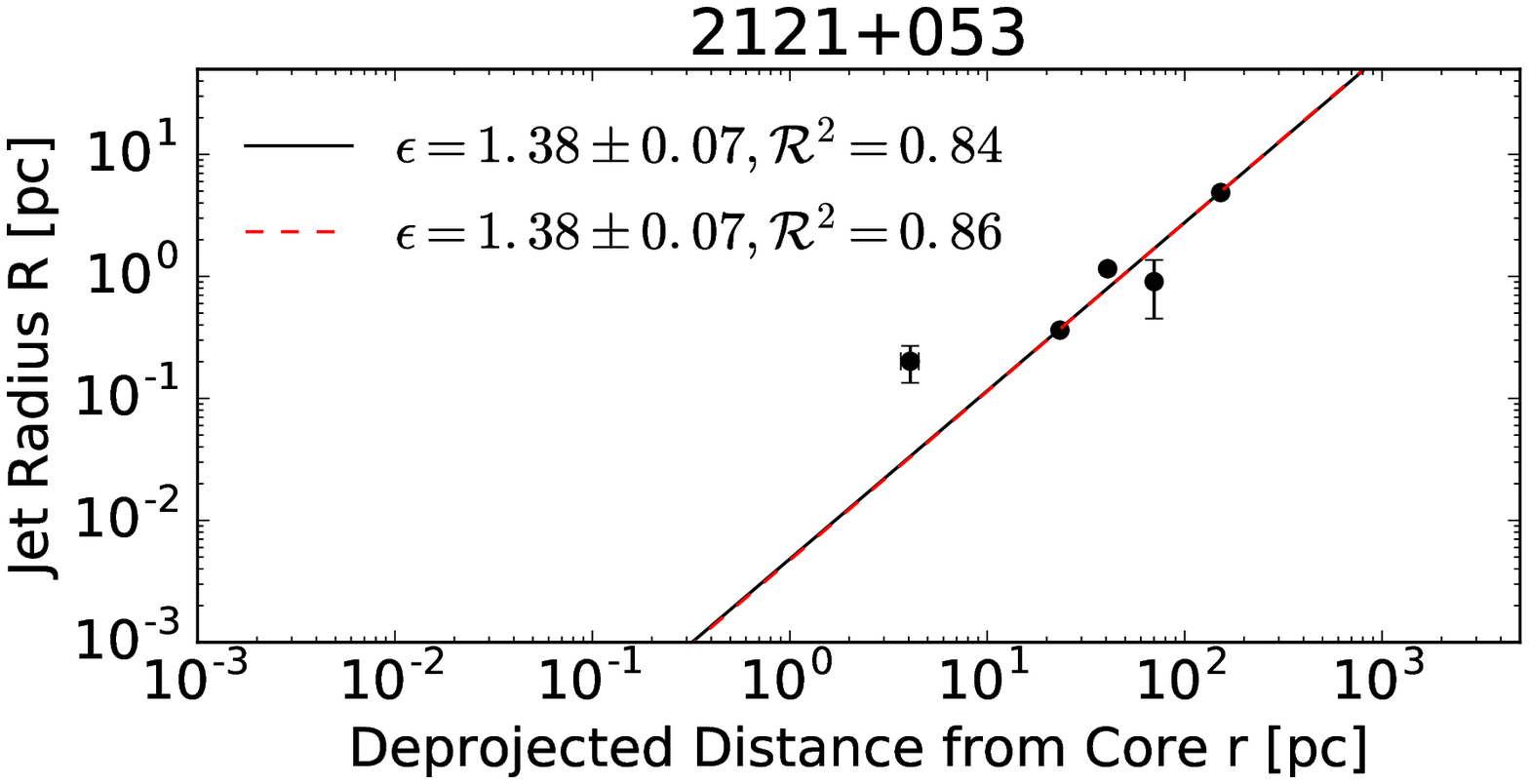}}\
\subfigure{\includegraphics[trim=0cm 0cm 0cm 2.5cm, clip=true,width=0.24\textwidth]{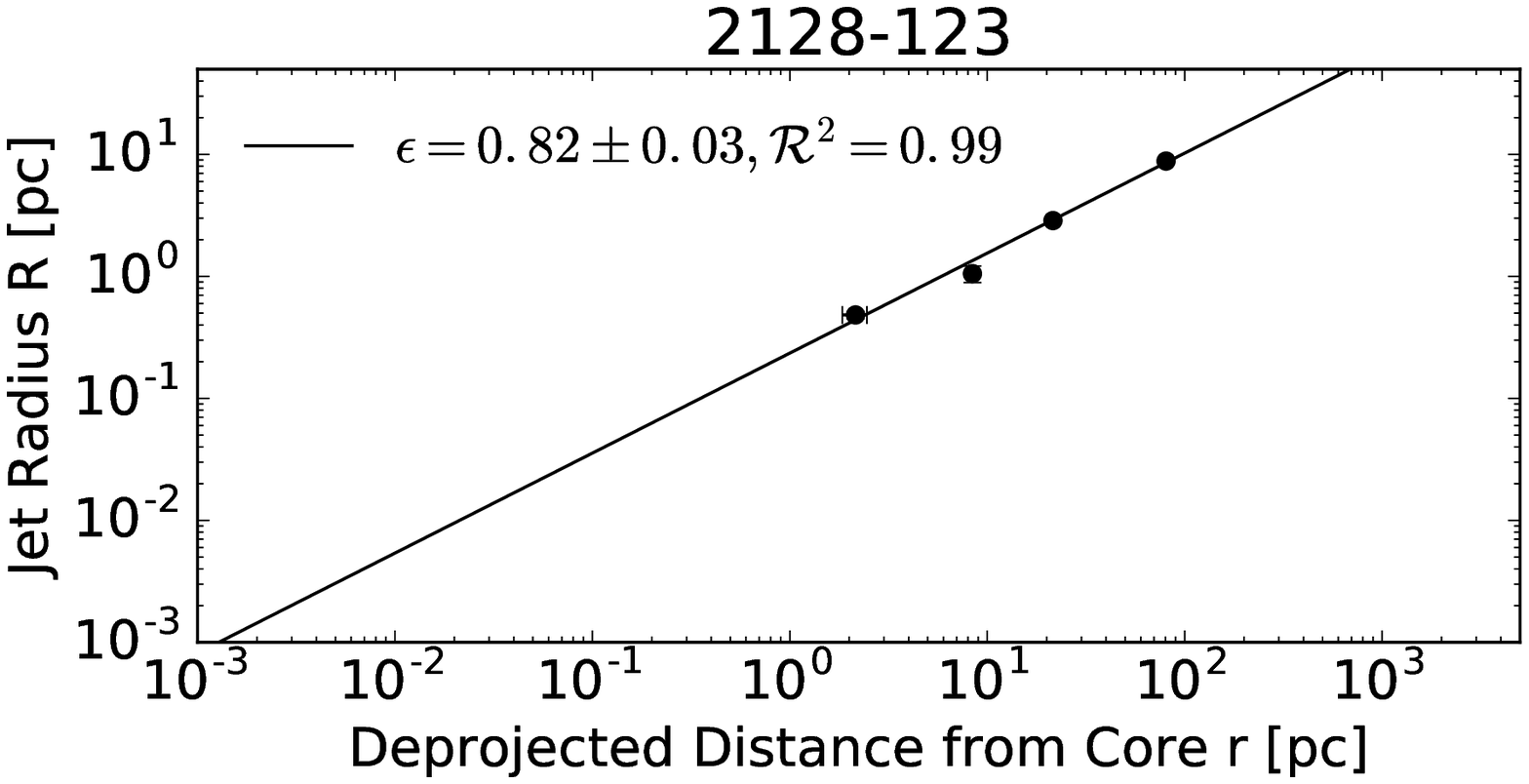}}\
\subfigure{\includegraphics[trim=0cm 0cm 0cm 2.5cm, clip=true,width=0.24\textwidth]{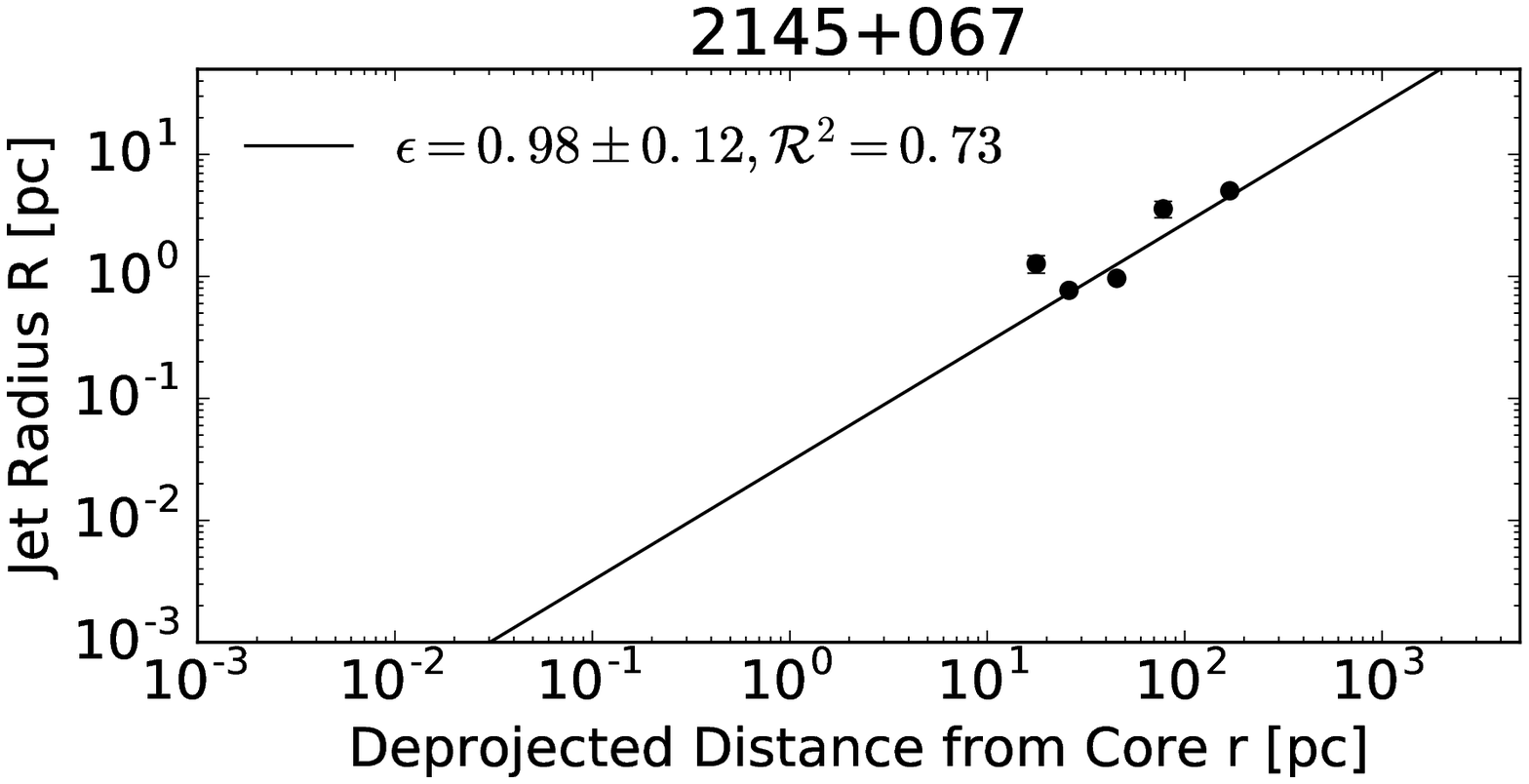}}\
\subfigure{\includegraphics[trim=0cm 0cm 0cm 2.5cm, clip=true,width=0.24\textwidth]{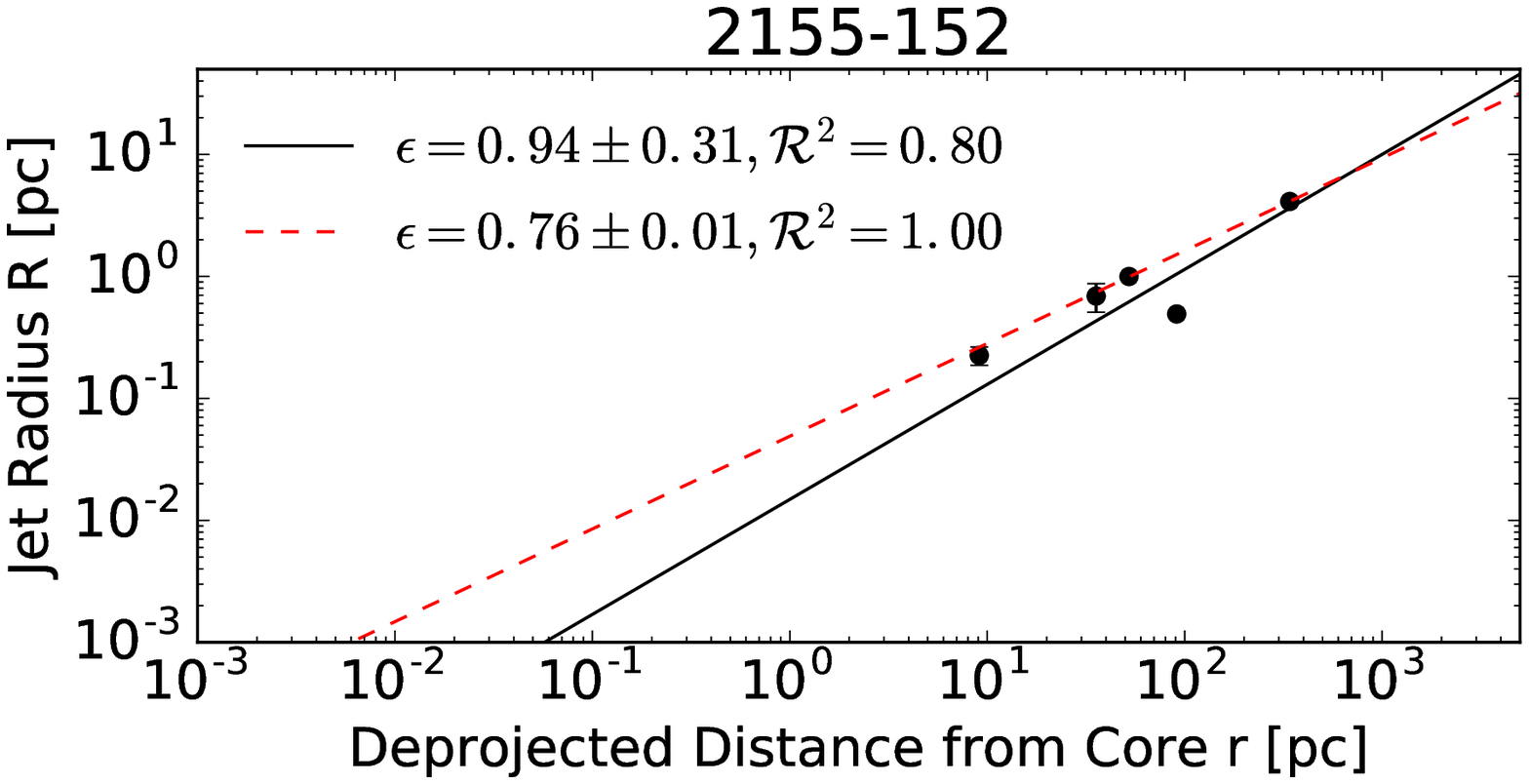}}\
\\
\vspace{-0.8cm} 
\subfigure{\includegraphics[trim=0cm 0cm 0cm 2.5cm, clip=true,width=0.24\textwidth]{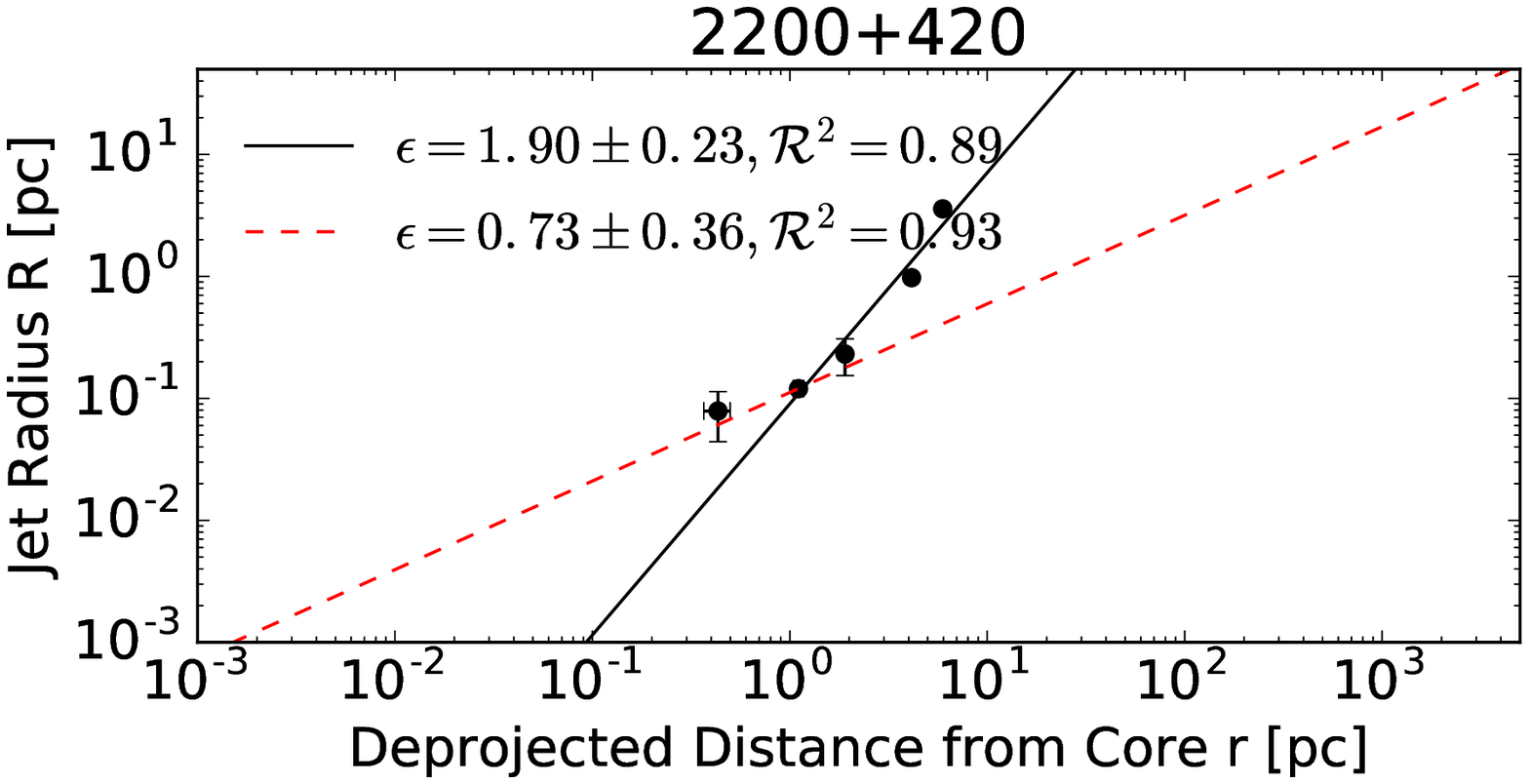}}\
\subfigure{\includegraphics[trim=0cm 0cm 0cm 2.5cm, clip=true,width=0.24\textwidth]{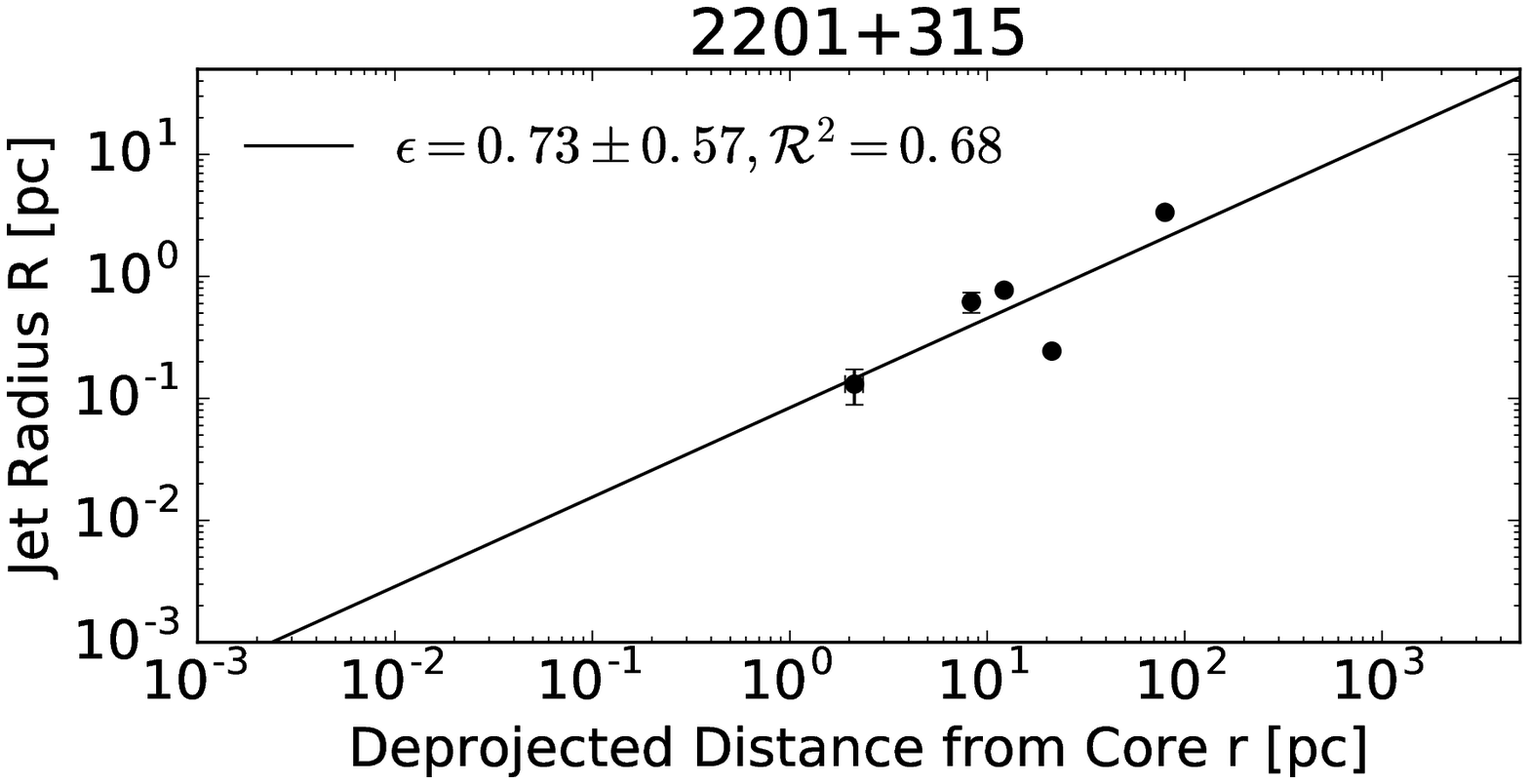}}\
\subfigure{\includegraphics[trim=0cm 0cm 0cm 2.5cm, clip=true,width=0.24\textwidth]{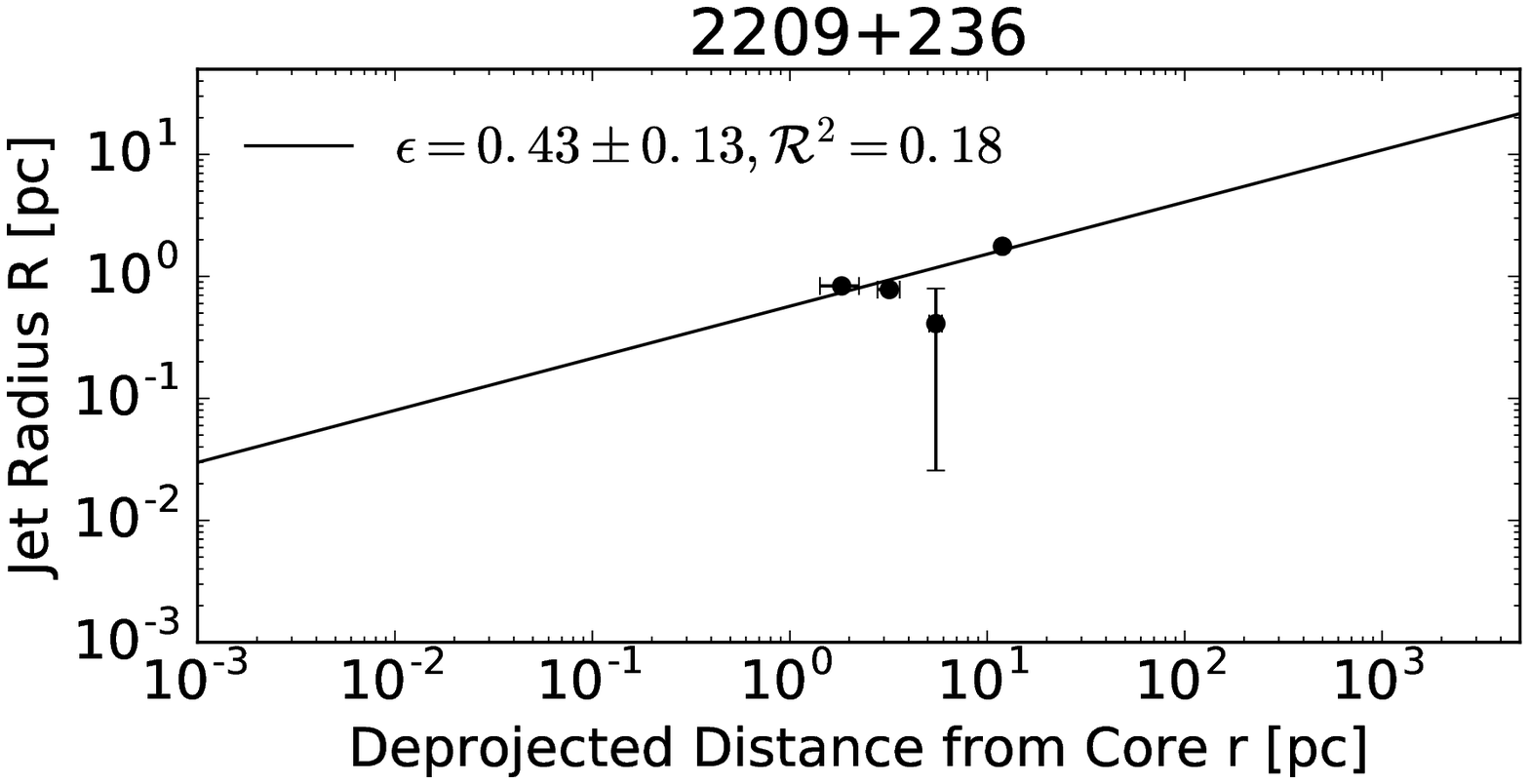}}\
\subfigure{\includegraphics[trim=0cm 0cm 0cm 2.5cm, clip=true,width=0.24\textwidth]{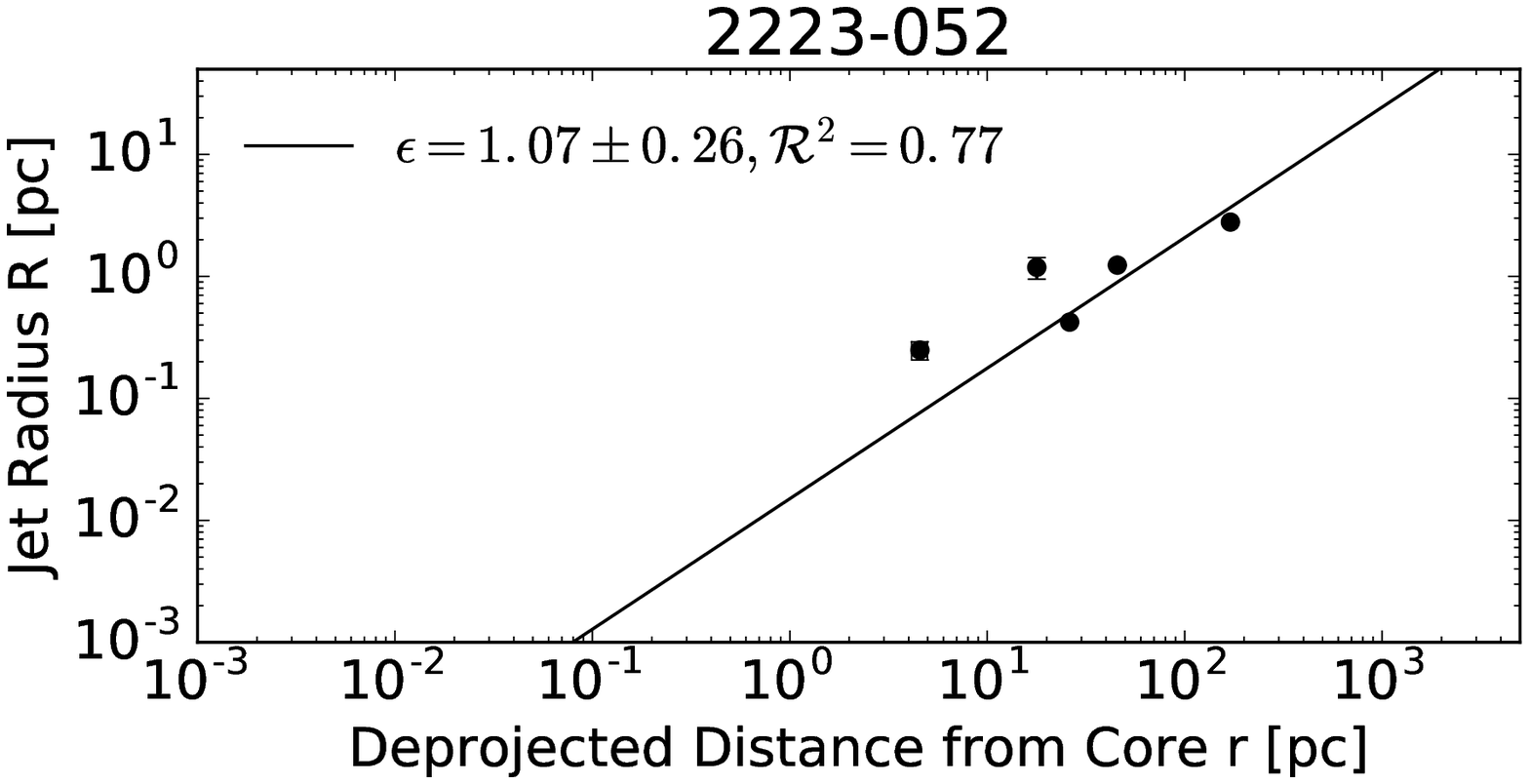}}\
\\
\vspace{-0.8cm} 
\subfigure{\includegraphics[trim=0cm 0cm 0cm 2.5cm, clip=true,width=0.24\textwidth]{fig52.eps}}\
\subfigure{\includegraphics[trim=0cm 0cm 0cm 2.5cm, clip=true,width=0.24\textwidth]{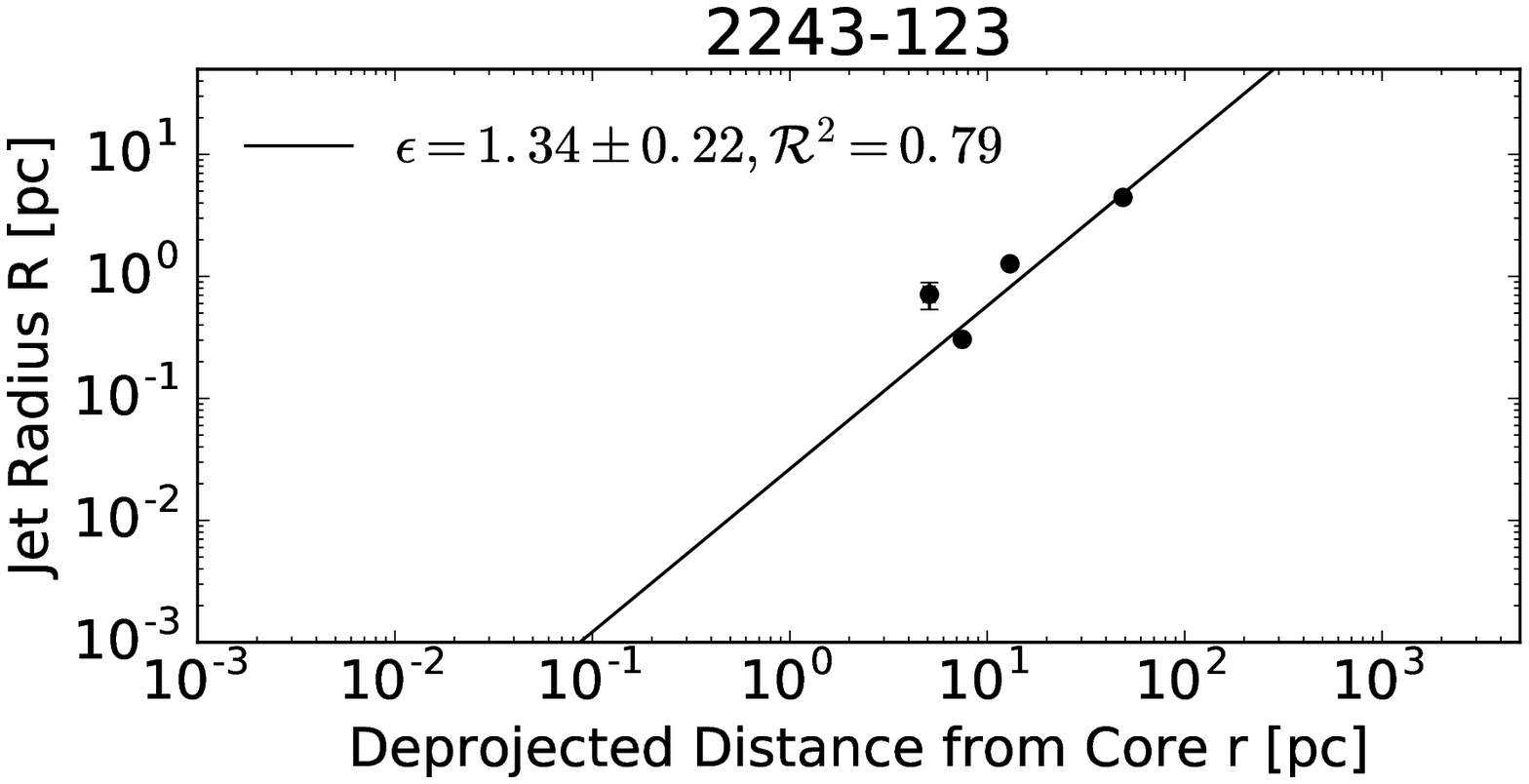}}\
\subfigure{\includegraphics[trim=0cm 0cm 0cm 2.5cm, clip=true,width=0.24\textwidth]{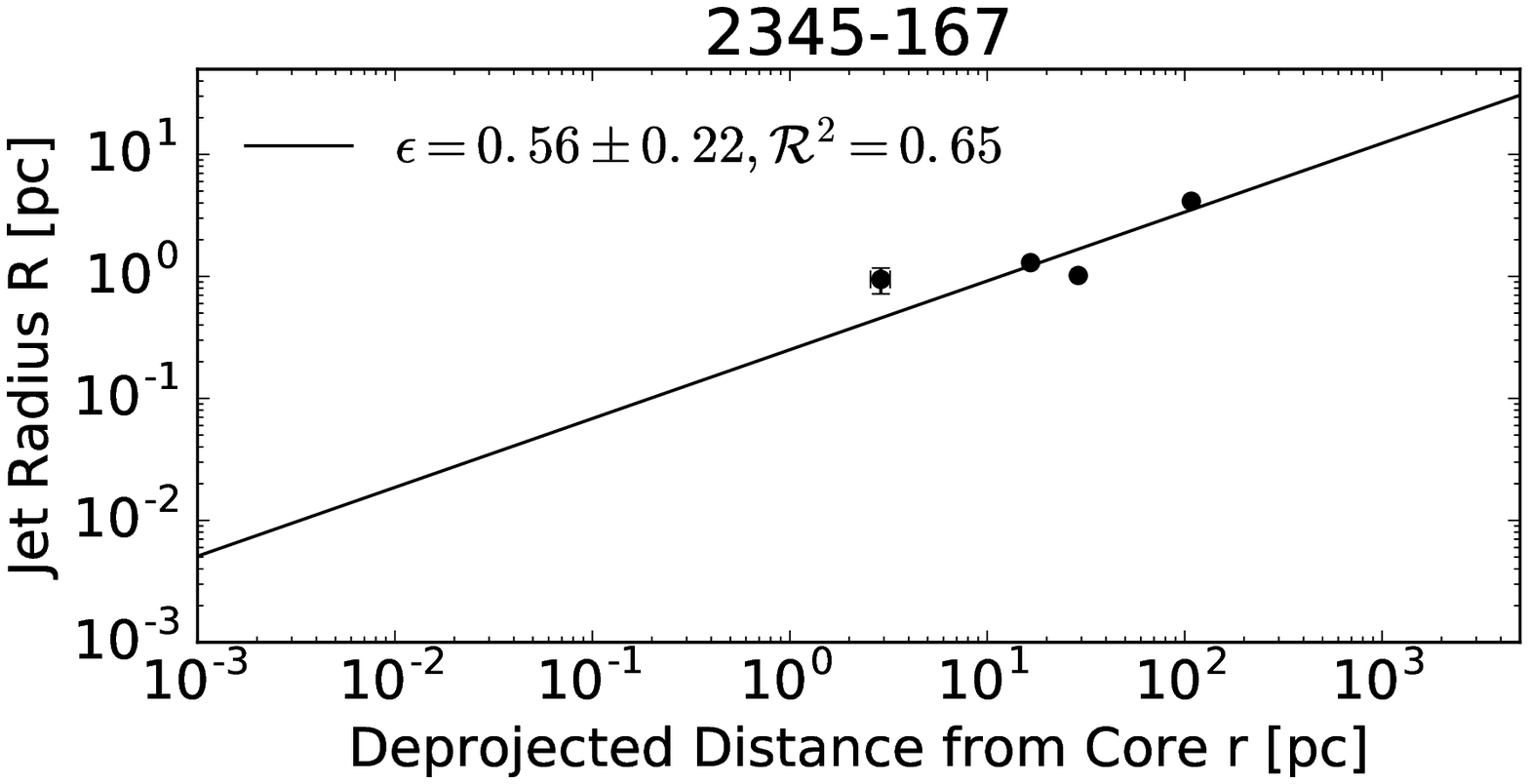}}\
\subfigure{\includegraphics[trim=0cm 0cm 0cm 2.5cm, clip=true,width=0.24\textwidth]{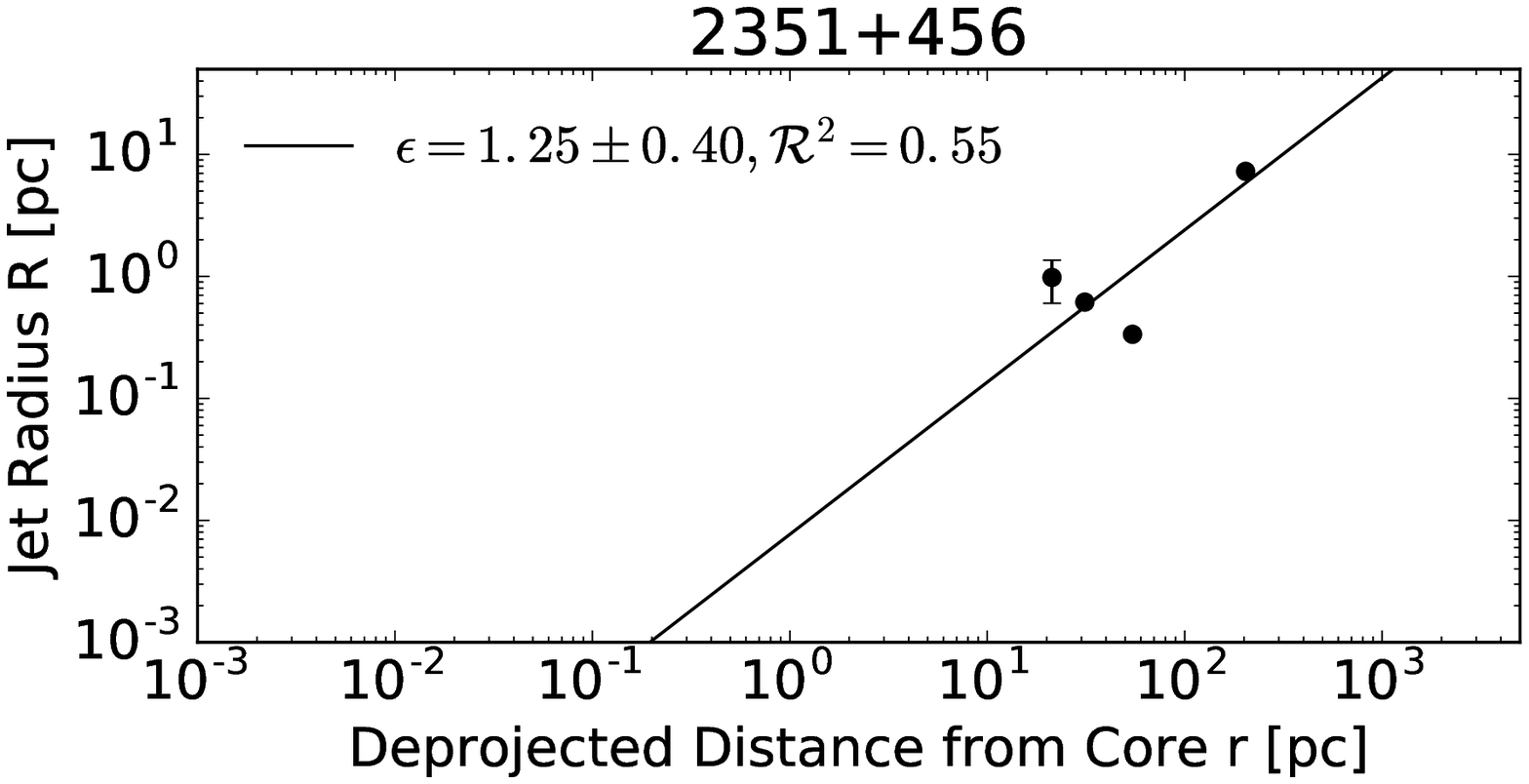}}\
\flushleft{\textsc{Fig. \ref{fitsAppendix}}. --- Continued}

\end{figure*}


\end{document}